\documentclass[preprint,12pt]{elsarticle}
\makeatletter
\def\ps@pprintTitle{%
 \let\@oddhead\@empty
 \let\@evenhead\@empty
 \def\@oddfoot{\footnotesize\itshape Tomislav Mari\'{c}, preprint, 27 October 2026}%
 \let\@evenfoot\@oddfoot}
\makeatother
\usepackage[a4paper,left=25mm,right=25mm,top=25mm,bottom=25mm]{geometry}
\usepackage{amsmath}
\usepackage{amssymb}
\usepackage{graphicx}
\usepackage{booktabs}
\usepackage{array}
\usepackage[labelfont=bf,font=small]{caption}
\usepackage{subcaption}
\usepackage{xcolor}
\usepackage{tikz}
\usetikzlibrary{arrows.meta,calc}
\tikzset{wetted/.style={draw=black!60, double=yellow!85!orange, double distance=2.4pt,
                        line width=0.9pt, line cap=round, line join=round}}
\usepackage{pgfplots}
\pgfplotsset{compat=1.18}
\usepgfplotslibrary{groupplots}

\usepackage[hidelinks]{hyperref}
\usepackage{xurl}
\usepackage{cleveref}
\allowdisplaybreaks
\graphicspath{{./}}

\definecolor{domaincol}{RGB}{90,90,90}
\definecolor{ifacecol}{RGB}{200,40,40}
\definecolor{facecol}{RGB}{230,120,30}
\definecolor{wettedcol}{RGB}{200,150,0}
\definecolor{framecol}{RGB}{60,60,60}

\newcommand{\x}{\mathbf{x}}
\newcommand{\n}{\mathbf{n}}
\newcommand{\Sf}{\mathbf{S}_f}
\newcommand{\Sfhat}{\hat{S}_f}
\newcommand{\nSk}{\mathbf{n}_{\tilde{\Sigma}_k}}
\newcommand{\ASk}{A_{\tilde{\Sigma}_k}}

\begin{document}
\frenchspacing

\begin{frontmatter}

\title{Dual Area--Volume-of-Fluid (DAVOF): co-advecting volume and face-area fractions for a closed-form, cell-local interface normal and position on unstructured meshes}

\author[mma]{Tomislav Mari\'{c}\corref{corr}}
\cortext[corr]{Corresponding author}
\ead{maric@mma.tu-darmstadt.de}
\address[mma]{Mathematical Modeling and Analysis, Department of Mathematics, Technical University of Darmstadt, Peter-Gr\"unberg-Stra{\ss}e 10, 64287 Darmstadt, Germany}

\begin{abstract}
Geometric Volume-of-Fluid (VOF) on unstructured polyhedral meshes reconstructs the interface orientation from the volume fractions of a stencil of neighbours; on irregular cells and in the presence of small numerical errors (wisps) it degrades, and the curvature built on it drives the parasitic currents of capillary flows. We transport, consistently to the volume fraction of every cell, the wetted fraction of every face area. From the area and volume fractions, closed-form equations recover the interface normal and position of a cell, for the first time in the context of VOF methods: the normal from one sum over the cell's own faces, the divergence theorem on the liquid sub-volume, and the plane's position from a second, without neighbours or iteration. Both identities are exact for a plane on hexahedral, perturbed and polyhedral cells. On a sphere and an ellipsoid, on all three mesh classes, the normal has the order of the face fractions as fractions, and the position is second order for every state and mesh; a quadric through one ring of recovered normals gives a converging curvature where the reference models show none, provided the sample errors are smooth on a fixed physical length. In the map in which every cell rebuilds its clipping surface from its own state, the paraboloid fitted to the face fractions is the identity and an anchored one drifts to first order per period. Transported with the plane clipping, for example using the isoAdvector scheme, the ring normal, the same identity on the union of the cell's ring, stabilises the state and carries a sphere over 3.3 diameters bounded and conservative, first order in the face fractions as the clipping law predicts, at plicRDF's position and shape errors; the stabilisation of the cell's own normal in the loop is a research goal of the method. A preliminary transport implementation with one paraboloid per cell, its frame from the normal, its curvature tensor from the ring and its position from the volume fraction, carries the sphere bounded and conservative over the resolution ladder with face fractions of order 2.8 per period in the frame of the exact normal and 2.1 with the ring quadric's normal over $N = 32$ to $128$, the highest orders reached so far. The loop has the order of the normal it is fed, so the research goal these measurements set is a transported normal of second order under refinement; the ring-state fit stated in the design is its first candidate. This document is the basis on which the further development of the method is documented.
\end{abstract}


\end{frontmatter}

\section{Introduction}
\label{sec:intro}

Capillary microfluidic devices, such as Lab-on-a-Chip diagnostic systems, inkjet print heads and fuel cells, depend on a liquid--gas interface that advances through narrow and geometrically complex channels, where surface tension dominates over gravity and inertia. Whether a cavity fills, a droplet forms at the desired place or a meniscus pins at an edge is decided by the shape and the motion of that interface. Simulation is the primary design tool for these devices, because the interface cannot be observed inside the device in three dimensions and parametric experiments are slow and costly. The task for the method developer is to make these simulations more accurate and more reliable, with fewer parameters that must be tuned to a family of problems.

On unstructured polyhedral meshes, the natural discretization for complex three-dimensional geometries, the standard tool for tracking the interface is the geometric Volume-of-Fluid (VOF) method \citep{MaricKotheBothe2020}. Every cell stores the fraction of its volume filled by liquid, the volume fraction $\alpha_k$. In every cell that contains a piece of the interface, the interface is represented by one plane, the piecewise-linear interface calculation (PLIC) of \citet{Youngs1982} and \citet{RiderKothe1998}, and the volume of liquid fluxed through each face in a time step is cut from that plane: the unsplit geometric VOF (UGVOF) of \citet{MaricMarschallBothe2018}, the algebraic schemes \citep{Xie2017,Qian2018} and the hybrid isoAdvector--plicRDF method \citep{RoenbyEtAl2016,ScheuflerRoenby2019,Dianat2017}, which has become the standard in OpenFOAM. The PLIC plane needs an orientation, the interface normal, and the volume fraction of one cell does not carry it. Every VOF method therefore reconstructs the normal from the volume fractions of a neighbourhood: a gradient \citep{Youngs1982} or a stencil minimisation, LVIRA \citep{Puckett1991}, ELVIRA \citep{PilliodPuckett2004} and plicRDF \citep{ScheuflerRoenby2019}, iterated in the better schemes. On Cartesian meshes these reconstructions are second order; on irregular polyhedral cells they degrade, and the gradient-based ones do not converge at all \citep{PilliodPuckett2004}. The curvature, differenced from the reconstructed normals, is one order worse, and the surface-tension force built on it produces parasitic currents, a numerical flow that does not vanish under refinement. In the Lab-on-a-Chip filling study of \citet{NagelEtAl2024,NagelEtAl2025MicrocavityFilling} with the isoAdvector--plicRDF method, these currents forced a ten-fold artificially increased viscosity and left the lowest capillary numbers unsimulated.

The issues above comprise three open problems: (1) a cell-local recovery of the interface orientation that does not degrade on irregular cells; (2) a converging curvature from the state on such cells; (3) an accuracy that is kept under transport, not only in a static reconstruction. The method proposed here, the dual area--volume-of-fluid (DAVOF) state (\cref{fig:coreideas}), builds on the following observation. A VOF method already computes, in its flux step, the wetted part of every face area at the new time; isoAdvector does so explicitly \citep{RoenbyEtAl2016} and discards it. If the wetted fraction of the face area, $\alpha_f$, is kept as part of the transported state, then the wetted parts of the faces of a cell enclose, together with the interface piece, the liquid sub-volume of the cell, and the divergence theorem on that sub-volume gives the area vector of the interface piece as minus the sum of the wetted face area vectors: one sum over the cell's own faces, without data from neighbouring cells. A second application of the divergence theorem gives the position of the interface plane in closed form from $\alpha_k$ and $\{\alpha_f\}$. The identity itself is the aperture relation of cut-cell embedded-boundary methods for static boundaries \citep{JohansenColella1998}; no VOF method has carried the wetted face areas as a transported state, which is what makes the recovery cell-local in the time step.

This differs from moment-of-fluid (MOF) methods \citep{Dyadechko2008,Jemison2013,SpainhourShashkov2025}, which transport the liquid centroid of the cell but no face information; from face-based interface positioning on polyhedral cells \citep{KromerBothe2022}, which assumes the normal as given; from isoAdvector- and plicRDF-type methods \citep{Dianat2017,RoenbyEtAl2016,ScheuflerRoenby2019,AsgharEtAl2025Wetting}, which use area information only during the flux evaluation; and from improved but still nonlocal normal and curvature reconstructions on polyhedra \citep{IveyMoin2015,KromerLeottaBothe2023}. Per cell, classical UGVOF and isoAdvector/plicRDF transport one value and recover the normal from a stencil, iteratively in plicRDF; MOF-type methods transport four values and recover the normal in the cell; DAVOF transports $1 + F/2$ on average with planar faces, $F$ the number of faces, and recovers the normal in one closed-form sum. The transport of MOF-type and DAVOF alike draws on the neighbouring cells.

The paper is organised as follows. \Cref{sec:state} defines the state and derives the two identities, their exactness and their sensitivity, including non-planar faces. \Cref{sec:order} gives the two estimates that relate the order of the face fractions to the order of the recovered normal. \Cref{sec:static} reports the static recovery on a plane, a sphere and an ellipsoid against the PLIC reconstructions of OpenFOAM's geometric VOF library, and the curvature from one ring of recovered normals. \Cref{sec:onering} asks, with a synthetic study, when a one-ring curvature can converge. \Cref{sec:regen} measures the regeneration map, in which each cell rebuilds its clipping surface from its own state and cuts its faces again, once and iterated, and identifies its fixed points. \Cref{sec:transport} states the design of the face update that these measurements support and the open questions. \Cref{sec:concl} concludes. Everything reported here was computed with the open leia OpenFOAM module \citep{LeiaRepo2026}; the studies, their configurations and every table are archived with the paper \citep{LeiaRepo2026} (data and code availability statement at the end).

\section{The dual area--volume-of-fluid state}
\label{sec:state}

\begin{figure}[htbp]
\centering
\begin{tikzpicture}[x=1.5cm,y=1.5cm,>=Stealth,font=\scriptsize]
  \colorlet{fluidcol}{blue!18}
  \colorlet{gascol}{gray!11}
  \colorlet{ifacecol}{blue!65!black}
  \colorlet{normcol}{red!65!black}
  \colorlet{wettedcol}{orange!80!black}
  \colorlet{gasbdrycol}{teal!58!black}
  \colorlet{bdrycol}{black!50}
  \colorlet{sfminuscol}{magenta!65!black}
  \begin{scope}[xshift=-3.55cm]
    \coordinate (A1) at (-0.72,-0.62);
    \coordinate (A2) at ( 0.48,-0.76);
    \coordinate (A3) at ( 1.10, 0.00);
    \coordinate (A4) at ( 0.90, 0.84);
    \coordinate (A5) at (-0.10, 1.00);
    \coordinate (A6) at (-0.90, 0.66);
    \coordinate (APe) at (-0.87, 0.52);
    \coordinate (APf) at ( 1.07, 0.30);
    \coordinate (Amid) at ($(APe)!0.5!(APf)$);
    \fill[gascol] (A1)--(A2)--(A3)--(A4)--(A5)--(A6)--cycle;
    \begin{scope}
      \clip (A1)--(A2)--(A3)--(A4)--(A5)--(A6)--cycle;
      \fill[fluidcol] (APe)--(A1)--(A2)--(A3)--(APf)--cycle;
    \end{scope}
    \draw[bdrycol, line width=1.1pt, line join=round] (A1)--(A2)--(A3)--(A4)--(A5)--(A6)--cycle;
    \draw[ifacecol, line width=1.9pt, dashed] (APe)--(APf);
    \fill[ifacecol] (APe) circle (1.4pt);
    \fill[ifacecol] (APf) circle (1.4pt);
    \draw[ifacecol, line width=1.0pt, fill=white] (Amid) circle (2.0pt);
    \node[blue!62!black] at (-0.06,-0.20) {$\alpha_k$};
    \draw[normcol, ->, dashed, line width=1.1pt] (Amid) -- ++(0.06,0.42);
    \node[normcol, anchor=west] at ($(Amid)+(0.12,0.32)$) {$\mathbf{n}_{\tilde\Sigma_k}=?$};
    \node[black, anchor=north] at (0.10,-1.14) {Classical UGVOF: $\alpha_k$ only};
    \node[black, align=center, anchor=south] at (0.10,1.32) {normal needs\\neighbor stencil};
    \draw[bdrycol, line width=0.7pt]
      (A1)--(-0.775,-0.895) (A1)--(-0.997,-0.662)
      (A2)--( 0.711,-0.919) (A2)--( 0.409,-1.031)
      (A3)--( 1.355, 0.115) (A3)--( 1.295,-0.201)
      (A4)--( 0.972, 1.111) (A4)--( 1.179, 0.865)
      (A5)--(-0.316, 1.178) (A5)--(-0.007, 1.264)
      (A6)--(-1.176, 0.610) (A6)--(-1.041, 0.902);
  \end{scope}
  \begin{scope}[xshift=+3.55cm]
    \coordinate (B1) at (-0.72,-0.62);
    \coordinate (B2) at ( 0.48,-0.76);
    \coordinate (B3) at ( 1.10, 0.00);
    \coordinate (B4) at ( 0.90, 0.84);
    \coordinate (B5) at (-0.10, 1.00);
    \coordinate (B6) at (-0.90, 0.66);
    \coordinate (BPe) at (-0.87, 0.52);
    \coordinate (B7) at ( 1.07, 0.30);
    \coordinate (BPf) at ( 1.07, 0.30);
    \coordinate (Bmid) at ($(BPe)!0.5!(BPf)$);
    \fill[gascol] (B1)--(B2)--(B3)--(B7)--(B4)--(B5)--(B6)--cycle;
    \begin{scope}
      \clip (B1)--(B2)--(B3)--(B7)--(B4)--(B5)--(B6)--cycle;
      \fill[fluidcol] (BPe)--(B1)--(B2)--(B3)--(BPf)--cycle;
    \end{scope}
    \draw[bdrycol, line width=0.7pt] (B1)--(B2)--(B3)--(B7)--(B4)--(B5)--(B6)--cycle;
    \draw[wetted] (BPe)--(B1)--(B2)--(B3)--(BPf);
    \draw[gasbdrycol, line width=1.9pt, line cap=round, line join=round] (BPf)--(B4)--(B5)--(B6)--(BPe);
    \draw[ifacecol, line width=2.0pt] (BPe)--(BPf);
    \fill[ifacecol] (BPe) circle (1.4pt);
    \fill[ifacecol] (BPf) circle (1.4pt);
    \draw[ifacecol, line width=1.0pt, fill=white] (Bmid) circle (2.0pt);
    \draw[normcol, ->, line width=1.6pt]
      (Bmid) -- ++(0.04,0.48)
      node[pos=0.55, right=3pt, normcol]{$\mathbf{n}_{\tilde\Sigma_k}$};
    \node[blue!62!black] at (-0.06,-0.20) {$\alpha_k$};
    \node[sfminuscol, anchor=east] at (-0.72,-0.10) {$\alpha_{f_1}$};
    \node[sfminuscol, anchor=north] at (-0.10,-0.82) {$\alpha_{f_2}$};
    \node[sfminuscol, anchor=west] at (0.88,-0.34) {$\alpha_{f_3}$};
    \node[sfminuscol, anchor=west] at (1.15, 0.13) {$\alpha_{f_4}{=}1$};
    \node[black, anchor=north] at (0.10,-1.14) {DAVOF: $(\alpha_k,\{\alpha_f\})$};
    \node[black, align=center, anchor=south] at (0.10,1.18) {normal from\\local face data};
  \end{scope}
\end{tikzpicture}
\caption{The core contrast. \textbf{Left}: classical unsplit geometric VOF stores only $\alpha_k$, so the interface normal must be reconstructed from a multi-cell stencil. \textbf{Right}: DAVOF additionally transports the face-area fractions $\alpha_f$ (wetted face parts in gold), from which it recovers the interface normal of the cell by the closed-form sum \cref{eq:avof-normal}.}
\label{fig:coreideas}
\end{figure}

The setting is that of geometrical VOF (\cref{fig:discretization}). The phase indicator $\chi(\x,t)$ is one in the liquid $\Omega^-(t)$ and zero in the gas $\Omega^+(t)$; the interface is $\Sigma(t) = \partial\Omega^-(t)$. The domain is decomposed into polyhedral cells $\Omega_k$ of size $h$ with faces $S_f$, $f \in F_k$, each face with the outward area vector $\Sf = |S_f|\,\n_f$. The volume fraction of a cell and the area fraction of a face are
\begin{equation}
\alpha_k(t) = \frac{1}{|\Omega_k|}\int_{\Omega_k}\chi(\x,t)\,dV, \qquad
\alpha_f(t) = \frac{1}{|S_f|}\int_{S_f}\chi(\x,t)\,dS,
\label{eq:fractions}
\end{equation}
so that $\alpha_f |S_f| = |S_f^-(t)|$ is the wetted part of the face area, the \emph{area of fluid} of the face. The DAVOF state of a cell is the pair $(\alpha_k, \{\alpha_f\}_{f \in F_k})$; the face fractions are shared between the two cells of a face, so a mesh of $M$ cells with $F$ planar faces each carries $M(1 + F/2)$ values on average.

\begin{figure}[htbp]
\centering
\begin{tikzpicture}[scale=1.2, >=Stealth, font=\small]
  \colorlet{fluidcol}{blue!18}
  \colorlet{gascol}{gray!9}
  \colorlet{ifacecol}{blue!65!black}
  \colorlet{normcol}{red!65!black}
  \colorlet{facecol}{orange!80!black}
  \colorlet{wettedcol}{orange!80!black}
  \colorlet{gasbdrycol}{teal!58!black}
  \colorlet{bdrycol}{black!50}
  \colorlet{domaincol}{black!70}
  \coordinate (V1) at (-1.32,-0.70);
  \coordinate (V2) at ( 0.50,-0.90);
  \coordinate (V3) at ( 1.52, 0.00);
  \coordinate (V4) at ( 1.40, 0.94);
  \coordinate (V5) at ( 0.06, 1.18);
  \coordinate (V6) at (-1.20, 0.76);
  \coordinate (Pe) at (-1.21, 0.615); 
  \coordinate (Pf) at ( 1.471,0.386); 
  \fill[gascol] (-2.80,-1.90) rectangle (2.80,1.80);
  \fill[fluidcol]
    (-2.80,0.30) -- (Pe) -- (Pf) -- (2.80,0.10)
    -- (2.80,-1.90) -- (-2.80,-1.90) -- cycle;
  \fill[gray!13] (Pe)--(Pf)--(V4)--(V5)--(V6)--cycle;
  \draw[domaincol, line width=0.9pt] (-2.80,-1.90) rectangle (2.80,1.80);
  \draw[ifacecol!45, line width=1.2pt, dashed] (-2.80,0.30) -- (Pe);
  \draw[ifacecol,    line width=2.2pt]          (Pe) -- (Pf);
  \draw[ifacecol!45, line width=1.2pt, dashed] (Pf) -- (2.80,0.10);
  \draw[bdrycol, line width=0.6pt] (V1)--(V2)--(V3)--(V4)--(V5)--(V6)--cycle;
  \draw[gasbdrycol, line width=1.8pt] (Pf)--(V4)--(V5)--(V6)--(Pe);
  \draw[wetted] (Pe)--(V1)--(V2)--(V3);
  \fill[ifacecol] (Pe) circle (2.0pt);
  \fill[ifacecol] (Pf) circle (2.0pt);
  \draw[facecol, line width=2.8pt] (V3)--(V4);
  \draw[wetted] (V3)--(Pf);
  \draw[normcol, ->, line width=1.4pt]
    (1.46,0.47) -- ++(0.50,0.064) node[right=2pt,black]{$\mathbf{n}_f$};
  \draw[normcol, ->, line width=1.4pt]
    (-2.00,0.46) -- ++(-0.10,0.56) node[right=4pt,black]{$\mathbf{n}_\Sigma$};
  \node[domaincol, font=\footnotesize] at (-2.44,1.60) {$\Omega$};
  \node[ifacecol, font=\footnotesize] at (-0.22,0.88) {$\tilde{\Sigma}_k(t)$};
  \draw[->, ifacecol!65, line width=0.5pt] (-0.22,0.80) -- (-0.04,0.52);
  \node[blue!62!black] at (-2.16,-0.82) {$\Omega^-(t)$};
  \node[blue!55!black, font=\footnotesize] at (-2.16,-1.12) {$\chi=1$};
  \node[black!62] at (2.10,1.44) {$\Omega^+(t)$};
  \node[black!52, font=\footnotesize] at (2.10,1.18) {$\chi=0$};
  \node[black!70] at (0.14,-0.18) {$\Omega_k$};
  \node[black!60, font=\footnotesize] at (0.50,-1.22) {$\partial\Omega_k$};
  \draw[->, black!40, line width=0.4pt] (0.50,-1.12) -- (0.50,-0.92);
  \node[facecol, font=\footnotesize] at (1.84,0.82) {$S_f$};
  \draw[->, facecol!65, line width=0.4pt] (1.84,0.76) -- (1.51,0.64);
  \node[wettedcol, font=\footnotesize, anchor=west] at (2.00,0.06) {$S_f^-(t)$};
  \draw[->, wettedcol!65, line width=0.4pt] (1.98,0.06) -- (1.52,0.17);
\end{tikzpicture}
\caption{Geometric setting in a two-dimensional cross-section: the polyhedral control volume $\Omega_k$ in the domain $\Omega$ with liquid $\Omega^-(t)$ ($\chi=1$, blue) and gas $\Omega^+(t)$ ($\chi=0$). The interface is drawn as PLIC segments, with the face $S_f$ (orange), its wetted part $S_f^-(t)$ (gold) and the outward normal $\mathbf{n}_f$. The Gauss identity on the liquid sub-volume $\Omega_k^-(t)$, enclosed by the wetted face parts and the PLIC segment, gives the interface area-normal $\mathbf{n}_{\tilde{\Sigma}_k}|A_{\tilde{\Sigma}_k}|$ as minus the sum of the wetted face area vectors $\alpha_f\mathbf{S}_f$, \cref{eq:avof-normal}; no data from neighbouring cells are used.}
\label{fig:discretization}
\end{figure}

\subsection{The normal from the wetted face areas}

Consider the liquid sub-volume $\Omega_k^-(t) = \Omega_k \cap \Omega^-(t)$ of a cell (\cref{fig:discretization}). Its boundary consists of the wetted parts of the cell faces and of the interface piece $\Omega_k \cap \Sigma(t)$. For any constant vector $\mathbf{C}$ the divergence theorem gives $\int_{\partial\Omega_k^-}\mathbf{C}\cdot\n\,dS = \int_{\Omega_k^-}\nabla\cdot\mathbf{C}\,dV = 0$, hence the outward normals of a closed surface integrate to zero,
\begin{equation}
\int_{\partial\Omega_k^-(t)}\n\,dS = \int_{\Omega_k \cap \Sigma(t)}\n\,dS + \sum_{f \in F_k}\int_{S_f}\chi(\x,t)\,\n\,dS = 0.
\label{eq:closed}
\end{equation}
If the interface piece in the cell is approximated by one plane $\tilde{\Sigma}_k$ with unit normal $\nSk$ and polygon area $|\ASk|$, the first integral is $\nSk|\ASk|$, and on planar faces the second is a sum of the wetted face area vectors:
\begin{equation}
\nSk\,|\ASk| = -\sum_{f \in F_k}\int_{S_f}\chi(\x,t)\,\n\,dS = -\sum_{f \in F_k}\alpha_f(t)\,\Sf.
\label{eq:avof-normal}
\end{equation}
The right-hand side involves the cell's own face fractions and its face area vectors only. Its magnitude is the area of the interface piece, its direction the normal. For a state that is realizable by one plane, i.e.\ a planar interface cut through the cell, \cref{eq:avof-normal} is exact on every cell with planar faces, convex or not, because the identity is the divergence theorem on a polyhedron; on warped faces it holds with the wetted area vector in place of $\alpha_f\Sf$ (\cref{sec:warped}).

\subsection{The position of the plane}

The divergence theorem applied to $\x$ on the same sub-volume gives its volume, $3|\Omega_k^-| = \int_{\partial\Omega_k^-}\x\cdot\n\,dS$. On the plane $\tilde{\Sigma}_k$ the product $\x\cdot\nSk$ is constant, so with the signed distance $p_k$ of the plane from the cell centroid $\x_k$,
\begin{equation}
p_k(t) := (\x - \x_k)\cdot\nSk(t), \qquad \x \in \tilde{\Sigma}_k(t),
\label{eq:planeoffset}
\end{equation}
and with \cref{eq:avof-normal} to cancel the terms in $\x_k$, the position of the plane follows in closed form,
\begin{equation}
p_k(t) = \frac{3|\Omega_k|\,\alpha_k(t) - \sum_{f \in F_k}\alpha_f(t)\,(\x_f - \x_k)\cdot\Sf}{\big\|\sum_{f \in F_k}\alpha_f(t)\,\Sf\big\|},
\label{eq:avof-position}
\end{equation}
where $\x_f$ is the face centroid; on a planar face $(\x - \x_k)\cdot\n_f$ is constant, so the face integral $\int_{S_f}\chi\,(\x - \x_k)\cdot\n\,dS = \alpha_f(\x_f - \x_k)\cdot\Sf$ is exact, and nothing beyond $\alpha_k$, the $\alpha_f$ and the mesh geometry enters. In two dimensions the factor 3 becomes 2, the space dimension. A plane has one positional degree of freedom, so \cref{eq:avof-position} determines the position completely, and for a state realizable by one plane \cref{eq:avof-normal,eq:avof-position} return that plane exactly. The identities were verified symbolically for the corner cut of a cube and of a square and in exact rational arithmetic for randomly oriented planes cutting a cube in every configuration (corner, edge and slab cuts), with the wetted face areas, the liquid volume from an independent tetrahedral decomposition and the polygon computed exactly; both identities hold with zero residual (the script is part of the archived data \citep{LeiaRepo2026}).

\subsection{The face fractions as derivatives of the liquid volume}
\label{sec:derivatives}

\begin{figure}[htbp]
\centering
\begin{tikzpicture}[>=Stealth, font=\small]
  \colorlet{fluidcol}{blue!18}
  \colorlet{gaincol}{blue!45}
  \colorlet{ifacecol}{blue!65!black}
  \colorlet{normcol}{red!65!black}
  \colorlet{facecol}{orange!80!black}
  \colorlet{bdrycol}{black!50}
  \begin{scope}[scale=3.0]
    \fill[fluidcol] (0,0) -- (1,0) -- (1,0.8) -- (0,0.3) -- cycle;
    \fill[gaincol] (1,0) -- (1.15,0) -- (1.15,0.8) -- (1,0.8) -- cycle;
    \fill[red!30] (1,0.8) -- (1.15,0.8) -- (1.15,0.875) -- cycle;
    \draw[bdrycol, line width=0.6pt] (0,0) rectangle (1,1);
    \draw[bdrycol, dashed, line width=0.6pt] (1,0) -- (1.15,0) -- (1.15,1) -- (1,1);
    \draw[ifacecol, line width=1.6pt] (0,0.3) -- (1,0.8);
    \draw[ifacecol, dashed, line width=0.9pt] (1,0.8) -- (1.15,0.875);
    \draw[facecol, line width=3pt] (1,0) -- (1,1);
    \draw[wetted] (1,0) -- (1,0.8);
    \draw[->, black, line width=0.8pt] (1,-0.1) -- (1.15,-0.1) node[midway, below=1pt, font=\footnotesize] {$s$};
    \draw[normcol, ->, line width=1pt] (1.15,0.45) -- (1.38,0.45) node[right=1pt, font=\footnotesize] {$\n_f$};
    \node[font=\footnotesize, anchor=north] at (0.55,-0.2) {(a) $\partial|\Omega_k^-|/\partial s_f = \alpha_f|S_f|$};
    \node[blue!62!black, font=\footnotesize] at (0.42,0.2) {$\Omega_k^-$};
    \node[ifacecol, font=\footnotesize, anchor=south east] at (0.6,0.6) {$\Sigma$};
    \node[facecol, font=\footnotesize, anchor=east] at (0.96,0.93) {$S_f$};
  \end{scope}
  \begin{scope}[xshift=4.9cm, scale=3.0]
    \fill[fluidcol] (0,0) -- (1,0) -- (1,0.8) -- (0,0.3) -- cycle;
    \fill[gaincol] (0,0.3) -- (1,0.8) -- (1,0.912) -- (0,0.412) -- cycle;
    \draw[bdrycol, line width=0.6pt] (0,0) rectangle (1,1);
    \draw[ifacecol, line width=1.6pt] (0,0.3) -- (1,0.8);
    \draw[ifacecol, dashed, line width=1.1pt] (0,0.412) -- (1,0.912);
    \draw[wetted] (0,0) -- (1,0);
    \draw[wetted] (1,0) -- (1,0.8);
    \draw[wetted] (0,0) -- (0,0.3);
    \draw[normcol, ->, line width=1pt] (0.5,0.55) -- (0.388,0.774) node[pos=1, above=1pt, font=\footnotesize] {$\nSk$};
    \draw[<->, black, line width=0.6pt] (0.72,0.66) -- (0.675,0.75) node[midway, right=1pt, font=\footnotesize] {$c$};
    \node[font=\footnotesize, anchor=north] at (0.5,-0.2) {(b) $\partial|\Omega_k^-|/\partial c = |\ASk|$};
    \node[blue!62!black, font=\footnotesize] at (0.42,0.18) {$\Omega_k^-$};
  \end{scope}
  \begin{scope}[xshift=8.9cm, yshift=-0.75cm, scale=1.5]
    \fill[fluidcol] (0,0) -- (3,0) -- (3,1.08) arc[start angle=69.8, end angle=110.2, radius=4.35] -- cycle;
    \draw[bdrycol, line width=0.6pt] (0,0) grid[step=1] (3,3);
    \draw[ifacecol, line width=1.4pt] (3,1.08) arc[start angle=69.8, end angle=110.2, radius=4.35];
    \draw[wetted] (0,1) -- (3,1);
    \draw[wetted] (0,1) -- (0,1.08);
    \draw[wetted] (1,1) -- (1,1.32);
    \draw[wetted] (2,1) -- (2,1.32);
    \draw[wetted] (3,1) -- (3,1.08);
    \draw[ifacecol, dashed, line width=1.1pt] (0,1.08) -- (1,1.32);
    \draw[ifacecol, dashed, line width=1.1pt] (1,1.32) -- (2,1.32);
    \draw[ifacecol, dashed, line width=1.1pt] (2,1.32) -- (3,1.08);
    \draw[normcol, ->, line width=0.9pt] (0.5,1.2) -- ++(-0.094,0.39);
    \draw[normcol, ->, line width=0.9pt] (1.5,1.32) -- ++(0,0.4);
    \draw[normcol, ->, line width=0.9pt] (2.5,1.2) -- ++(0.094,0.39);
    \node[font=\footnotesize, anchor=north] at (1.5,-0.1) {(c) the ring's data};
    \node[black!60, font=\footnotesize] at (1.5,2.5) {$\Omega^+$};
    \node[blue!62!black, font=\footnotesize] at (1.5,0.45) {$\Omega^-$};
  \end{scope}
\end{tikzpicture}
\caption{The face fractions as derivatives of the liquid volume, in two dimensions. (a) Displacing the planar face $S_f$ (orange) outward by $s$ along $\n_f$ adds to the cell the liquid under the wetted part of the face (dark blue), $\alpha_f|S_f|\,s$, and a wedge of second order in $s$ at the trace of the interface (red). (b) Displacing the interface piece by $c$ along its normal adds the slab over the patch, $|\ASk|\,c$; the position identity \cref{eq:avof-position} and the volume anchor of \cref{sec:tpf} use this derivative. (c) One ring of cells crossed by a curved interface (solid): the measured ring fit of \cref{sec:curv} takes the recovered planes of the ring cells (dashed chords and their normals, each the identity \cref{eq:avof-normal} on that cell's faces) as its data; the fit of \cref{sec:transport} would take the wetted face areas (gold) and the volumes of the ring cells themselves, each a derivative of the volume by (a), without forming a normal in any cell.}
\label{fig:derivatives}
\end{figure}

Both identities describe how the liquid volume of the cell responds to a displacement of its boundary, and that view says what a surface fitted to the state is fitted to. Let the planar face $S_f$ be displaced outward along its normal $\n_f$ by a distance $s$ (\cref{fig:derivatives}a). The cell gains the slab $S_f \times [0, s]$, and the liquid it gains is the wetted part of the face times $s$, up to a wedge at the trace of the interface on the face whose volume is of second order in $s$, because the phase indicator is constant along $\n_f$ over the distance $s$ everywhere except within $O(s)$ of that trace:
\begin{equation}
\frac{\partial |\Omega_k^-|}{\partial s_f} = |S_f^-| = \alpha_f\,|S_f| .
\label{eq:dvds}
\end{equation}
The face fraction is the derivative of the liquid volume with respect to the position of its face, and the state $(\alpha_k, \{\alpha_f\})$ is the liquid volume of the cell together with its gradient with respect to the $F$ face positions, a Hermite datum of the volume. The same displacement of the interface piece along its normal by $c$ adds the slab over the patch (\cref{fig:derivatives}b), $\partial|\Omega_k^-|/\partial c = |\ASk|$, the derivative that the explicit position \cref{eq:avof-position} inverts and that the volume anchor of \cref{sec:tpf} uses in its root iteration. The two derivatives together give \cref{eq:avof-normal}: translating the cell by $\delta\x$ moves every face by $\n_f\cdot\delta\x$ and changes the liquid volume by $\sum_f \alpha_f\Sf\cdot\delta\x$; seen from the cell, the same translation moves the interface by $-\delta\x$, which for a plane is a normal displacement by $-\nSk\cdot\delta\x$ (a tangential shift leaves a plane in place), so $\sum_f \alpha_f\Sf\cdot\delta\x = -|\ASk|\,\nSk\cdot\delta\x$ for every $\delta\x$. For a curved piece the tangential shift changes the surface, which is the sagitta term of \cref{sec:defects}.

A quadric fitted to the state is therefore fitted to integrals of the phase indicator, the volume and the wetted areas, never to samples of a distance: the surface is the boundary of the liquid, the liquid is the side below it, and every residual of the fit is a volume or a wetted area under the surface. The six-coefficient surface of \cref{sec:full} is this fit on one cell, the residuals being the cell's own $F$ wetted areas and its volume. The derivative of a residual with respect to a coefficient is, by \cref{eq:dvds} applied to the surface, a moment of the wetted patch projected along the normal: the projected area for the offset, the first moments for the two slopes and the second moments for the curvature tensor, so the Jacobian of such a fit is composed of the quantities the volume rule integrates. \Cref{sec:transport} states which data the transported surface takes from this and which it should.

\subsection{Sensitivity of the position and the plane defect}
\label{sec:defects}

\begin{figure}[htbp]
\centering
\begin{tikzpicture}[scale=5.0, >=Stealth, font=\small]
  \colorlet{fluidcol}{blue!18}
  \colorlet{sagcol}{blue!45}
  \colorlet{ifacecol}{blue!65!black}
  \colorlet{bdrycol}{black!50}
  \fill[fluidcol] (0,0) -- (1,0) -- (1,0.2483) arc[start angle=56.25, end angle=123.75, radius=0.9] -- cycle;
  \fill[sagcol] (0,0.2483) -- (1,0.2483) arc[start angle=56.25, end angle=123.75, radius=0.9] -- cycle;
  \draw[bdrycol, line width=0.6pt] (0,0) rectangle (1,1);
  \draw[ifacecol, line width=1.6pt] (1,0.2483) arc[start angle=56.25, end angle=123.75, radius=0.9];
  \draw[black, dashed, line width=1.1pt] (0,0.2483) -- (1,0.2483);
  \draw[red!65!black, dash dot, line width=1.1pt] (0,0.3512) -- (1,0.3512);
  \draw[red!65!black, dotted, line width=1.3pt] (0,0.4541) -- (1,0.4541);
  \node[ifacecol, anchor=south, font=\footnotesize] at (0.5,0.405) {$\Sigma$};
  \node[anchor=west, font=\footnotesize] at (1.02,0.2483) {chord $\tilde\Sigma_k$};
  \node[red!65!black, anchor=west, font=\footnotesize] at (1.02,0.3512) {volume match: $+\,\delta V/|\ASk|$};
  \node[red!65!black, anchor=west, font=\footnotesize] at (1.02,0.4541) {\cref{eq:avof-position}: $+\,d\,\delta V/|\ASk|$};
  \node[sagcol!60!black, anchor=north, font=\footnotesize] at (0.5,0.24) {sagitta volume $\delta V$};
  \node[blue!62!black, font=\footnotesize] at (0.5,0.1) {$\Omega_k^-$};
  \node[black!60, font=\footnotesize] at (0.5,0.85) {$\Omega_k^+$};
  \draw[<->, black!70, line width=0.5pt] (0.08,0.2483) -- (0.08,0.3512);
  \draw[<->, black!70, line width=0.5pt] (0.16,0.2483) -- (0.16,0.4541);
\end{tikzpicture}
\caption{The plane defect, in a two-dimensional cell ($d = 2$). The curved interface $\Sigma$ (solid) encloses, above its chord $\tilde\Sigma_k$ (dashed, the plane with the recovered normal through the chord), the sagitta volume $\delta V$ (dark blue), $O(h)$ as a fraction of the cell. The state of $\Sigma$ carries the full liquid volume, so a plane that matches $\alpha_k$ by its cut volume sits $\delta V/|\ASk|$ above the chord (dash-dotted), while the explicit position \cref{eq:avof-position} evaluated on the same state sits $d\,\delta V/|\ASk|$ above it (dotted), $d$ the space dimension, because its face terms are the chord's and do not follow the volume. The difference between $\alpha_k$ and the fraction cut by the chord is the plane defect; drawn with the arc of radius $0.9h$.}
\label{fig:defects}
\end{figure}

For a state that is not realizable by one plane, two cases arise (\cref{fig:defects}). The exact state of a curved interface is not realizable by one plane: the fraction cut by the recovered plane then differs from $\alpha_k$ by the sagitta volume, the volume between the interface and its chord surface over the cell, which is $O(h)$ as a fraction, and the explicit position still converges at second order (\cref{sec:static}). We call that difference, $\alpha_k$ against the fraction cut by the recovered plane, the \emph{plane defect}; it is $O(h)$ even for the exact state of a curved interface and is a diagnostic of \cref{eq:avof-position}. A state that no single plane realizes has a \emph{realizability defect}, which the plane defect measures; a second measure against a curved clipping surface, the paraboloid defect, is introduced in \cref{sec:transport}. The second case is a state whose volume fraction and face fractions have been advanced by different discrete operations: a defect $\delta V$ in the liquid volume moves the plane of \cref{eq:avof-position} by $3\,\delta V/|\ASk|$, three times (the space dimension times) the shift $\delta V/|\ASk|$ that a volume-matching positioning applies, because the face terms in the numerator do not follow the volume. The explicit position is exact for consistent data and sensitive to inconsistency by the space dimension. This sensitivity is measured directly in \cref{sec:static} and decides, in \cref{sec:regen}, which plane a transport loop may use.

\subsection{Non-planar faces}
\label{sec:warped}

On a non-planar (warped) face the wetted area vector $\int_{S_f}\chi\,\n\,dS$ is not parallel to the face area vector $\Sf$, and a scalar fraction $\alpha_f$ does not determine it. The identities hold with the wetted area vector and its first moment $\int_{S_f}\chi\,(\x - \x_k)\cdot\n\,dS$ as the face quantities, which reduce to $\alpha_f\Sf$ and $\alpha_f(\x_f - \x_k)\cdot\Sf$ on a planar face. In the static tests of \cref{sec:static} nothing is stored per face: every face is circulated as the fan of triangles $(\x_f, \mathbf{p}_i, \mathbf{p}_{i+1})$ over its vertices, each triangle's wetted fraction times its area vector is summed into the cell normal, and the first moment of every wet triangle part into the position. Measured on hexahedra perturbed vertex by vertex, where the faces warp: the scalar face fraction loses the plane (normal error $8\,\%$ at a vertex displacement of $0.1h$ and $17\,\%$ at $0.2h$), while the fan sums recover the normal to $10^{-13}$ and the position to $10^{-11}h$ on the same meshes, and keep the plane at round-off on the full perturbed and polyhedral ladders of \cref{tab:plane}. On hexahedra perturbed column-wise, where the faces stay planar (skewness 0.41), every state is exact to $10^{-12}$. A transported state on non-planar faces therefore carries, per face, the wetted area vector and its first moment, six scalars, which reduce to the single fraction $\alpha_f$ on a planar face.

\section{The order of the recovered normal}
\label{sec:order}

Two estimates relate the accuracy of the face fractions to the accuracy of the recovered interface; both are measured in \cref{sec:static}. First, from \cref{eq:avof-normal}, an error $\delta\alpha_f$ of the face fractions perturbs the recovered normal by
\begin{equation}
|\delta\nSk| \le \frac{\sum_{f \in F_k}|\delta\alpha_f|\,|S_f|}{|\ASk|} = O(h^q) \qquad\text{for } |\delta\alpha_f| = O(h^q),
\label{eq:normal-propagation}
\end{equation}
since $|\ASk| = O(h^2)$ and $|S_f| = O(h^2)$: the normal has the same order of accuracy as the face fractions \emph{as fractions}, that is, in the error of $\alpha_f$ relative to the face area. The magnitude of the sum, the area, is not affected at leading order, because the error of the wetted area vector of one face is a vector in the direction of $\n_f$ and the errors of opposite faces cancel in the magnitude but not in the direction.

Second, a face fraction computed by cutting the face against a discrete interface with position error $d_\Sigma$ on the face displaces the crossing curve by $d_\Sigma$ along a length $O(h)$, so that the error of the fraction is
\begin{equation}
\begin{aligned}
|\delta\alpha_f| &= O(d_\Sigma/h): \\
d_\Sigma = O(h^2)\ \text{(any per-cell plane)} &\;\Rightarrow\; q = 1, \qquad
d_\Sigma = O(h^3) \;\Rightarrow\; q = 2,
\end{aligned}
\label{eq:clipping-law}
\end{equation}
where $q$ denotes the order of the face fractions as fractions. Any single plane per cell, however positioned, lies $O(h^2)$ off a curved interface and therefore gives first-order face fractions and a first-order normal; a surface $O(h^3)$ close to the interface, such as a paraboloid with the right curvature, gives second order. \Cref{eq:normal-propagation,eq:clipping-law} together state the requirement on any construction of face fractions, static or transported: $O(h^2)$ accuracy as a fraction.

\section{Static recovery: plane, sphere, ellipsoid}
\label{sec:static}

\subsection{Setup}

The identities were implemented in the leia OpenFOAM module \citep{LeiaRepo2026}. Given $(\alpha_k, \{\alpha_f\})$ of an implicit surface, the normal \cref{eq:avof-normal} and the position \cref{eq:avof-position} are evaluated in every interface cell, the plane is cut through the cell, which gives its polygon, the polygon centroid and the fraction it cuts. Four constructions of the state are used. (i) \emph{Exact} fractions of a sphere: the closed-form circle--polygon intersection for every face and adaptive quadrature for every cell. (ii) The \emph{Detrixhe--Aslam} state (DA), the area and volume fractions of the Detrixhe--Aslam construction \citep{DetrixheAslam2016}: the exact signed distance at the cell centre, the face centres and the mesh points, in the Detrixhe--Aslam formulas on the tetrahedra $(\x_k, \x_f, \mathbf{p}_i, \mathbf{p}_{i+1})$ and on the face triangle fans with the same vertex values; the formulas are exact for a linear function on each simplex, so this state describes one $C^0$ piecewise-linear surface, the chord surface, which lies $O(h^2)$ off the curved interface. (iii) The same cell fractions with \emph{quadratic faces}: the exact signed distance also at the edge midpoints of every fan triangle and the quadratic interpolant cut exactly on every triangle (the sliver between chord and conic arc by Gauss--Legendre quadrature); these are second-order face fractions, $d_\Sigma = O(h^3)$, paired with the chord cell fraction, so the state is inconsistent by the sagitta volume on purpose; we call it the \emph{Detrixhe--Aslam-Q} state (DA-Q in the tables; Q for the quadratic face fractions) or the \emph{chord state}. (iv) The \emph{plane indicator} of the level-set solver: one least-squares plane per cell, the face fraction the average of the two adjacent cells' planes. For comparison, the PLIC reconstructions plicRDF, gradAlpha and isoAlpha of OpenFOAM's \texttt{geometricVoF} library \citep{ScheuflerRoenby2019,ScheuflerRoenby2023TwoPhaseFlow} are run on the identical $\alpha_k$ (plicRDF with 10 iterations from the least-squares gradient normal, the library's defaults), and their plane centroids are scored with the same measures. Errors are evaluated over the interface cells, $|\nSk|\,|\ASk| > 10^{-3}|\Omega_k|^{2/3}$ (the wisp threshold): the normal error $|\nSk - \n_\Sigma|$ with the exact normal at the recovered polygon centroid, the position error $d$ as the distance of the polygon centroid from the exact surface, and the total area against the exact one. $L_1$ is the mean, $L_2$ the root mean square and $L_\infty$ the maximum over the interface cells; orders are least-squares slopes of the error against the cell size over the four meshes $N = 20, 40, 80, 160$ cells per side, with the three finest in parentheses where given.

\label{sec:meshes}Every ladder is run on three mesh classes. \emph{Hexahedral}: the Cartesian block mesh. \emph{Perturbed hexahedral}: the same mesh with every vertex displaced by a random vector of up to $0.1h$ (a fixed seed), so that every face is warped and the fan sums of \cref{sec:warped} carry the state. \emph{Polyhedral}: the cfMesh polyhedral mesher on the same box with the maximum cell size $h\,2^{-1/3}$, whose cells at the interface measure $h$, so that the polyhedral ladder has the interface resolution of the hexahedral one (784 against 772 interface cells of the plane at $N = 20$; maximum skewness 2.2). The exact-sphere construction exists on the hexahedral mesh only (its face formula is the circle--rectangle intersection); the Detrixhe--Aslam and the DA-Q states and the comparators run on all three. Pre-registered before the non-hexahedral runs: the plane at round-off on every mesh class; the DA-Q normal second order with a larger constant than on the hexahedra, the Detrixhe--Aslam normal first order, the position second order; plicRDF at about 1.1 everywhere, gradAlpha non-convergent.

\subsection{Plane: exactness}
\label{sec:plane}

A plane with normal $(0.3, 0.5, 0.8124)$ through $(2.13, 1.71, 1.97)\,\mathrm{mm}$ in an $8^3$ box of $4\,\mathrm{mm}$ side is recovered to round-off from every construction: normal error $L_\infty$ of $5\cdot10^{-14}$ (Detrixhe--Aslam) and $1\cdot10^{-12}$ (quadratic faces), position error $10^{-16}\,\mathrm{m}$ at a cell size of $5\cdot10^{-4}\,\mathrm{m}$, the difference between $\alpha_k$ and the plane cut and the realizability of the state by the recovered plane at $10^{-14}$. On the same field plicRDF's normal error is $1.9\cdot10^{-4}$, gradAlpha's and isoAlpha's about $10^{-1}$. On unstructured meshes with planar faces, the face-fraction state recovers the interface normal of a plane exactly, without any iteration, and with it the position and the curvature. The test is the regression gate of the module \citep{LeiaRepo2026} and fails at $10^{-12}$. \Cref{tab:plane} repeats it on the ladder $N = 20, 40, 80, 160$ and on the three mesh classes of \cref{sec:meshes}: hexahedral, hexahedral with every vertex displaced by up to $0.1h$ (warped faces, the fan sums of \cref{sec:warped}) and polyhedral (cfMesh, with the interface resolution of the hexahedral ladder). Both surface-based states stay at round-off on every mesh and every $N$: normal error $10^{-14}$ in $L_2$ and $3\cdot10^{-13}$ at most, position $5\cdot10^{-17}\,\mathrm{m}$, curvature below $10^{-9}\,\mathrm{m^{-1}}$; an error at round-off has no order, and those cells are left empty. plicRDF's normal error on the same fields is $2.5\cdot10^{-4}$ on the hexahedra, $5.3\cdot10^{-3}$ on the perturbed and $3.5\cdot10^{-3}$ on the polyhedral mesh, with maxima of $0.02$, $0.38$ and $0.47$, and it does not converge under refinement on any of them, although the plane is exactly representable; gradAlpha and isoAlpha stay at $10^{-1}$ everywhere.

\begin{table}[htbp]
\centering
\caption{The plane of \cref{sec:plane} on the ladder $N = 20, 40, 80, 160$ and on the three mesh classes (hexahedral; every vertex displaced by up to $0.1h$; cfMesh polyhedral with the interface resolution of the hexahedral ladder). Columns as in \cref{tab:sphere}; errors at $N = 160$. An order is not printed where the error is at round-off (the normal below $10^{-11}$, the position below $10^{-14}\,\mathrm{m}$, the curvature below $10^{-9}\,\mathrm{m^{-1}}$). DA: the Detrixhe--Aslam area and volume fractions; DA-Q: the same cell fractions with quadratic face fractions (constructions (ii) and (iii) of \cref{sec:static}). The comparators depend on $\alpha_k$ alone, which both states share, so their rows appear once per mesh block. Bold: a DAVOF value better than every comparator of its mesh class.}
\label{tab:plane}
\footnotesize
\setlength{\tabcolsep}{3pt}
\resizebox{\linewidth}{!}{
\begin{tabular}{lllcccccccc}
\toprule
mesh & $(\alpha_k,\alpha_f)$ & method & $p_1(\mathbf{n})$ & $p_2(\mathbf{n})$ & $L_2(\mathbf{n})$ & $L_\infty(\mathbf{n})$ & $p(\mathbf{x})$ & $L_2(d)$ [m] & $p(\kappa)$ & $L_2(\kappa)/L_2(\kappa_\Sigma)$ \\
\midrule
hexahedral & DA-Q & DAVOF & -- & -- & \textbf{1.78e-14} & \textbf{3.32e-13} & -- & \textbf{4.57e-17} & -- & -- \\
 & DA & DAVOF & -- & -- & \textbf{9.81e-15} & \textbf{3.27e-13} & -- & \textbf{4.48e-17} & -- & -- \\
\midrule
 & both states ($\alpha_k$ shared) & plicRDF & 0.41 (0.13) & -0.29 (-0.55) & 2.51e-04 & 1.89e-02 & 0.12 & 8.32e-12 & -- & -- \\
 &  & gradAlpha & 0.11 (0.08) & 0.13 (0.10) & 7.46e-02 & 3.95e-01 & 1.31 & 2.99e-08 & -- & -- \\
 &  & isoAlpha & 0.38 (0.30) & 0.39 (0.37) & 4.18e-02 & 5.41e-01 & 1.45 & 3.63e-08 & -- & -- \\
\midrule
perturbed hexahedral (0.1 h) & DA-Q & DAVOF & -- & -- & \textbf{1.89e-14} & \textbf{3.61e-13} & -- & \textbf{4.63e-17} & -- & -- \\
 & DA & DAVOF & -- & -- & \textbf{1.16e-14} & \textbf{3.46e-13} & -- & \textbf{4.61e-17} & -- & -- \\
\midrule
 & both states ($\alpha_k$ shared) & plicRDF & 0.06 (-0.02) & 0.24 (-0.36) & 5.26e-03 & 3.75e-01 & 0.95 & 2.65e-07 & -- & -- \\
 &  & gradAlpha & 0.10 (0.08) & 0.12 (0.10) & 7.56e-02 & 4.06e-01 & 0.96 & 2.58e-07 & -- & -- \\
 &  & isoAlpha & 0.21 (0.17) & 0.26 (0.23) & 6.08e-02 & 5.61e-01 & 1.02 & 2.16e-07 & -- & -- \\
\midrule
polyhedral (cfMesh) & DA-Q & DAVOF & -- & -- & \textbf{1.72e-14} & \textbf{2.94e-13} & -- & \textbf{2.52e-17} & -- & -- \\
 & DA & DAVOF & -- & -- & \textbf{8.05e-15} & \textbf{2.11e-13} & -- & \textbf{2.51e-17} & -- & -- \\
\midrule
 & both states ($\alpha_k$ shared) & plicRDF & 0.81 (1.01) & 0.00 (0.82) & 3.47e-03 & 4.70e-01 & 1.47 & 5.89e-08 & -- & -- \\
 &  & gradAlpha & 0.17 (0.15) & 0.21 (0.19) & 6.99e-02 & 5.53e-01 & 1.58 & 8.13e-08 & -- & -- \\
 &  & isoAlpha & 0.30 (0.30) & 0.26 (0.30) & 8.14e-02 & 1.95e+00 & 1.40 & 1.35e-07 & -- & -- \\
\bottomrule
\end{tabular}
}
\end{table}

\subsection{Sphere and ellipsoid: convergence}

\begin{table}[htbp]
\centering
\caption{Sphere of radius $R = 1\,\mathrm{mm}$ in a box of side $4R$, centre $(2.0692, 2.1172, 1.9548)\,\mathrm{mm}$ off every mesh symmetry plane, meshes $N = 20, 40, 80, 160$ ($R/h = 5 \ldots 40$). Three mesh classes (\cref{sec:meshes}): hexahedral, hexahedral with every vertex displaced by up to $0.1h$, and cfMesh polyhedral with the interface resolution of the hexahedral ladder; the exact-sphere state exists on the hexahedral mesh only. Least-squares orders $p$ over the four meshes (the three finest in parentheses) and errors on the finest; $p_1$ and $p_2$ are the orders of the normal in $L_1$ and $L_2$, $L_\infty(\mathbf{n})$ its maximum error, $p(\mathbf{x})$ the order of the position; DA: the Detrixhe--Aslam area and volume fractions, DA-Q: the same cell fractions with quadratic face fractions (\cref{sec:static}); the PLIC schemes are run on the identical $\alpha_k$ of each construction, and where both states share $\alpha_k$ their rows appear once. Bold: a DAVOF value better than every comparator of its mesh class.}
\label{tab:sphere}
\footnotesize
\setlength{\tabcolsep}{3pt}
\resizebox{\linewidth}{!}{
\begin{tabular}{lllcccccccc}
\toprule
mesh & $(\alpha_k,\alpha_f)$ & method & $p_1(\mathbf{n})$ & $p_2(\mathbf{n})$ & $L_2(\mathbf{n})$ & $L_\infty(\mathbf{n})$ & $p(\mathbf{x})$ & $L_2(d)$ [m] & $p(\kappa)$ & $L_2(\kappa)/L_2(\kappa_\Sigma)$ \\
\midrule
hexahedral & exact sphere & DAVOF & \textbf{2.05 (2.00)} & \textbf{1.98 (1.88)} & \textbf{1.15e-04} & \textbf{5.53e-03} & 2.00 & 4.35e-08 & -- & -- \\
 & exact sphere & plicRDF & 1.07 (1.04) & 1.06 (1.05) & 3.87e-03 & 1.67e-02 & 2.00 & 4.35e-08 & -- & -- \\
 & exact sphere & gradAlpha & 0.07 (0.04) & 0.07 (0.02) & 6.64e-02 & 1.29e-01 & 1.95 & 4.98e-08 & -- & -- \\
 & exact sphere & isoAlpha & 0.51 (0.33) & 0.52 (0.36) & 2.08e-02 & 7.85e-02 & 1.99 & 4.35e-08 & -- & -- \\
\midrule
 & DA-Q & DAVOF & \textbf{2.05 (2.00)} & \textbf{1.95 (1.86)} & \textbf{1.19e-04} & \textbf{5.53e-03} & \textbf{2.01} & 7.53e-08 & -- & -- \\
 & DA-Q & plicRDF & 1.06 (1.04) & 1.06 (1.05) & 3.86e-03 & 1.65e-02 & 2.00 & 7.94e-08 & -- & -- \\
 & DA-Q & gradAlpha & 0.06 (0.04) & 0.06 (0.02) & 6.64e-02 & 1.29e-01 & 1.98 & 8.38e-08 & -- & -- \\
 & DA-Q & isoAlpha & 0.51 (0.34) & 0.52 (0.37) & 2.08e-02 & 7.87e-02 & 2.00 & 7.92e-08 & -- & -- \\
\midrule
 & DA & DAVOF & 1.02 (1.00) & 1.01 (1.00) & \textbf{2.43e-03} & \textbf{1.05e-02} & \textbf{2.03} & \textbf{1.68e-08} & -- & -- \\
 & DA & plicRDF & 1.06 (1.04) & 1.06 (1.05) & 3.86e-03 & 1.65e-02 & 2.00 & 7.94e-08 & -- & -- \\
 & DA & gradAlpha & 0.06 (0.04) & 0.06 (0.02) & 6.64e-02 & 1.29e-01 & 1.98 & 8.38e-08 & -- & -- \\
 & DA & isoAlpha & 0.51 (0.34) & 0.52 (0.37) & 2.08e-02 & 7.87e-02 & 2.00 & 7.92e-08 & -- & -- \\
\midrule
perturbed hexahedral (0.1 h) & DA-Q & DAVOF & \textbf{2.12 (2.49)} & \textbf{2.40 (4.11)} & \textbf{9.49e-05} & \textbf{3.43e-03} & \textbf{2.07} & \textbf{7.77e-08} & -- & -- \\
 & DA & DAVOF & \textbf{0.99 (0.98)} & \textbf{0.97 (0.98)} & \textbf{2.57e-03} & \textbf{1.23e-02} & \textbf{2.04} & \textbf{1.77e-08} & -- & -- \\
\midrule
 & both states ($\alpha_k$ shared) & plicRDF & 0.90 (0.83) & 0.93 (0.87) & 5.25e-03 & 1.45e-01 & 1.50 & 2.37e-07 & -- & -- \\
 &  & gradAlpha & 0.06 (0.04) & 0.05 (0.03) & 6.71e-02 & 1.45e-01 & 1.51 & 2.40e-07 & -- & -- \\
 &  & isoAlpha & 0.17 (0.09) & 0.20 (0.11) & 4.75e-02 & 1.97e-01 & 1.57 & 2.03e-07 & -- & -- \\
\midrule
polyhedral (cfMesh) & DA-Q & DAVOF & \textbf{2.14 (2.04)} & \textbf{2.19 (2.08)} & \textbf{2.17e-05} & \textbf{1.63e-03} & \textbf{2.02} & \textbf{1.33e-08} & -- & -- \\
 & DA & DAVOF & 1.05 (0.99) & 1.04 (0.99) & \textbf{6.46e-04} & \textbf{5.07e-03} & \textbf{2.01} & \textbf{9.22e-09} & -- & -- \\
\midrule
 & both states ($\alpha_k$ shared) & plicRDF & 1.05 (1.04) & 1.06 (1.05) & 2.08e-03 & 1.07e-02 & 2.00 & 6.07e-08 & -- & -- \\
 &  & gradAlpha & 0.05 (0.02) & 0.06 (0.03) & 4.64e-02 & 1.03e-01 & 1.99 & 6.16e-08 & -- & -- \\
 &  & isoAlpha & 0.06 (0.01) & 0.06 (0.01) & 5.37e-02 & 1.17e-01 & 1.97 & 6.44e-08 & -- & -- \\
\bottomrule
\end{tabular}
}
\end{table}

\begin{table}[htbp]
\centering
\caption{Triaxial ellipsoid with half-axes $(1.0, 0.8, 0.6)\,\mathrm{mm}$ in a $4\,\mathrm{mm}$ box, $N = 20, 40, 80, 160$ (tightest curvature radius $1.8h \ldots 14.4h$), on the three mesh classes of \cref{tab:sphere}. Orders $p$ over the four meshes (three finest in parentheses), errors at $N = 160$ ($h = 25\,\mu\mathrm{m}$); $L_2(d)$ is the root-mean-square distance of the recovered polygon centroid from the ellipsoid; $\kappa = \kappa_1 + \kappa_2$ is the total curvature of the quadric fitted to one ring of recovered centroids and normals (\cref{sec:curv}), its error relative to the root mean square of the exact curvature over the interface cells. DA and DA-Q as in \cref{tab:sphere}. The comparators depend on $\alpha_k$ alone, which both states share, so their rows appear once. Bold: a DAVOF value better than every comparator of its mesh class.}
\label{tab:ellipsoid}
\footnotesize
\setlength{\tabcolsep}{3pt}
\resizebox{\linewidth}{!}{
\begin{tabular}{lllcccccccc}
\toprule
mesh & $(\alpha_k,\alpha_f)$ & method & $p_1(\mathbf{n})$ & $p_2(\mathbf{n})$ & $L_2(\mathbf{n})$ & $L_\infty(\mathbf{n})$ & $p(\mathbf{x})$ & $L_2(d)$ [m] & $p(\kappa)$ & $L_2(\kappa)/L_2(\kappa_\Sigma)$ \\
\midrule
hexahedral & DA-Q & DAVOF & \textbf{1.96 (1.97)} & \textbf{1.78 (1.77)} & \textbf{3.01e-04} & \textbf{1.22e-02} & 2.00 & \textbf{1.01e-07} & 1.55 & 4.55e-03 \\
 & DA & DAVOF & 1.05 (1.00) & 1.00 (1.00) & \textbf{3.63e-03} & \textbf{2.61e-02} & \textbf{2.04} & \textbf{2.32e-08} & 1.31 & 6.75e-03 \\
\midrule
 & both states ($\alpha_k$ shared) & plicRDF & 1.19 (1.10) & 1.10 (1.06) & 5.36e-03 & 3.49e-02 & 2.00 & 1.06e-07 & -- & -- \\
 &  & gradAlpha & 0.09 (0.05) & 0.10 (0.04) & 6.80e-02 & 1.29e-01 & 1.98 & 1.10e-07 & -- & -- \\
 &  & isoAlpha & 0.70 (0.51) & 0.65 (0.51) & 2.15e-02 & 9.29e-02 & 2.00 & 1.06e-07 & -- & -- \\
\midrule
perturbed hexahedral (0.1 h) & DA-Q & DAVOF & \textbf{2.02 (2.00)} & \textbf{1.87 (1.91)} & \textbf{2.81e-04} & \textbf{1.06e-02} & \textbf{1.99} & \textbf{1.04e-07} & 1.53 & 4.65e-03 \\
 & DA & DAVOF & 1.04 (1.03) & 1.01 (1.02) & \textbf{3.80e-03} & \textbf{3.01e-02} & \textbf{2.06} & \textbf{2.44e-08} & 1.27 & 7.23e-03 \\
\midrule
 & both states ($\alpha_k$ shared) & plicRDF & 1.06 (0.98) & 1.02 (1.01) & 6.43e-03 & 1.39e-01 & 1.64 & 2.39e-07 & -- & -- \\
 &  & gradAlpha & 0.10 (0.06) & 0.11 (0.05) & 6.86e-02 & 1.45e-01 & 1.63 & 2.46e-07 & -- & -- \\
 &  & isoAlpha & 0.29 (0.14) & 0.31 (0.17) & 4.81e-02 & 1.93e-01 & 1.78 & 2.06e-07 & -- & -- \\
\midrule
polyhedral (cfMesh) & DA-Q & DAVOF & \textbf{1.97 (1.97)} & \textbf{1.79 (1.80)} & \textbf{8.36e-05} & \textbf{4.23e-03} & 1.98 & \textbf{1.77e-08} & 1.83 & 2.33e-03 \\
 & DA & DAVOF & \textbf{1.26 (1.17)} & 1.06 (1.02) & \textbf{8.89e-04} & \textbf{9.25e-03} & \textbf{2.01} & \textbf{1.22e-08} & 1.58 & 2.60e-03 \\
\midrule
 & both states ($\alpha_k$ shared) & plicRDF & 1.18 (1.12) & 1.11 (1.10) & 2.73e-03 & 2.02e-02 & 2.00 & 8.03e-08 & -- & -- \\
 &  & gradAlpha & 0.12 (0.06) & 0.13 (0.07) & 4.55e-02 & 1.15e-01 & 1.99 & 8.10e-08 & -- & -- \\
 &  & isoAlpha & 0.09 (0.02) & 0.09 (0.02) & 5.36e-02 & 1.16e-01 & 1.98 & 8.28e-08 & -- & -- \\
\bottomrule
\end{tabular}
}
\end{table}

\begin{figure}[htbp]
\centering
\includegraphics[width=0.8\linewidth]{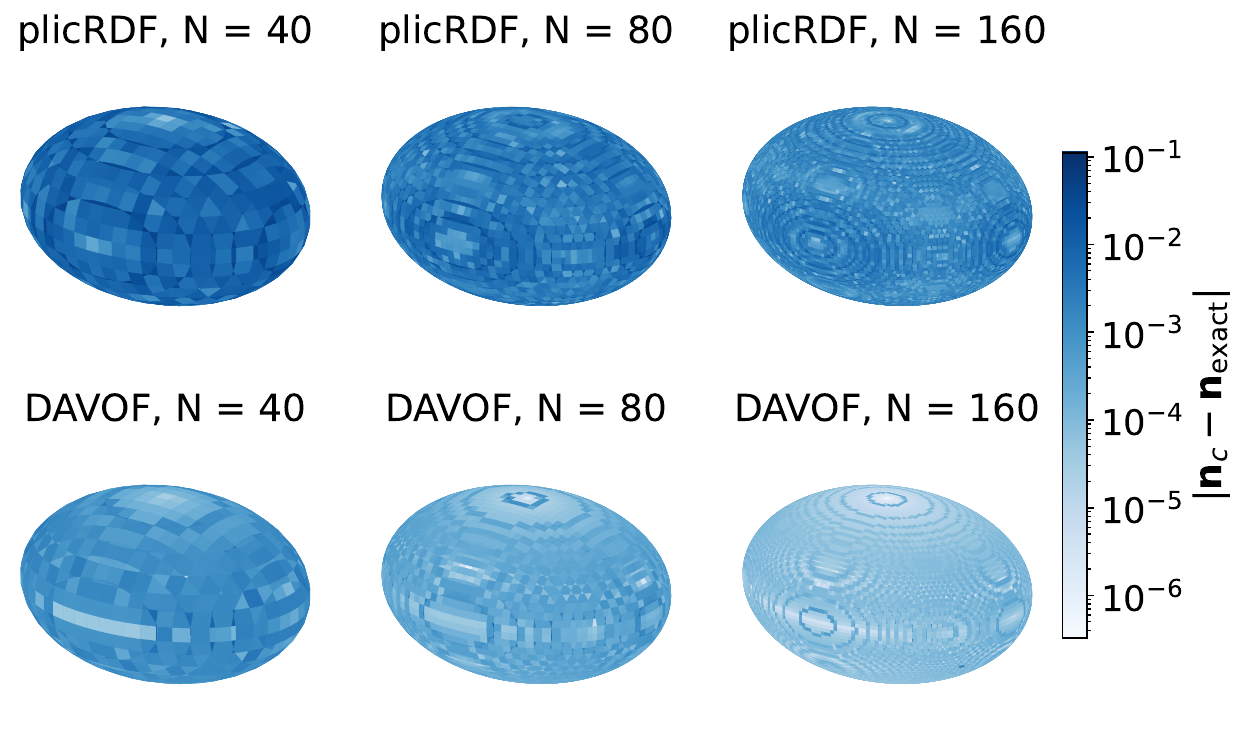}
\caption{The ellipsoid test: the recovered interfaces, one polygon per interface cell, of plicRDF (top) and of the recovery \cref{eq:avof-normal,eq:avof-position} from the DA-Q state (bottom) at $N = 40, 80, 160$, coloured by the normal error on a logarithmic scale.}
\label{fig:davof-gates}
\end{figure}

\Cref{tab:sphere,tab:ellipsoid} and \cref{fig:davof-gates} summarise the two convergence tests. All predictions were pre-registered before the runs; the falsified ones are stated as such.

\emph{Second-order input gives a second-order normal.} With fractions accurate to second order as fractions, the exact sphere fractions and the quadratic faces on the ellipsoid, the normal \cref{eq:avof-normal} is second order: 1.98 in $L_2$ on the sphere with the exact fractions, with an error 34 times smaller than that of plicRDF, whose order remains about 1.1 on the exact fractions (1.07 in $L_1$, 1.06 in $L_2$); 1.96 in $L_1$ and 1.78 in $L_2$ on the ellipsoid, where plicRDF on the identical $\alpha_k$ gives 1.19 and 1.10 at an $L_2$ error 18 times larger; isoAlpha converges at order 0.5 to 0.7 and gradAlpha, the centred-gradient normal, does not converge, in line with \citet{PilliodPuckett2004}. The quadratic faces reproduce the error of the exact fractions within $4\,\%$ on every sphere mesh. The $L_2$ order 1.78 of the DA-Q normal on the ellipsoid is set by 47 of 18642 interface cells near the two poles; the maximum normal error is $1.22\cdot10^{-2}$ against $3.49\cdot10^{-2}$ for plicRDF, a factor of 3 where the root mean square gives 18.

\emph{First-order input gives a first-order normal, and the reason is known.} With the Detrixhe--Aslam state the normal is first order (1.05 in $L_1$ and 1.00 in $L_2$ on the ellipsoid, 1.02 and 1.01 on the sphere), and so is it with the plane indicator (about 1.1 on the ellipsoid, not in the tables): both locate the interface to $O(h^2)$ on the faces, the fractions are therefore accurate to $O(h)$ as fractions, and by \cref{eq:normal-propagation} the direction of the sum inherits that order while its magnitude, the area, stays second order (the total area converges at order 2.00 for every construction; the column is omitted from the tables). One polynomial degree more in the face cut, \cref{eq:clipping-law}, restores second order without iteration. The pre-registered second order for the Detrixhe--Aslam state was falsified by this measurement, and \cref{eq:normal-propagation,eq:clipping-law} were written to explain it.

\emph{The position is second order for every initialisation.} On the exact sphere state it coincides with the volume-matched position of plicRDF to $0.02\,\%$; on the Detrixhe--Aslam state of the ellipsoid its error is 4.6 times smaller than that of the volume-matched plane of the same $\alpha_k$ (4.7 on the sphere). With the quadratic faces its error is four times larger than with the Detrixhe--Aslam state on the ellipsoid ($1.01\cdot10^{-7}$ against $2.32\cdot10^{-8}\,\mathrm{m}$), because that construction keeps the chord cell fraction: the state is inconsistent by the sagitta volume, the plane defect of \cref{sec:defects}, to which \cref{eq:avof-position} is three times as sensitive as a volume match. This is the first measured realizability defect of the state; a transported state must pair second-order face fractions with a consistent cell fraction. Every row of \cref{tab:ellipsoid}, the comparators included, has a position order between 1.98 and 2.04: the position does not discriminate between the reconstructions, the normal does.

\emph{The orders are mesh-independent, and the constant does not grow off the hexahedral mesh.} On the perturbed hexahedra and on the polyhedra the DA-Q normal is second order (2.12 and 2.14 in $L_1$ on the sphere, 2.02 and 1.97 on the ellipsoid), the Detrixhe--Aslam normal first order (0.99 to 1.26) and the position second order for every state. The error constant is the hexahedral one on the perturbed mesh (ellipsoid $L_2$ $2.81\cdot10^{-4}$ against $3.01\cdot10^{-4}$; the $L_2$ order rises from 1.78 to 1.87 because the perturbation breaks the alignment of the poles with the grid planes that sets the hexahedral $L_2$) and 3.6 to 5.5 times smaller on the polyhedra (sphere $2.17\cdot10^{-5}$ against $1.19\cdot10^{-4}$, ellipsoid $8.36\cdot10^{-5}$ against $3.01\cdot10^{-4}$): a polyhedral cell has more faces, so more face fractions enter the sum \cref{eq:avof-normal}, and its point-neighbour ring holds more samples for the curvature fit of \cref{sec:curv}, whose order rises to 1.83 on the polyhedra at half the error. The pre-registered larger constant off the hexahedral mesh is falsified in the favourable direction. The comparators lose on these meshes: plicRDF falls to order 0.90 on the perturbed sphere with a maximum normal error of 0.14 (nine times its hexahedral maximum) and keeps 1.05 to 1.18 on the polyhedra at 30 to 100 times the DAVOF error; gradAlpha and isoAlpha do not converge on the polyhedra (0.05 to 0.13), and the position order of all three comparators drops to 1.5 to 1.8 on the perturbed mesh, where the DAVOF position keeps 2.0. One defect case appears on the perturbed sphere at $N = 20$: a cell with $\alpha_k = 0$, whose chord tetrahedra contain no liquid while the quadratic cut wets its warped fan triangles, passes the wisp threshold and is scored with a normal error of 1.2; it drops out under refinement ($L_1$ unaffected; the $L_2$ order 2.40 over the four meshes is this cell's departure) and is the realizability defect of the chord state in its sharpest form, a cell a transported method must detect (\cref{sec:defects}).

\subsection{Curvature from one ring of recovered normals}
\label{sec:curv}

A quadric, a smooth quadratic surface, fitted per interface cell to the recovered polygon centroids and normals of its ring, the neighbouring interface cells that share a point with it, gives the curvature tensor of the state; the total curvature $\kappa = \kappa_1 + \kappa_2$ follows in closed form from the fitted coefficients \citep{GoldmanCurvature2005}. With the quadratic faces its error on the ellipsoid converges at order 1.55 (pairwise 1.83, 1.49, 1.26, falling toward first order), with a relative $L_2$ error of $4.6\cdot10^{-3}$ at $N = 160$; with the Detrixhe--Aslam state at 1.31. Gaussian and principal curvatures converge at the same orders, and the values transferred to cell centres and faces by the parallel-surface relations keep these orders (1.59, 1.36 and 1.63 for the tensor, the cell and the face delivery). We attach no order threshold to this result, because the curvature models of the reference implementation, the paraboloid fit and the reconstructed distance function, show zero-order convergence, i.e.\ no convergence at all, in the static benchmarks of \citet{ScheuflerRoenby2023TwoPhaseFlow} (their Fig.~15), on hexahedral as on triangular-prism grids; the height-function method needs cuboid cells (the mesh-decoupled variant of \citet{OwkesDesjardins2015} lifts that with a Cartesian auxiliary grid). The result is the convergence itself, obtained from the state and one ring. Two more predictions were falsified in the favourable direction: the curvature of the first-order state converges too, its normal error being a smooth tilt whose gradient converges, and the fit is insensitive within $3\,\%$ to the centroid rows. The curvature is not the sharpest detector of a lost fraction order; the normal is (1.00 against 1.78).

\section{When can a one-ring curvature converge?}
\label{sec:onering}

\begin{figure}[htbp]
\centering
\includegraphics[width=0.9\linewidth]{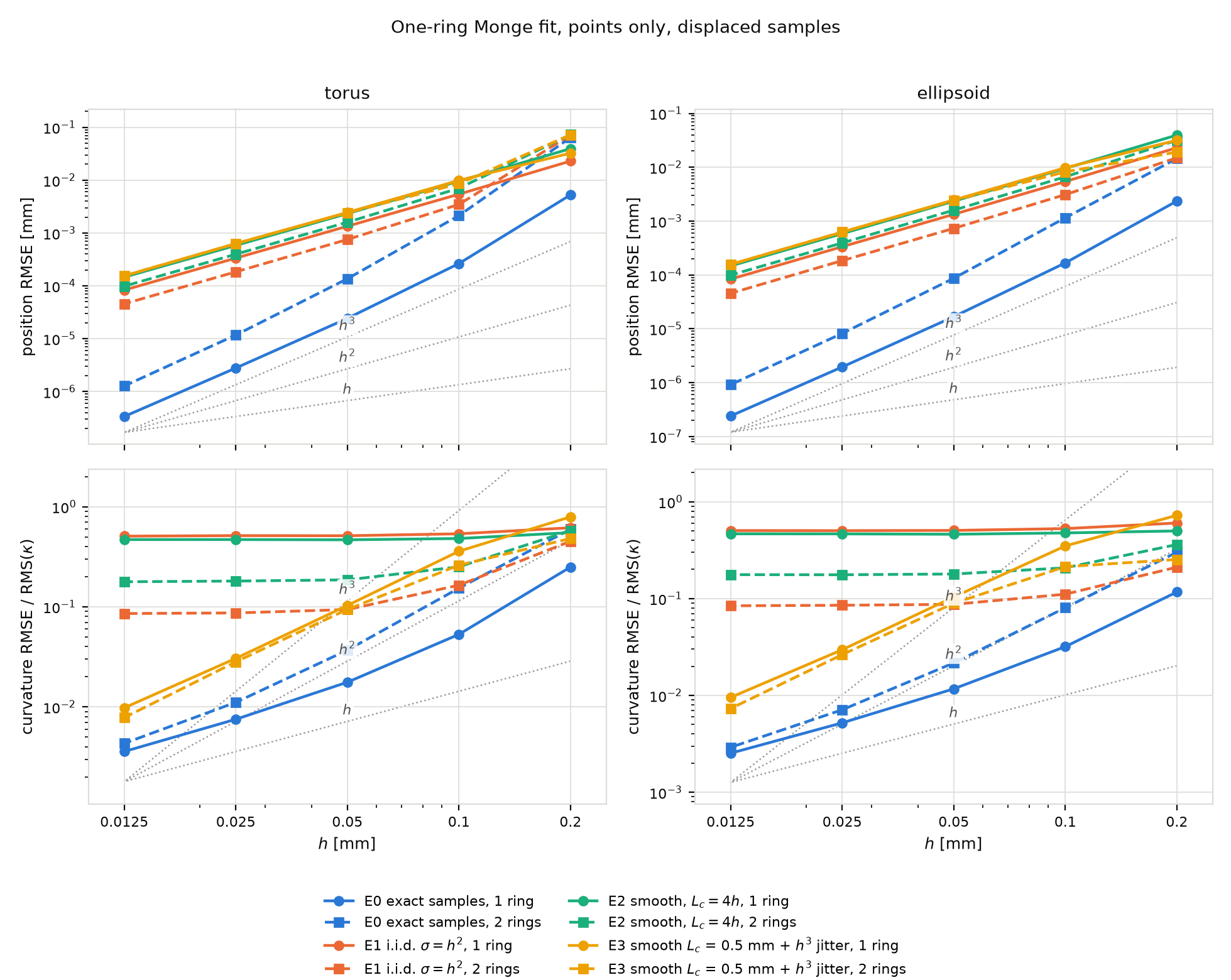}
\caption{The one-ring fit study: curvature and position errors against the cell size for the error models E0 to E3, one and two rings, on the torus (left) and the ellipsoid (right).}
\label{fig:onering}
\end{figure}

The result of \cref{sec:curv} is first order asymptotically, and the reason lies in the stencil. A quadric fitted to samples within one ring of cells, a patch of width $O(h)$, differences the samples over a length $O(h)$: with second-order normals the tensor is first order, and with sample positions carrying independent errors $\sigma = O(h^2)$ the curvature error scales as $\sigma\sqrt{45/n}/L^2$ for a patch of half-width $L$ and $n$ samples, which is $O(1)$ for $L = O(h)$. Second order from such a fit is possible only with a patch of fixed physical half-width, or with sample errors that are smooth on such a length. A synthetic study measures this (\cref{tab:onering}, \cref{fig:onering}): a Cartesian grid of spacing $h = 4\,\mathrm{mm}/N$, $N = 20 \ldots 320$, cut cells by the sign change of the implicit function at the cell vertices, one sample per cut cell at the closest-point projection of the cell centre displaced by a random offset of up to $0.3h$ (a symmetric Cartesian stencil cancels part of the cubic term: with undisplaced centres the order rises to 1.23, so the displacement is kept as the honest case), a patch of the cut cells of the $3\times3\times3$ block around the centre cell (about 15 samples), a Monge patch of six coefficients in the frame of the exact normal (the form that is well posed with one ring), 20 random orientations and 4000 patch centres per level, on a torus with radii 1 and $0.4\,\mathrm{mm}$ (Gaussian curvature of both signs) and on the ellipsoid of \cref{tab:ellipsoid}. Error models applied to the sample positions along the normal: E0 none; E1 independent Gaussian errors of standard deviation $h^2$; E2 a smooth random field with correlation length $4h$ and amplitude $h^2$; E3 a smooth field of amplitude $h^2$ with a fixed correlation length of $0.5\,\mathrm{mm}$ plus independent jitter $h^3$.

\begin{table}[htbp]
\centering
\caption{The one-ring quadratic fit on the torus and the ellipsoid: root-mean-square curvature error relative to the root mean square of the exact curvature, and the position error (root-mean-square distance of the fitted surface from the exact one at the centre, in mm), against the cell size; orders over the three finest levels. E2 shows the same orders as E1 (curvature error 0.47 of the exact root mean square, order 0.0). Two rings (about 50 samples): E0 1.55, E1 0.07, E3 1.80. E1 does not converge because its sample errors, of size $h^2$ and independent per cell, are differenced by the fit over a patch of width $O(h)$: the curvature error $\sigma\sqrt{45/n}/L^2$ of the text is $O(1)$ for $\sigma = h^2$ and $L = O(h)$ at every refinement. Scripts in the archived data \citep{LeiaRepo2026}.}
\label{tab:onering}
\small
\begin{tabular}{lccccc}
\toprule
$h$ [mm] & torus E0 $\kappa$ & torus E1 $\kappa$ & torus E3 $\kappa$ & torus E3 position & ellipsoid E3 $\kappa$ \\
\midrule
0.2 & $2.50\times10^{-1}$ & $6.16\times10^{-1}$ & $7.92\times10^{-1}$ & $3.24\times10^{-2}$ & $7.24\times10^{-1}$ \\
0.1 & $5.30\times10^{-2}$ & $5.36\times10^{-1}$ & $3.59\times10^{-1}$ & $9.90\times10^{-3}$ & $3.50\times10^{-1}$ \\
0.05 & $1.76\times10^{-2}$ & $5.13\times10^{-1}$ & $1.03\times10^{-1}$ & $2.45\times10^{-3}$ & $1.04\times10^{-1}$ \\
0.025 & $7.52\times10^{-3}$ & $5.15\times10^{-1}$ & $3.05\times10^{-2}$ & $6.31\times10^{-4}$ & $2.96\times10^{-2}$ \\
0.0125 & $3.58\times10^{-3}$ & $5.08\times10^{-1}$ & $9.79\times10^{-3}$ & $1.55\times10^{-4}$ & $9.55\times10^{-3}$ \\
\midrule
order & $1.15$ & $0.01$ & $1.70$ & $1.99$ & $1.72$ \\
\bottomrule
\end{tabular}

\end{table}

With exact samples the one-ring curvature converges at first order (1.15 on the torus, 1.10 on the ellipsoid; two rings 1.55 and 1.45; the ellipsoid and two-ring orders from the study's order table in the released data) and the position at third order. Sample errors of size $h^2$ that are independent per cell (E1) or correlated over four cells (E2) give no curvature convergence at all, an error of about half the root mean square of the exact curvature, as the scaling above predicts for $\sigma = h^2$ and $L = O(h)$: the fit differences errors of the size of the sagitta over the length of one cell, and the quotient does not fall under refinement; the position stays second order (1.99 for E3 in \cref{tab:onering}, the same for E1 and E2). Only errors that are smooth over a fixed physical length (E3) give an order near two (1.70 on the torus, 1.72 on the ellipsoid, 1.80 with two rings). The convergence of a one-ring curvature is therefore decided by the smoothness of the sample error at the mesh scale, not by the fit; the recovered planes of \cref{tab:ellipsoid}, whose one-ring fit (centroids and normals) converged at 1.55, are the smooth case; adding the exact normals to the synthetic fit changed neither its orders nor its errors. For a curvature model on a VOF iso-surface, whose errors are at the mesh scale, this means that the regularization must smooth over a fixed physical length, and that length competes with the features of the geometry: in the microcavities of \citet{NagelEtAl2025MicrocavityFilling}, $0.07$ and $0.14\,\mathrm{mm}$ deep at $33\,\mu\mathrm{m}$ cells, a length of $0.5\,\mathrm{mm}$ would smooth the feature away. The scaling of the regularization length with the mesh and its bias are the questions a curvature model from VOF data has to answer; a quadric fitted with the DAVOF normal as its frame needs only the first-order tensor for the clipping surface of \cref{sec:transport}, which one ring provides.

\section{The regeneration map and its fixed points}
\label{sec:regen}

The question for transport is whether a face update keeps $q = 2$. Before any transport code exists, one map can be measured: the \emph{regeneration}, in which every interface cell recovers from its own state the normal \cref{eq:avof-normal}, the plane \cref{eq:avof-position} and the one-ring curvature tensor of \cref{sec:curv}, builds a clipping surface from them and cuts its own faces again, which yields new face fractions. Applied once, this is the zero-velocity limit of the face update of \cref{sec:transport}; applied $K$ times to its own output, it is the zero-velocity limit of a transport loop, and a number of applications proportional to $1/h$ is one transport period at fixed Courant number. The face-fraction error is measured as a fraction, in $L_1$ over the (cell, face) pairs of the interface cells, against a twice-refined reference cut of the exact surface; the owner-neighbour mismatch of a shared face, which each cell cuts from its own surface, is reported separately.

\subsection{One application: second order with the face-anchored paraboloid}

Five clipping surfaces (\emph{arms}) were measured on the sphere, on the exact state and on the chord state (construction (iii)), and on the ellipsoid (chord state), $N = 20 \ldots 160$, with the orders over the three finest meshes. The plane \cref{eq:avof-position} regenerates the fractions at $q = 1$ (0.98 to 1.01), as \cref{eq:clipping-law} predicts. A paraboloid with the recovered normal and the ring tensor whose vertex sits on that plane is $O(h^2)$ off the interface and gives $q = 1$ as well. The \emph{face-anchored} paraboloid, whose offset along the normal is set so that its wetted area over the cell's faces equals the state's (one scalar from the $F$ face fractions), regenerates the fractions at $q = 2$ on every state and surface tested: 1.98 on the exact sphere, 2.00 on the chord sphere, 2.09 on the ellipsoid (pairwise 2.12, 2.18, 2.00), with an $L_1$ error at $N = 160$ that is 300 and 190 times below the plane's on the sphere and the ellipsoid. The owner-neighbour mismatch of the cell-local regeneration is $O(h^2)$ for the face anchoring (2.00, 2.01) and $O(h)$ for the plane. Two pre-registered predictions were falsified and explain the design: a paraboloid matched to the cell volume $\alpha_k$ with the Detrixhe--Aslam tetrahedral quadrature is $q = 2$ on the chord state (1.97, 2.04), because the quadrature carries the same chord defect as the cell fraction and the defects cancel, but $q = 1$ on the exact state (1.00); and a paraboloid matched to its own refined cut volume is $q = 1$ on the exact state too (1.05), because its twice-refined linear cut keeps one sixteenth of the chord defect. A volume anchor is consistent only in the quadrature that defined the cell fraction; the face anchor needs no pairing. The regenerated normal and position are second order for every arm, because every clipping surface's mean normal is second order; the fraction order, not the normal, is the result of a zero-step test.

\subsection{Iterated: first order per period, and the fixed points}

Iterating the map (sphere and ellipsoid, $N = 20 \ldots 160$, $K = 500$, and $K = 5000$ on the two coarsest sphere meshes) exposes what one application does not. First, the explicit position \cref{eq:avof-position} inside the loop amplifies the realizability defect by a factor of $-2$ per application (the plane moves by three times the defect, \cref{sec:defects}, so the defect changes sign and doubles: $1 - 3 = -2$), and a loop built on it explodes within ten iterations: the plane a transport loop uses must be the one positioned by $\alpha_k$ (the explicit position is its initial guess and a diagnostic), which with each cell's own view of its faces is an exact fixed point of the map, constant to every digit over 5000 iterations, at $q = 1$. Second, the face-anchored paraboloid is not a fixed point: the fraction error grows linearly with the iteration count, with an increment per application of the size of the first regeneration's error, $O(h^2)$, and after $K \propto 1/h$ applications the error is first order per period, 0.89, 0.90, 0.89 between successive meshes on the sphere and 1.03, 0.89, 0.88 on the ellipsoid (\cref{tab:iterated}); at fixed large $K$ the error is independent of $h$. The normal rotates by $O((h/R)^2)$ per application, because the paraboloid's mean normal equals its axis only when the vertex sits at the patch centroid. Anchoring instead the three components of the cell's wetted area vector (offset plus in-plane vertex position, by Newton iteration) lowers the constant by a factor of 5.5 to 6.9 but drifts at the same order (0.78, 0.94, 0.98 and 1.06, 1.00, 0.92), and the Newton iteration fails in a constant 13 to 18\,\% of the interface cells on every mesh, where the in-plane vertex has no root. With a single-valued face view, in which the two cells of a face agree on one fraction by averaging, every arm saturates or diverges. Third, the one-step cut against faces displaced by up to $0.5h$ along the normal, the closest zero-velocity stand-in for a displaced pre-image (the face traced backward through the flow, \cref{sec:transport}), is $q = 2$ for both paraboloid anchorings (orders 2.2 to 2.7) and $q = 1$ for the plane: it is the repetition that costs the order, not the displacement.

\begin{table}[htbp]
\centering
\caption{The iterated regeneration: $L_1$ face-fraction error as a fraction after $k = 2N$ applications, one transport period at fixed Courant number, for $N = 20, 40, 80, 160$, and the order between successive meshes ($K = 500$; the same orders hold at $k = 4N$ where it lies inside the run). Cell view: every cell cuts its own faces from its own surface; for the $\alpha_k$-positioned plane the error after $2N$ applications equals the error after one. From the leia tables \texttt{davof\_regen\_iter\_*}.}
\label{tab:iterated}
\footnotesize
\setlength{\tabcolsep}{4pt}
\begin{tabular}{llccccl}
\toprule
clipping surface & surface & $N = 20$ & $40$ & $80$ & $160$ & order per period \\
\midrule
plane positioned by $\alpha_k$ & sphere & 3.92e-03 & 1.98e-03 & 9.92e-04 & 4.96e-04 & fixed point, $q = 1$ \\
 & ellipsoid & 5.48e-03 & 2.82e-03 & 1.40e-03 & 7.05e-04 & fixed point, $q = 1$ \\
face-anchored paraboloid & sphere & 5.52e-03 & 2.98e-03 & 1.60e-03 & 8.67e-04 & 0.89, 0.90, 0.89 \\
 & ellipsoid & 1.29e-02 & 6.30e-03 & 3.40e-03 & 1.85e-03 & 1.03, 0.89, 0.88 \\
area-vector-anchored paraboloid & sphere & 8.21e-04 & 4.80e-04 & 2.50e-04 & 1.26e-04 & 0.78, 0.94, 0.98 \\
 & ellipsoid & 2.34e-03 & 1.12e-03 & 5.58e-04 & 2.95e-04 & 1.06, 1.00, 0.92 \\
\bottomrule
\end{tabular}
\end{table}

\subsection{The six-coefficient surface: the identity of the map}
\label{sec:full}

The anchored paraboloids drift because neither passes through the cell's own face fractions: every re-cut moves the fractions toward a foreign set, with one sign, $1/h$ times per period. A paraboloid whose six coefficients in the frame of the normal (offset, two slopes, the three components of the curvature tensor) are fitted to the cell's $F$ face fractions reproduces them by construction and makes the zero-velocity map the identity. The fit is a damped Gauss--Newton iteration from the face-anchored paraboloid, the residual being the $|S_f|$-weighted difference between each face's cut and the cell's own fraction, with the ring tensor entering as the damping prior (the surface usually crosses four or five of the six faces of a hexahedron, so the system is under-determined and the damped step keeps the null-space components at their start). Measured on the sphere and on the ellipsoid, $N = 20 \ldots 160$, $K = 500$ (\cref{tab:full}): the fit converges in all but 0.0 to 0.6\,\% of the interface cells, 0.1\,\% on the three finer meshes; the regeneration reproduces the input state's own fractions, so that after one application and after 500 the $L_1$ error equals the input state's error against the reference to three digits on every mesh and both surfaces, in the cell view and in the single-valued face view alike, where every anchored arm saturates or diverges; the owner--neighbour mismatch is $10^{-5}$ to $10^{-8}$; the regenerated normal is the state's normal except in the non-converged cells. One prediction fell: the displaced-face cut has a constant 1.2 to 1.5 times the face-anchored arm's, at the same second order (2.2 to 2.8 between meshes); the interpolating surface is exact on the faces and slightly worse away from them. This surface is therefore the fixed point of the zero-velocity map at every resolution tested; the transport question becomes the remainder of the fit at the displaced pre-image, $O(h^2)$ per step, and whether its sum over a period is $O(h^2)$ or $O(h)$.

\begin{table}[htbp]
\centering
\caption{The six-coefficient surface, cell view, $K = 500$: the input state's own $L_1$ face-fraction error against the reference ($E_0$), the error after 500 applications, the share of interface cells in which the fit did not converge, the owner--neighbour mismatch after one application, and the one-step cut against faces displaced by $-0.5h$ with the face-anchored arm's value in parentheses. From the leia tables \texttt{davof\_regen\_iter\_*} and the per-case files.}
\label{tab:full}
\footnotesize
\begin{tabular}{lcccccc}
\toprule
surface & $N$ & $E_0$ & $E(500)$ & non-converged & mismatch & displaced $-0.5h$ \\
\midrule
sphere & 20 & 9.62e-05 & 9.99e-05 & 0.6\,\% & 1.2e-05 & 7.4e-04 (5.6e-04) \\
 & 40 & 1.25e-05 & 1.28e-05 & 0.1\,\% & 7.1e-07 & 1.13e-04 (8.7e-05) \\
 & 80 & 2.03e-06 & 2.03e-06 & 0.1\,\% & 8.1e-09 & 1.67e-05 (1.42e-05) \\
 & 160 & 7.48e-07 & 7.47e-07 & 0.1\,\% & 6.5e-09 & 3.78e-06 (3.07e-06) \\
ellipsoid & 20 & 1.49e-04 & 1.51e-04 & 0.0\,\% & 1.9e-06 & 1.96e-03 (1.69e-03) \\
 & 40 & 3.30e-05 & 3.31e-05 & 0.2\,\% & 1.6e-07 & 3.39e-04 (2.62e-04) \\
 & 80 & 5.82e-06 & 5.83e-06 & 0.2\,\% & 4.6e-08 & 6.76e-05 (4.65e-05) \\
 & 160 & 1.38e-06 & 1.38e-06 & 0.1\,\% & 1.9e-08 & 1.50e-05 (1.03e-05) \\
\bottomrule
\end{tabular}
\end{table}

\section{Transport: the closed loop under uniform translation}
\label{sec:tpf}

This section measures the closed loop: the face fractions set the normal, the normal sets the clipping surface, and the clipping surface sets the next face fractions. Two loops were built in a fork of the TwoPhaseFlow library of \citet{ScheuflerRoenby2023TwoPhaseFlow}, inside the isoAdvector--plicRDF framework \citep{RoenbyEtAl2016,ScheuflerRoenby2019} that the comparators use: the plane-clipped loop, in which the clipping surface is the PLIC plane positioned by $\alpha_k$ and the volume fluxes are isoAdvector's, and the dual advection with one surface per cell, in which a paraboloid serves the volume flux and the face update alike. Both are measured on a translated plane and on a translated sphere. The results are stated here with the tables and figures that carry the numbers; the implementation record of every measurement, in the same order, is \ref{app:transport}.

\subsection{Setup}
\label{sec:tpf-setup}

The state is initialised by the constructions of \cref{sec:static}. In the plane-clipped loop a reconstruction scheme forms the normal \cref{eq:avof-normal} from the stored face fractions and positions the PLIC plane by $\alpha_k$ with the library's bisection; isoAdvector sweeps every face of an interface cell with the moving plane, the submerged area of the face at the end of the step is the new face fraction, and a face shared by two interface cells takes the upwind cell's plane, consistent with its volume flux. By \cref{eq:clipping-law} this plane clipping gives first-order face fractions, $q = 1$; it is measured first on purpose, as the baseline for the paraboloid. The time step is fixed at a Courant number $\mathrm{Co} = |\mathbf{U}|\Delta t/h$; a run is recorded as diverged when $\alpha$ leaves $[-0.5, 1.5]$ or the number of interface cells exceeds ten times its initial value. The translated plane is the plane of \cref{sec:plane}, carried over $0.2L$ along its normal and along $(1, 0, 0)$, $N = 20$ to $160$; the translated sphere has $R = 0.15$ in a $2 \times 1 \times 1$ box and is carried by $\mathbf{U} = (1, 0, 0)$ over one box length, 3.3 diameters, $N = 32$, $64$, $128$ cells across the box height ($R/h = 4.8$ to $19$). \Cref{tab:tpf-sources} lists the normal sources that drive the loop and the controls. Every prediction was pre-registered in the study configuration before the run; the ones that failed are reported where they failed.

\begin{table}[htbp]
\centering
\caption{The normal sources of the closed loop, the controls and the comparators. Every row runs on the identical initial state; plicRDF and isoAlpha run on the same $\alpha_k$ (the Detrixhe--Aslam cell fraction, not the one their own initialisation would give).}
\label{tab:tpf-sources}
\footnotesize
\begin{tabular}{p{0.26\linewidth}p{0.66\linewidth}}
\toprule
normal in the loop & definition \\
\midrule
own normal & \cref{eq:avof-normal} on the cell's own faces, as it stands \\
ring normal & \cref{eq:avof-normal} on the union of the cell's ring: $\mathbf{n}_k = \sum_j \mathbf{m}_j / |\sum_j \mathbf{m}_j|$ over the point neighbours $j$ (the interior faces cancel); exact for a plane, never divided by a small area, the mean normal of the patch \\
ring quadric & the normal at the cell of a weighted least-squares Monge patch through the ring's recovered planes (\cref{sec:curv}) \\
exact normal (control) & the normal of the translated surface; the loop is open \\
plicRDF normal (control) & the plicRDF normal from $\alpha_k$ in the DAVOF update \\
\midrule
plicRDF, isoAlpha & the library's own methods on the same $\alpha_k$; they carry no face fractions \\
\bottomrule
\end{tabular}
\end{table}

\subsection{The cell's own normal in the loop, and its stabilisation as a research goal}
\label{sec:tpf-sliver}

In the loop each face takes its fraction from the surface of its upwind cell, so the fractions of one cell come from several surfaces, which agree only in exact arithmetic. Linearised about the exact plane, the turn of the cell's area vector $\mathbf{m}_k$ is the neighbours' displacement at the traces of the faces they set, weighted by the trace length times its lever arm of about $h/2$, divided by the cell's own patch area $A$. The own faces' weights sum to one, so the loop has no restoring force, and a foreign weight exceeds one where $A$ is small against the lever arm: in a sliver cell, an interface cell whose wetted area is a small fraction of $h^2$. The stencil reconstructions of classical VOF average the volume fractions of a neighbourhood before anything is divided and never meet this. Sliver cells are not rare (\cref{tab:tpf-sliver}); a rule that treats them specially acts on about one cell in eight at every resolution. Measured cell by cell on one plane run at $N = 80$, the linearised sum predicts the turn (correlation $0.99$, median ratio $1.0$), cells fed only their own surface turn by their own tilt, and the loop grows as one mode. The own-normal runs start at round-off and grow geometrically from the second step, by $1.10$ to $1.26$ per step at $\mathrm{Co} = 0.2$ and $1.6$ to $8.8$ at $0.5$: the linear instability of an exact fixed point, independent of the clipping surface. The ring normal divides the same mismatch by about $9h^2$ and contracts, at the price of being the mean normal of a patch of width $3h$. The research goal that follows is a stabilisation of the cell's own normal that keeps its exactness on a plane and contracts the mismatch in sliver cells; the sliver threshold alone is not it (\cref{sec:tpf-ladder}), and the ring-state fit of \cref{sec:transport}, which forms no normal in any cell, is the first candidate.

\begin{table}[htbp]
\centering
\caption{Sliver cells on the translated sphere: the share of interface cells with $|A_{\tilde\Sigma_k}| < 0.1 h^2$ at the first step and over a run.}
\label{tab:tpf-sliver}
\footnotesize
\begin{tabular}{lrrr}
\toprule
 & $N = 32$ & $N = 64$ & $N = 128$ \\
\midrule
first step & $11.5\,\%$ & $13.6\,\%$ & $11.3\,\%$ \\
over a run & \multicolumn{3}{c}{$8$ to $14\,\%$} \\
\bottomrule
\end{tabular}
\end{table}

\subsection{The plane-clipped loop}
\label{sec:tpf-plane-loop}

\begin{figure}[htbp]
\centering
\includegraphics[width=0.78\linewidth]{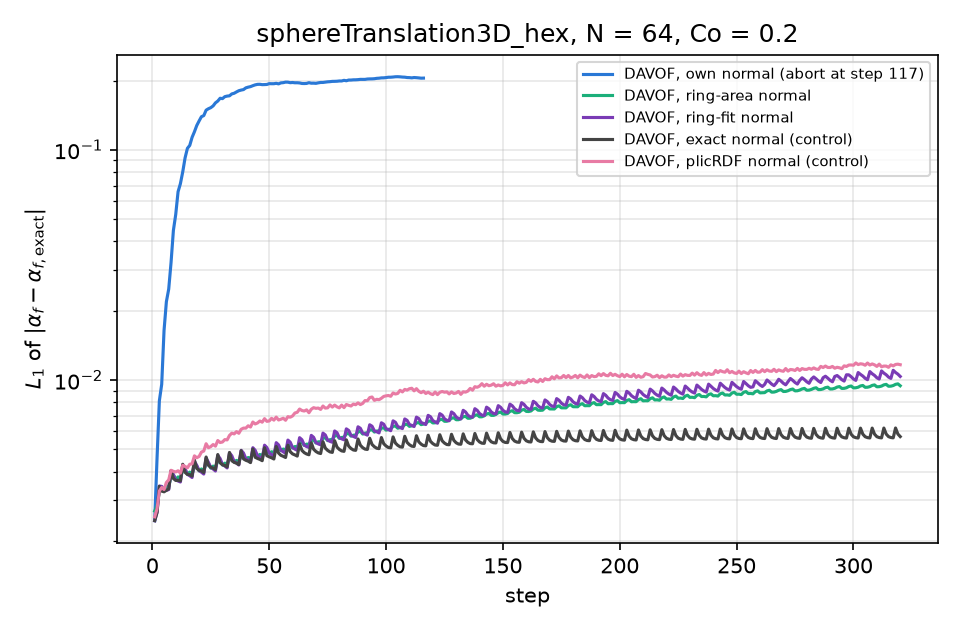}
\caption{The face-fraction error $L_1(\alpha_f)$ against the step for the translated sphere at $N = 64$ and $\mathrm{Co} = 0.2$ (320 steps), plane-clipped loop. The own normal grows from the first steps and is stopped at step 117; the ring normal and the ring quadric grow slowly from the $O(h/R)$ clipping error of the first step; the exact-normal control saturates; the plicRDF-normal control grows as the ring normal does at a higher level.}
\label{fig:tpf-history}
\end{figure}

\Cref{tab:tpf-plane} scores the translated plane, \cref{tab:tpf-sphere} the translated sphere, \cref{fig:tpf-history} the error against the step. On the plane the ring normal, the ring quadric and the exact normal keep the face fractions and the normal at round-off in all 32 runs each; the own normal holds the gate at $N = 20$ and $40$ for $\mathrm{Co} \le 0.2$, and its stabilisation at finer meshes and larger Courant numbers is the goal of \cref{sec:tpf-sliver}; the plicRDF normal is not exact on the corner cells of a plane and serves as the control only. On the sphere the ring normal is the first stable DAVOF transport loop: bounded to $10^{-14}$ and conservative to round-off at every resolution and Courant number, where the own normal is stopped after 45 to 232 steps, and the ring quadric, bounded but with its fit failing in up to 1700 cells per step under the plane flux, needs the surface's own flux of \cref{sec:tpf-dual}, under which it converges in every cell. The face fractions transport at first order with the ring normal and with the exact normal alike: the order is the plane clipping's, $q = 1$ by \cref{eq:clipping-law}, not the normal's. Against plicRDF on the same $\alpha_k$ the ring normal is first order at $1.1$ to $1.35$ times plicRDF's error, the centroid is second order at plicRDF's value to three digits, the shape error is plicRDF's, and the cost is below plicRDF's (\cref{tab:tpf-cost}). At fixed $N$ the ring normal's errors rise as the Courant number falls (\cref{tab:tpf-courant}), because the per-step clipping error accumulates over the $1/\mathrm{Co}$ times more steps; the pre-registered prediction of the opposite direction was falsified. One measurement remains: the plicRDF-normal control and plicRDF itself differ in the fluxes beyond the downwind faces and in the plane positioned along the helper's normal, and which of the two sets the control's convergence on the sphere is to be measured (\ref{app:transport}).

\begin{table}[htbp]
\centering
\caption{The translated plane in TwoPhaseFlow: the plane of \cref{sec:plane} carried over $0.2L$ along its normal and along $(1, 0, 0)$, $N = 20, 40, 80, 160$, $\mathrm{Co} = 0.05$ to $0.5$. For every normal source and Courant number the number of the eight runs (two directions, four meshes) in which the face fractions and the normal stay at round-off over the whole run (both run maxima of $L_1$ below $10^{-12}$); the last column is the run maximum of $L_1(\alpha_f)$ at $N = 160$ and $\mathrm{Co} = 0.5$ on the normal arm (for plicRDF and isoAlpha, which carry no face fractions, the normal error). Table body generated by \texttt{make\_table\_advection.py} from the study table.}
\label{tab:tpf-plane}
\footnotesize
\setlength{\tabcolsep}{4pt}
\resizebox{\linewidth}{!}{
\begin{tabular}{lrrrrr}
\toprule
normal & Co $= 0.05$ & Co $= 0.1$ & Co $= 0.2$ & Co $= 0.5$ & $\max_t L_1(\alpha_f)$, $N = 160$, Co $= 0.5$ \\
\midrule
DAVOF, ring normal & 8/8 & 8/8 & 8/8 & 8/8 & 6.24e-14 \\
DAVOF, own normal & 4/8 & 4/8 & 1/8 & 1/8 & 3.64e-01 \\
DAVOF, exact normal (control) & 8/8 & 8/8 & 8/8 & 8/8 & 5.48e-14 \\
DAVOF, plicRDF normal & 0/8 & 0/8 & 0/8 & 0/8 & 9.89e-02 \\
DAVOF, ring quadric & 8/8 & 8/8 & 8/8 & 8/8 & 5.72e-14 \\
plicRDF & -- & -- & -- & -- & 7.07e-02 (normal) \\
isoAlpha & -- & -- & -- & -- & 1.76e-01 (normal) \\
\bottomrule
\end{tabular}
}
\end{table}

\begin{table}[htbp]
\centering
\caption{The translated sphere in TwoPhaseFlow: $R = 0.15$ in a $2 \times 1 \times 1$ box carried by $\mathbf{U} = (1, 0, 0)$ over one box length, 3.3 diameters, $N = 32, 64, 128$ cells across the box height ($R/h = 4.8$ to $19$), $\mathrm{Co} = 0.1, 0.2, 0.5$. Errors at the end of the run on the finest mesh and the least-squares order $p$ over the three meshes: $L_1(\alpha_f)$ the face-fraction error as a fraction over the faces of the interface cells, $L_1(\mathbf{n})$ and $L_1(\mathbf{x})$ the normal and centroid errors of the recovered planes against the translated sphere, $E_{\mathrm{shape}}$ the volume of the symmetric difference between the transported and the exact liquid. The own normal aborts at the listed steps (interface cells above ten times the initial count) for $N = 32/64/128$. Every other run is bounded to $10^{-14}$ and conservative to round-off. Table body generated by \texttt{make\_table\_advection.py}.}
\label{tab:tpf-sphere}
\footnotesize
\setlength{\tabcolsep}{3.5pt}
\resizebox{\linewidth}{!}{
\begin{tabular}{llrrrrrrr}
\toprule
normal & Co & $L_1(\alpha_f)$ & $p$ & $L_1(\mathbf{n})$ & $p$ & $L_1(\mathbf{x})$ & $p$ & $E_{\mathrm{shape}}$ \\
\midrule
DAVOF, ring normal & 0.1 & 7.72e-03 & 0.80 & 1.26e-02 & 0.94 & \textbf{1.34e-04} & 1.86 & \textbf{3.90e-05} \\
 & 0.2 & 5.76e-03 & 0.81 & 1.09e-02 & 0.95 & \textbf{1.02e-04} & 1.86 & \textbf{3.05e-05} \\
 & 0.5 & 2.96e-03 & 1.00 & 8.12e-03 & \textbf{1.02} & 4.88e-05 & 1.99 & \textbf{1.58e-05} \\
\midrule
DAVOF, own normal & 0.1 & \multicolumn{7}{l}{aborted at step 205/232/177} \\
 & 0.2 & \multicolumn{7}{l}{aborted at step 111/117/97} \\
 & 0.5 & \multicolumn{7}{l}{aborted at step 46/56/45} \\
\midrule
DAVOF, exact normal (control) & 0.1 & 3.84e-03 & 0.85 & \textbf{5.55e-04} & \textbf{1.43} & \textbf{7.93e-05} & \textbf{1.92} & \textbf{2.19e-05} \\
 & 0.2 & 3.32e-03 & 0.82 & \textbf{4.70e-04} & \textbf{1.48} & \textbf{6.76e-05} & 1.93 & \textbf{1.94e-05} \\
 & 0.5 & 3.46e-03 & 0.99 & \textbf{5.54e-04} & \textbf{1.90} & \textbf{4.76e-05} & 1.98 & \textbf{1.58e-05} \\
\midrule
DAVOF, plicRDF normal & 0.1 & 2.19e-02 & 0.16 & 5.17e-02 & 0.05 & 5.11e-04 & 1.03 & 6.75e-05 \\
 & 0.2 & 1.68e-02 & 0.15 & 4.10e-02 & -0.01 & 3.09e-04 & 1.18 & 6.26e-05 \\
 & 0.5 & 8.57e-03 & 0.24 & 2.70e-02 & -0.11 & 1.33e-04 & 1.26 & 3.40e-05 \\
\midrule
\midrule
DAVOF, ring quadric & 0.1 & 7.97e-03 & 0.82 & \textbf{7.80e-03} & \textbf{1.09} & \textbf{1.39e-04} & 1.88 & 3.98e-05 \\
 & 0.2 & 6.02e-03 & 1.86 & \textbf{5.49e-03} & \textbf{2.61} & 1.05e-04 & \textbf{3.49} & 3.10e-05 \\
 & 0.5 & 3.38e-03 & 1.06 & \textbf{1.50e-03} & \textbf{1.60} & \textbf{4.71e-05} & 1.98 & \textbf{1.57e-05} \\
\midrule
\midrule
\midrule
\midrule
\midrule
plicRDF & 0.1 & -- & -- & 1.14e-02 & 1.03 & 1.40e-04 & 1.90 & 3.93e-05 \\
 & 0.2 & -- & -- & 9.20e-03 & 1.04 & 1.04e-04 & 1.94 & 3.08e-05 \\
 & 0.5 & -- & -- & 6.03e-03 & 0.97 & 4.87e-05 & 1.99 & 1.61e-05 \\
\midrule
isoAlpha & 0.1 & -- & -- & 2.29e-02 & 0.80 & 1.62e-04 & 1.87 & 4.58e-05 \\
 & 0.2 & -- & -- & 2.03e-02 & 0.74 & 1.20e-04 & 1.87 & 3.53e-05 \\
 & 0.5 & -- & -- & 1.65e-02 & 0.64 & 5.97e-05 & 1.85 & 1.71e-05 \\
\bottomrule
\end{tabular}
}
\end{table}

\begin{table}[htbp]
\centering
\caption{The Courant dependence at fixed resolution. Plane-clipped loop with the ring normal at $N = 64$: the error accumulates over the $1/\mathrm{Co}$ steps and rises as the Courant number falls. Dual advection with the ring-tensor paraboloid and the ring normal at $N = 128$: the error no longer accumulates and rises with the pre-image's displacement, i.e.\ with the Courant number. The exact-normal controls are flat in the Courant number in both loops.}
\label{tab:tpf-courant}
\footnotesize
\begin{tabular}{llrrr}
\toprule
loop & error & $\mathrm{Co} = 0.1$ & $\mathrm{Co} = 0.2$ & $\mathrm{Co} = 0.5$ \\
\midrule
plane-clipped, ring normal, $N = 64$ & $L_1(\alpha_f)$ & 1.22e-02 & 9.45e-03 & 6.06e-03 \\
 & $L_1(\mathbf{n})$ & 2.53e-02 & 2.23e-02 & 1.84e-02 \\
paraboloid by volume, ring normal, $N = 128$ & $L_1(\alpha_f)$ & 8.52e-04 & 8.71e-04 & 1.22e-03 \\
\bottomrule
\end{tabular}
\end{table}

\subsection{What a second-order clipping needs: one surface per cell}
\label{sec:tpf-sagitta}

The plane is $O(h^2)$ off the interface, so the leading-order repair that stays within the plane loop corrects the clipped area of every face by the sagitta between the interface's arc and the plane's chord on the face, from the ring curvature tensor of \cref{sec:curv} (the formula is in \ref{app:transport}). \Cref{tab:tpf-sagitta} shows what it does: in pure geometry it removes nearly all of the plane's face error; after one step from the exact state it removes $93\,\%$; over a run it keeps $27$ to $30\,\%$ of that gain and leaves the order at one; on the chord-initialised state it doubles the error from the first step. The per-face fields locate the cause in the volume anchor: the volume fluxes are the volumes under the moving plane, so within the $1/\mathrm{Co}$ steps in which a cell holds the interface $\alpha_k$ acquires a one-signed $O(h/R)$ deviation from the volume under the surface, and a correction that is right for the exact state overshoots. This is the finding of \cref{sec:regen} under transport: a volume anchor is consistent only in the quadrature that defined the cell fraction. The requirement that follows is measured, not assumed: the volume flux and the face update must be the volume and the boundary moments of one surface per cell.

\begin{table}[htbp]
\centering
\caption{The sagitta correction of the plane-clipped face areas (\texttt{faceClipping planeSagitta} of the fork), translated sphere, exact normal in the loop, $\mathrm{Co} = 0.2$ unless stated. The DA-Q state keeps the chord cell fraction of the Detrixhe--Aslam construction; the exact state has $\alpha_k$ equal to the volume under the sphere. Orders over $N = 32$, $64$, $128$.}
\label{tab:tpf-sagitta}
\footnotesize
\begin{tabular}{p{0.36\linewidth}p{0.28\linewidth}p{0.28\linewidth}}
\toprule
test & plane clipping & with the correction \\
\midrule
pure geometry: random cubes cut by a sphere, exact tensor and volume, $h/R = 0.21$ to $0.05$ & -- & $92$ to $97\,\%$ of the plane's face error removed \\
one step, exact state, $N = 32$, $\mathrm{Co} = 0.02$ & $L_1(\alpha_f) = 6.34\cdot10^{-3}$ & $4.40\cdot10^{-4}$ ($93\,\%$ removed; $90\,\%$ at $\mathrm{Co} = 0.2$) \\
one step, DA-Q state, $N = 32$ & $2.78\cdot10^{-3}$ & $6.22\cdot10^{-3}$ \\
run, DA-Q state, $N = 32/64/128$ & $1.04\cdot10^{-2}$, $5.68\cdot10^{-3}$, $3.32\cdot10^{-3}$ (order $0.82$) & $1.03\cdot10^{-2}$, $5.53\cdot10^{-3}$, $3.23\cdot10^{-3}$ ($0.83$) \\
run, exact state, $N = 32/64/128$ & $1.15\cdot10^{-2}$, $6.24\cdot10^{-3}$, $3.64\cdot10^{-3}$ & $8.42\cdot10^{-3}$, $4.33\cdot10^{-3}$, $2.58\cdot10^{-3}$ ($0.85$); $27$ to $30\,\%$ of the one-step gain kept \\
\bottomrule
\end{tabular}
\end{table}

\subsection{One surface per cell: the dual advection}
\label{sec:tpf-dual}

One time step of the dual advection: (1) the normal and the position of every interface cell follow from $(\alpha_k, \{\alpha_f\})$ by \cref{eq:avof-normal,eq:avof-position}; (2) the curvature tensor of the cell is taken from the ring of recovered normals (\cref{sec:curv}); (3) a surface is built in the frame of the normal; (4) $\alpha_k$ is advanced by the liquid volume of the prism swept between each face's pre-image $f - \mathbf{U}\Delta t$ and the face under the surface of the upwind cell; (5) each $\alpha_f$ is advanced by the pre-image's liquid fraction under the same surface. The surface is a quadric, so along any segment it is a quadratic polynomial with exact roots, and the pre-image cut is the exact face cut of \cref{sec:static} up to its Gauss rule. A face shared by two interface cells takes the upwind cell's surface; a face whose two cells carry no surface takes the closest neighbouring one. Three surfaces were run on the translated sphere at $N = 32$ and $\mathrm{Co} = 0.2$ over 160 steps (\cref{tab:tpf-surfaces}). The exact sphere as every cell's surface is the gate of the machinery, not a test of the method, and it holds: the update reproduces the exact state to round-off and transports an inconsistent initial state unchanged. The six-coefficient surface of \cref{sec:full}, which interpolates the cell's own face fractions, sets the first research goal of the update, a surface that uses more of the cell's own fractions than the normal and stays stable: under transport the faces of a cell come from different surfaces, so the six fractions are not the fractions of any one paraboloid, and a surface that interpolates them carries their inconsistency to the next pre-images amplified; the zero-velocity map is the identity on this surface, a neutral fixed point, and the displaced pre-images push it past one. The ring-state fit of \cref{sec:transport}, which keeps this information without forming a normal in any cell, is the candidate. The surface that is stable takes nothing from the cell's own faces beyond the normal: the frame of the normal, the curvature tensor of the ring, zero slopes, and the offset from the conserved $\alpha_k$ as in classical VOF, with a second-order surface in place of the plane and with the surface's own volume flux. The plane loop is stable with the ring normal for the same reason: its one degree of freedom beyond the normal, the position, comes from $\alpha_k$. The volume anchor, the quadrature and volume rules and the parallel exchange are described in \ref{app:transport}.

\begin{table}[htbp]
\centering
\caption{The three surfaces of the dual advection on the translated sphere, $N = 32$, $\mathrm{Co} = 0.2$, 160 steps, exact-sphere initial state, split quadrature for the volumes (\ref{app:transport}). Errors at the end of the run; the plane-clipped loop on the same state in parentheses.}
\label{tab:tpf-surfaces}
\footnotesize
\begin{tabular}{p{0.30\linewidth}p{0.17\linewidth}p{0.45\linewidth}}
\toprule
surface & normal in the frame & outcome \\
\midrule
exact sphere (gate) & -- & face fractions exact to $10^{-14}$ from the second step on; $\alpha$ bounded to $10^{-17}$, conservative to $3\cdot10^{-17}$; the chord deficit of a DA-Q initial state ($8\cdot10^{-3}$ of $V$) is transported unchanged \\
\midrule
six-coefficient paraboloid fitted to the cell's own face fractions (\cref{sec:full}) & exact / own / ring & first step better than the plane loop's by $20$ / $13$ / $2.6$ ($2.2\cdot10^{-4}$ against $4.4\cdot10^{-3}$ with the exact normal); then the error grows by $2$ to $2.5$ per step and the run is stopped at step 10 to 21, with every normal, at $\mathrm{Co} = 0.05$ as at $0.2$, at both quadrature tolerances; a volume shift of the offset after the fit changes nothing; the one-scalar face anchoring of \cref{sec:regen} is stopped at step 81 \\
\midrule
ring-tensor paraboloid positioned by $\alpha_k$ & exact & bounded to $10^{-20}$, conservative, surface defect at the anchor's tolerance; $L_1(\alpha_f) = 5.7\cdot10^{-4}$ ($1.15\cdot10^{-2}$); normal from the cell's own faces $2.8\cdot10^{-4}$ ($3.7\cdot10^{-3}$); centroid $4.1\cdot10^{-4}$ ($7.7\cdot10^{-4}$); maximum face error $2.4\cdot10^{-2}$ on a few grazing cells \\
 & ring & $L_1(\alpha_f) = 5.9\cdot10^{-3}$ ($1.78\cdot10^{-2}$); centroid $5.3\cdot10^{-4}$ ($1.35\cdot10^{-3}$); transported normal $3.3\cdot10^{-2}$, the ring normal's own bias \\
 & own & diverges through the sliver cells, as the plane loop does \\
\bottomrule
\end{tabular}
\end{table}

\subsection{The resolution ladder}
\label{sec:tpf-ladder}

\begin{figure}[htbp]
\centering
\includegraphics[width=\linewidth]{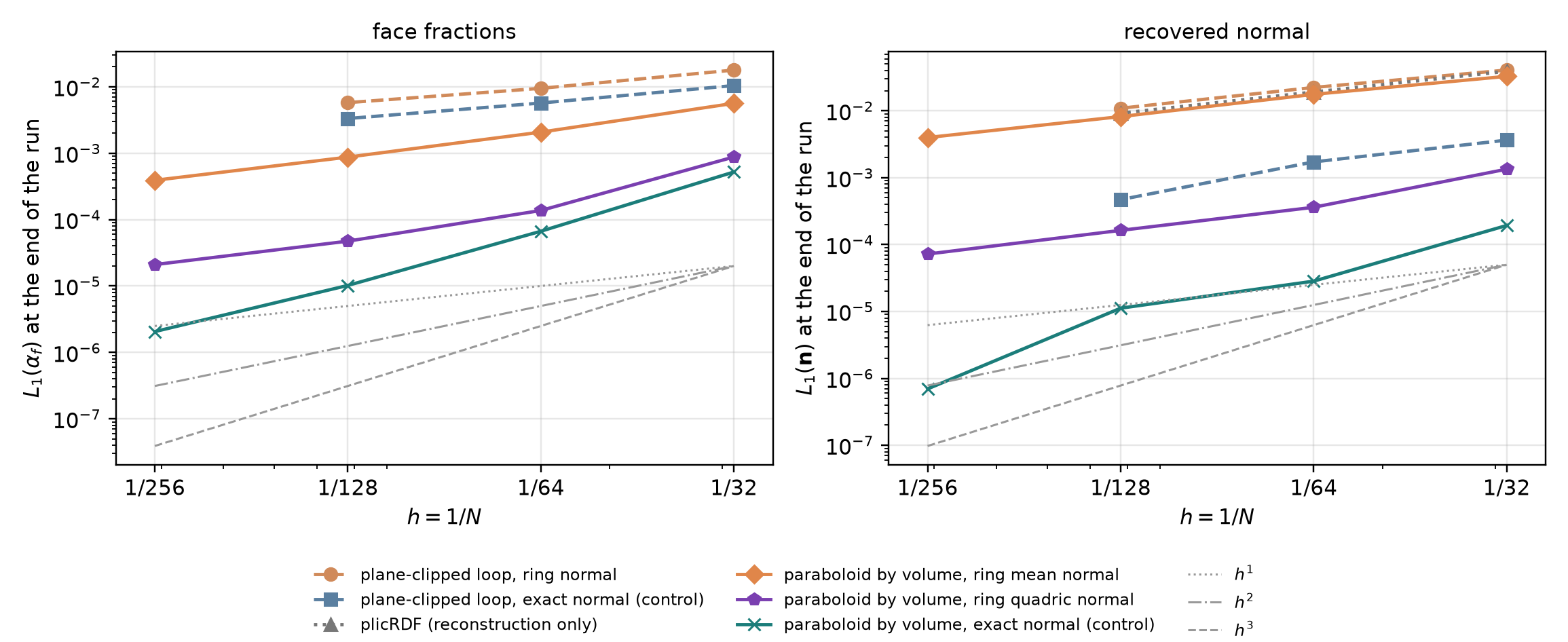}
\caption{The resolution ladders of the translated sphere at $\mathrm{Co} = 0.2$: the face-fraction error (left) and the error of the normal recovered from the transported fractions (right) at the end of the run, against the cell size, for the plane-clipped loop (dashed; plicRDF's own normal dotted) and the dual advection with the ring-tensor paraboloid positioned by $\alpha_k$ (solid), $N = 32$ to $256$. The computed-normal rows of the paraboloid are first order on the finest pair at constants $18$ apart; the exact-normal control keeps order $2.3$ to $2.8$. Generated by \texttt{make\_fig\_ladders.py} from the exported order tables.}
\label{fig:tpf-ladders}
\end{figure}

The ring-tensor paraboloid positioned by $\alpha_k$ was run over the resolutions and Courant numbers of \cref{tab:tpf-sphere} from the exact-sphere initial state, with the volume rule of \citet{KromerBothe2023Init}, and on the translated plane of \cref{tab:tpf-plane}. \Cref{fig:tpf-ladders} plots the ladders against the plane-clipped loop, \cref{tab:tpf-curved} carries the numbers, \cref{tab:tpf-curved-plane} the plane gate. Three results. First, the order of the closed loop is the order of the normal fed back: the face fractions are of order $2.84$ per period with the exact normal, $2.11$ with the ring quadric's normal and $1.35$ with the ring's mean normal, and the orders after the first step ($2.76$, $1.92$, $1.05$) are the same. The error at the end of the run is a constant factor of $1.8$ to $2.0$ above the error after the first step: the per-step remainder does not accumulate over the period, which answers the question of \cref{sec:transport} in the negative. Second, every row is bounded and conservative at every resolution and Courant number, the volume anchor converges in every cell, and on the plane the exact and the ring normal hold the gate in all 16 runs each; the own normal, and the variant that keeps it above the sliver threshold (bounded in one of nine runs, where its pre-registered prediction had it bounded in all), show that the stabilisation of the own normal is a design question of the whole loop, not a threshold (\cref{sec:tpf-sliver}). Third, the fourth mesh settles the asymptotic order of the computed normals (\cref{tab:tpf-n256}): from $N = 128$ to $256$ the ring quadric's row is of order $1.17$ in the face fractions and $1.18$ in the normal, the ring mean's $1.16$ and $1.04$, the exact normal's $2.30$; a fit $c_1 h + c_2 h^2$ through the two finest meshes puts $88\,\%$ of the ring quadric's error at $N = 256$ in the first-order term, and the ratio of the two computed-normal rows is constant ($18.4$ and $18.6$). The loop therefore delivers over a period the order of the normal it is fed, up to $2.84$ in the frame of the exact normal, and the fourth mesh sets the research goal of the method: a transported normal of second order under refinement; the ring quadric's normal brings a constant eighteen times below the ring mean's on the way, and the ring-state fit of \cref{sec:transport} is the first candidate. With the ring quadric's normal the transported normal at $N = 128$ is fifty times below the ring normal's and plicRDF's, the fit converges in every interface cell of every step, and the row's four pre-registered predictions held, the normal's order at the margin. The Courant dependence has reversed against the plane loop (\cref{tab:tpf-courant}): the error no longer accumulates over the steps and rises with the pre-image's displacement.

\begin{table}[htbp]
\centering
\caption{The dual advection with one surface per cell on the translated sphere of \cref{tab:tpf-sphere}, from the exact-sphere initial state, $N = 32, 64, 128$, $\mathrm{Co} = 0.1, 0.2, 0.5$: the exact sphere as every cell's surface (the control; its three rows are the normal measured against the translated sphere: the exact one the loop uses, the one recovered from the transported fractions by \cref{eq:avof-normal}, and the ring normal) and the ring-tensor paraboloid positioned by $\alpha_k$ with its frame from the exact normal, from the ring normal, from the ring quadric's normal, from the cell's own normal with the ring normal below $0.1h^2$, and from the cell's own normal. Per row and Courant number: $L_1(\alpha_f)$ at the end of the run on the finest mesh with the least-squares order $p$ over the three meshes and the order $p_1$ of the error after the first step; the normal and centroid errors with their orders; the shape error; the minimum of $\alpha$ over the run; the run maximum of the surface defect ($L_1$ over the interface cells of $\alpha_k$ against the volume under the surface); the CPU time per step on the finest mesh with the per-step diagnostics on. Rows that abort list the steps for the three meshes. Table body generated by \texttt{make\_table\_advection.py}.}
\label{tab:tpf-curved}
\footnotesize
\setlength{\tabcolsep}{3pt}
\resizebox{\linewidth}{!}{
\begin{tabular}{llrrrrrrrrrrr}
\toprule
surface, normal & Co & $L_1(\alpha_f)$ & $p$ & $p_1$ & $L_1(\mathbf{n})$ & $p$ & $L_1(\mathbf{x})$ & $p$ & $E_{\mathrm{shape}}$ & $\min_t \alpha$ & defect & s/step \\
\midrule
exact sphere, exact & 0.1 & 7.78e-13 & -- & -- & 1.07e-05 & 1.47 & 2.57e-05 & 1.98 & 2.30e-05 & -9.82e-15 & 9.28e-10 & 5.64 \\
 & 0.2 & 3.26e-13 & -- & -- & 1.07e-05 & 1.47 & 2.57e-05 & 1.98 & 2.30e-05 & -9.53e-15 & 2.98e-10 & 3.38 \\
 & 0.5 & 3.12e-15 & -- & -- & 1.07e-05 & 1.47 & 2.57e-05 & 1.98 & 2.30e-05 & -1.00e-14 & 3.32e-11 & 3.81 \\
\midrule
exact sphere, own & 0.1 & 7.78e-13 & -- & -- & 1.37e-04 & 1.40 & 2.57e-05 & 1.98 & 2.30e-05 & -9.82e-15 & 9.28e-10 & 5.61 \\
 & 0.2 & 3.26e-13 & -- & -- & 1.37e-04 & 1.40 & 2.57e-05 & 1.98 & 2.30e-05 & -9.53e-15 & 2.98e-10 & 5.69 \\
 & 0.5 & 3.12e-15 & -- & -- & 1.37e-04 & 1.40 & 2.57e-05 & 1.98 & 2.30e-05 & -1.00e-14 & 3.32e-11 & 6.36 \\
\midrule
exact sphere, ring & 0.1 & 7.78e-13 & -- & -- & 7.68e-03 & 0.98 & 2.57e-05 & 1.98 & 2.30e-05 & -9.82e-15 & 9.28e-10 & 5.60 \\
 & 0.2 & 3.26e-13 & -- & -- & 7.68e-03 & 0.98 & 2.57e-05 & 1.98 & 2.30e-05 & -9.53e-15 & 2.98e-10 & 5.84 \\
 & 0.5 & 3.12e-15 & -- & -- & 7.68e-03 & 0.98 & 2.57e-05 & 1.98 & 2.30e-05 & -1.00e-14 & 3.32e-11 & 3.72 \\
\midrule
ring-tensor paraboloid, exact & 0.1 & 1.10e-05 & 2.87 & 2.81 & 1.15e-05 & 2.10 & 2.57e-05 & 2.00 & 2.30e-05 & -8.51e-14 & 3.50e-16 & 3.83 \\
 & 0.2 & 1.02e-05 & 2.84 & 2.76 & 1.12e-05 & 2.05 & 2.57e-05 & 1.99 & 2.30e-05 & -7.12e-14 & 3.31e-16 & 4.47 \\
 & 0.5 & 1.11e-05 & 2.87 & 2.82 & 1.15e-05 & 2.31 & 2.57e-05 & 1.99 & 2.30e-05 & -2.00e-13 & 3.42e-16 & 4.94 \\
\midrule
ring-tensor paraboloid, ring & 0.1 & 8.52e-04 & 1.38 & 1.08 & 8.19e-03 & 1.00 & 2.89e-05 & 2.09 & 2.30e-05 & -3.73e-14 & 3.65e-16 & 4.30 \\
 & 0.2 & 8.71e-04 & 1.34 & 1.05 & 8.19e-03 & 1.00 & 2.85e-05 & 2.09 & 2.30e-05 & -5.07e-14 & 3.04e-16 & 4.35 \\
 & 0.5 & 1.22e-03 & 1.28 & 1.05 & 8.27e-03 & 0.99 & 2.83e-05 & 2.10 & 2.30e-05 & -1.62e-13 & 3.27e-16 & 4.91 \\
\midrule
ring-tensor paraboloid, ring quadric & 0.1 & 5.06e-05 & 2.24 & 1.88 & 1.67e-04 & 1.67 & 2.57e-05 & 2.00 & 2.30e-05 & -8.50e-14 & 3.52e-16 & 4.35 \\
 & 0.2 & 4.73e-05 & 2.11 & 1.92 & 1.63e-04 & 1.52 & 2.57e-05 & 2.00 & 2.30e-05 & -7.27e-14 & 3.31e-16 & 4.30 \\
 & 0.5 & 4.80e-05 & 1.93 & 2.23 & 1.61e-04 & 1.31 & 2.57e-05 & 1.99 & 2.30e-05 & -2.01e-13 & 3.42e-16 & 4.95 \\
\midrule
ring-tensor paraboloid, own, ring below $0.1h^2$ & 0.1 & \multicolumn{11}{l}{aborted at step 317 ($N=32$)/128 ($N=64$)/115 ($N=128$)} \\
 & 0.2 & \multicolumn{11}{l}{aborted at step 108 ($N=64$)/77 ($N=128$)} \\
 & 0.5 & \multicolumn{11}{l}{aborted at step 41 ($N=32$)/50 ($N=64$)/61 ($N=128$)} \\
\midrule
ring-tensor paraboloid, own & 0.1 & \multicolumn{11}{l}{aborted at step 66 ($N=32$)/76 ($N=64$)/82 ($N=128$)} \\
 & 0.2 & \multicolumn{11}{l}{aborted at step 46 ($N=32$)/63 ($N=64$)/68 ($N=128$)} \\
 & 0.5 & \multicolumn{11}{l}{aborted at step 45 ($N=32$)/51 ($N=64$)/54 ($N=128$)} \\
\bottomrule
\end{tabular}
}
\end{table}

\begin{table}[htbp]
\centering
\caption{The translated plane of \cref{tab:tpf-plane} with the ring-tensor paraboloid positioned by $\alpha_k$ and the dual advection, in the format of that table: per normal source and Courant number the number of the eight runs in which the face fractions and the normal stay at round-off over the whole run, and the run maximum of $L_1(\alpha_f)$ at $N = 160$ and $\mathrm{Co} = 0.5$ on the normal arm. Table body generated by \texttt{make\_table\_advection.py}.}
\label{tab:tpf-curved-plane}
\footnotesize
\setlength{\tabcolsep}{4pt}
\resizebox{0.8\linewidth}{!}{
\begin{tabular}{lrrr}
\toprule
surface, normal & Co $= 0.2$ & Co $= 0.5$ & $\max_t L_1(\alpha_f)$, $N = 160$, Co $= 0.5$ \\
\midrule
ring-tensor paraboloid, exact normal & 8/8 & 8/8 & 8.14e-14 \\
ring-tensor paraboloid, ring normal & 8/8 & 8/8 & 9.81e-14 \\
ring-tensor paraboloid, own normal & 1/8 & 0/8 & 1.18e-01 \\
\bottomrule
\end{tabular}
}
\end{table}

\begin{table}[htbp]
\centering
\caption{The fourth mesh: the ring-tensor paraboloid positioned by $\alpha_k$ at $N = 256$ ($33.6$ million cells, 96 processors, 1280 steps, $\mathrm{Co} = 0.2$; bounded and conservative, no fit fallback and no anchor failure) against $N = 128$, with the pairwise order. The exact-normal row's normal reading is the offset of the reference point, not an error (\ref{app:transport}), and is omitted. On 96 processors $7$ to $87$ of the $2\cdot10^5$ curved faces per step were left without a surface at processor corners, a gap of the parallel exchange absent in serial; against the serial digits at $N = 128$ its effect is $4\cdot10^{-5}$ to $6\cdot10^{-4}$ relative in the end values and $10^{-3}$ in the pairwise orders.}
\label{tab:tpf-n256}
\footnotesize
\begin{tabular}{llrrr}
\toprule
normal in the frame & error & $N = 128$ & $N = 256$ & order \\
\midrule
exact (control) & $L_1(\alpha_f)$ & 1.02e-05 & 2.06e-06 & 2.30 \\
 & $L_1(\mathbf{x})$ & 2.57e-05 & 6.27e-06 & 2.03 \\
ring quadric & $L_1(\alpha_f)$ & 4.73e-05 & 2.10e-05 & 1.17 \\
 & $L_1(\mathbf{n})$ & 1.63e-04 & 7.22e-05 & 1.18 \\
 & $L_1(\mathbf{x})$ & 2.57e-05 & 6.27e-06 & 2.03 \\
ring mean & $L_1(\alpha_f)$ & 8.71e-04 & 3.90e-04 & 1.16 \\
 & $L_1(\mathbf{n})$ & 8.19e-03 & 3.98e-03 & 1.04 \\
 & $L_1(\mathbf{x})$ & 2.85e-05 & 7.16e-06 & 1.99 \\
all rows & $E_{\mathrm{shape}}$ & 2.30e-05 & 5.74e-06 & 2.00 \\
\bottomrule
\end{tabular}
\end{table}

\subsection{Cost}
\label{sec:tpf-cost}

\Cref{tab:tpf-cost} compares the time per step with plicRDF on the same initial state, one core, $\mathrm{Co} = 0.2$, the error diagnostics reduced to the end of the run, three repetitions per case and the median taken; the three repetitions reproduce every error digit. The plane-clipped loop with the ring normal is cheaper than plicRDF, because its reconstruction is one sum of face fractions, one ring sum and one plane positioning per interface cell where plicRDF iterates its stencil ten times. The dual advection costs $1.2$ to $1.9$ times plicRDF after four changes to its first implementation, which had cost $3$ to $6$ times: planar faces integrated whole and only warped faces fanned, the diagnostics off unless asked for, the cell polyhedron built once per cell, and the volume anchor by the consecutive cubic spline positioning of \citet{Maric2021CCS} at $3.4$ volume evaluations per cell and step in place of $6.3$. The remaining difference is the Kromer--Bothe volume of every prism of the faces of the interface cells and their neighbours, where the plane loop has one plane cut per face.

\begin{table}[htbp]
\centering
\caption{Time per step on the translated sphere, one core, $\mathrm{Co} = 0.2$, medians of three repetitions with the diagnostics reduced to the end of the run (the node scatter of the medians is about $20\,\%$). The profile of one node: in the plane-clipped loop the DAVOF reconstruction takes $106\,\mathrm{s}$ of the $521\,\mathrm{s}$ advection time against $398$ of $811\,\mathrm{s}$ for plicRDF, the face update about $25\,\mathrm{s}$ more than plicRDF's flux; in the first implementation of the dual advection $57\,\%$ of the step was the volume anchor, $14\,\%$ two diagnostics of the anchor, $9\,\%$ the prism volumes, $7\,\%$ the face fractions, and the second-ring search, the ring normal and the tensor fit below $1\,\%$ each.}
\label{tab:tpf-cost}
\footnotesize
\begin{tabular}{lrrr}
\toprule
method & $N = 32$ & $N = 64$ & $N = 128$ \\
\midrule
plicRDF (plane-loop study) & $=$ DAVOF & $0.15\,\mathrm{s}$ & $1.72\,\mathrm{s}$ \\
plane-clipped loop, ring normal & $=$ plicRDF & $0.13\,\mathrm{s}$ & $1.31\,\mathrm{s}$ \\
\midrule
plicRDF (paraboloid study) & $0.031\,\mathrm{s}$ & $0.18\,\mathrm{s}$ & $1.8\,\mathrm{s}$ \\
paraboloid by volume, ring normal & $0.053\,\mathrm{s}$ ($1.7\times$) & $0.36\,\mathrm{s}$ ($1.9\times$) & $2.1\,\mathrm{s}$ ($1.2\times$) \\
same, first implementation & $6\times$ & $3.5\times$ & $3\times$ \\
\bottomrule
\end{tabular}
\end{table}

\subsection{Perturbed hexahedra}
\label{sec:tpf-perturbed}

The same two tests on hexahedra with every vertex displaced by up to $0.1h$ (warped faces, $\mathrm{Co} = 0.2$) measure which face state a warped face needs: the wetted area vector is the transportable quantity (\cref{sec:warped}). \Cref{tab:tpf-perturbed} lists the outcome. With the scalar fraction the exact-normal control locates the defect in what is stored per face, not in the normal; on the sphere the plane-clipped loop needs that face state as well, and plicRDF and isoAlpha, which stay bounded, carry normal errors ten times their static values on the same meshes, so distorted-mesh transport is a research goal for the framework as a whole, not for DAVOF alone. With the wetted area vector per face, re-anchored centroids in the tensor fit and the flux prism fanned about the face centre, the translated plane stays at round-off in 11 of 12 runs with the exact and with the ring normal. On the perturbed sphere the dual advection completes at every resolution; its boundedness there is the research goal that follows, with the face state that already keeps the plane at round-off. The hexahedral results above are untouched by this.

\begin{table}[htbp]
\centering
\caption{Transport on hexahedra with every vertex displaced by up to $0.1h$, $\mathrm{Co} = 0.2$. The face state is either the scalar fraction $\alpha_f$ or the wetted area vector $\mathbf{a}_f = \sum_t A^{\mathrm{wet}}_t \mathbf{n}_t$ over the face's fan triangles (branch \texttt{davof-warped} of the fork; with re-anchored centroids in the tensor fit and the flux prism fanned about the face centre).}
\label{tab:tpf-perturbed}
\footnotesize
\begin{tabular}{p{0.2\linewidth}p{0.17\linewidth}p{0.55\linewidth}}
\toprule
test & face state & outcome \\
\midrule
plane, plane-clipped loop & scalar fraction & no row passes the gate; exact normal: the normal at round-off, the fractions off by $1$ to $5\,\%$; ring normal: the fraction error grows to $0.03$ to $0.11$ within the $0.2L$ travel \\
plane, dual advection & scalar fraction & ring normal at $2\cdot10^{-3}$, flat in $N$ \\
plane, dual advection & area vector & at round-off in 11 of 12 runs ($N = 20$, $40$, $80$, both arms, exact and ring normal); the twelfth, the ring row at $N = 80$ on the oblique arm, carries single-face jumps to $10^{-8}$ at four steps on faces a vertex of which grazes the plane \\
\midrule
sphere, plane-clipped loop & scalar fraction & every DAVOF row diverges, the exact-normal control included (stopped after 144 to 315 steps); plicRDF and isoAlpha bounded with converging shape errors and normal errors of $4.9\cdot10^{-2}$ to $8.8\cdot10^{-2}$ (plicRDF, $N = 32$ to $128$) that do not converge \\
sphere, dual advection & either & completes without abort, not bounded ($\min\alpha$ down to $-3\cdot10^{-3}$ at $N = 64$, the anchor failing in up to $1163$ cells over a run); centroid error $2$ to $2.3$ times the hexahedral value, shape error below it \\
\bottomrule
\end{tabular}
\end{table}

\subsection{What the transport studies establish, and the research goals they set}
\label{sec:tpf-summary}

The closed loop needs a filtered normal. The cell-local normal is exact for a plane, but fed the fractions of several surfaces it turns by their mismatch times the neighbours' lever arms over its own patch area, with no restoring force, and amplifies round-off on the plane and the $O(h^2)$ mismatch of neighbouring surfaces on the sphere; the one-ring average divides the same mismatch by about $9h^2$ and contracts, at the price of the ring's mean normal. With it the plane-clipped loop is, as the clipping law predicts, a first-order method in the face fractions that matches plicRDF in position and shape, stays within $35\,\%$ of it in the normal and costs less. A second-order clipping needs the volume flux of the same surface, because the plane flux lets $\alpha_k$ leave the volume under the surface. With one surface per cell for the flux and the face update, the ring-tensor surface positioned by the conserved $\alpha_k$, with the normal from the faces and nothing else, is stable, bounded and conservative at every resolution and Courant number, with face fractions of the order of the normal fed back and a per-step error that does not accumulate over the period. Three research goals follow. The first is a transported normal of second order under refinement: the loop delivers the order of the normal it is fed, the one-ring constructions measured here are first order beyond $N = 128$ at constants eighteen apart, and the ring-state fit of \cref{sec:transport} is the first candidate. The second is a stable surface that uses more of the cell's own fractions than the normal, which the same fit addresses. The third is the boundedness of the dual advection on perturbed meshes, with the face quantities of \cref{sec:warped} in the update, which already keep the plane at round-off there.

\section{The paraboloid update: design and open questions}
\label{sec:transport}

\begin{figure}[htbp]
\centering
\begin{tikzpicture}[scale=1.7, >=Stealth, font=\small]
  \colorlet{fluidcol}{blue!18}
  \colorlet{gascol}{gray!10}
  \colorlet{ifacecol}{blue!65!black}
  \colorlet{normcol}{red!65!black}
  \colorlet{wettedcol}{orange!80!black}
  \colorlet{gasbdrycol}{teal!58!black}
  \colorlet{bdrycol}{black!50}
  \colorlet{facecol}{orange!80!black}
  \colorlet{sweptcol}{black!8}
  \coordinate (Sf_bot) at ( 0.00, -0.80);
  \coordinate (Sf_top) at ( 0.20,  0.90);
  \coordinate (LK1) at (-1.60, -0.82);
  \coordinate (LK2) at (-0.40, -0.98);
  \coordinate (LK5) at (-0.55,  1.02);
  \coordinate (LK6) at (-1.52,  0.88);
  \coordinate (RL2) at ( 1.78, -0.58);
  \coordinate (RL3) at ( 1.62,  0.92);
  \coordinate (RL4) at ( 0.48,  1.05);
  \coordinate (Pe_k) at (-1.56,  0.12);
  \coordinate (Pf_k) at ( 0.12,  0.22);
  \coordinate (Pf_l) at ( 0.07, -0.21);
  \coordinate (Pe_l) at ( 1.70,  0.17);
  \coordinate (HA) at (-0.48, -0.77);
  \coordinate (HB) at (-0.34,  0.93);
  \coordinate (HP) at (-0.40,  0.19);
  \fill[gascol] (LK1)--(LK2)--(Sf_bot)--(Sf_top)--(LK5)--(LK6)--cycle;
  \fill[gascol] (Sf_bot)--(RL2)--(RL3)--(RL4)--(Sf_top)--cycle;
  \fill[sweptcol] (Sf_bot)--(Sf_top)--(HB)--(HA)--cycle;
  \fill[fluidcol] (LK1)--(LK2)--(Sf_bot)--(Pf_k)--(Pe_k)--cycle;
  \fill[fluidcol] (Sf_bot)--(RL2)--(Pe_l)--(Pf_l)--cycle;
  \draw[bdrycol, line width=0.6pt]
    (LK1)--(LK2)--(Sf_bot)--(Sf_top)--(LK5)--(LK6)--cycle;
  \draw[bdrycol, line width=0.6pt]
    (Sf_bot)--(RL2)--(RL3)--(RL4)--(Sf_top)--cycle;
  \draw[wetted] (Pe_k)--(LK1)--(LK2)--(Sf_bot);
  \draw[gasbdrycol, line width=1.8pt] (Sf_top)--(LK5)--(LK6)--(Pe_k);
  \draw[wetted] (Sf_bot)--(RL2)--(Pe_l);
  \draw[gasbdrycol, line width=1.8pt] (Pe_l)--(RL3)--(RL4)--(Sf_top);
  \draw[ifacecol, line width=2.2pt] (Pe_k)--(Pf_k);
  \draw[ifacecol, line width=2.2pt] (Pf_l)--(Pe_l);
  \fill[ifacecol] (Pe_k) circle (1.4pt);
  \fill[ifacecol] (Pf_k) circle (1.4pt);
  \fill[ifacecol] (Pf_l) circle (1.4pt);
  \fill[ifacecol] (Pe_l) circle (1.4pt);
  \draw[facecol, line width=3.2pt] (Sf_bot)--(Sf_top);
  \draw[wetted]          (HA)--(HP);
  \draw[bdrycol!75, dashed, line width=1.6pt]  (HP)--(HB);
  \draw[->, gray!55!black, line width=1.0pt] (Sf_bot) -- (HA);
  \draw[->, gray!55!black, line width=1.0pt] (Sf_top) -- (HB);
  \draw[normcol, ->, line width=1.2pt]
    (0.12, 0.05) -- ++(0.46,-0.054) node[pos=1, right=1pt, normcol]{$\mathbf{n}_f$};
  \node[blue!62!black, font=\footnotesize] at (-1.08, -0.50) {$\Omega_k^-(t^n)$};
  \node[ifacecol,      font=\footnotesize] at (-1.06,  0.26) {$\tilde{\Sigma}_k(t^n)$};
  \node[black!55,      font=\footnotesize] at (-1.08,  0.76) {$\Omega_k^+(t^n)$};
  \node[blue!62!black, font=\footnotesize] at ( 1.10, -0.25) {$\Omega_l^-(t^n)$};
  \node[ifacecol,      font=\footnotesize] at ( 1.04,  0.18) {$\tilde{\Sigma}_l(t^n)$};
  \node[black!55,      font=\footnotesize] at ( 1.10,  0.68) {$\Omega_l^+(t^n)$};
  \node[facecol, font=\footnotesize, anchor=west] at (0.38, -0.56) {$S_f\,(t^{n+1})$};
  \draw[->, facecol!70, line width=0.5pt] (0.37,-0.56) -- (0.04,-0.56);
  \node[black!60, font=\footnotesize, anchor=west] at (-0.92, 0.60) {$\hat{S}_f$};
  \draw[->, black!45, line width=0.5pt] (-0.76,0.60) -- (-0.38,0.60);
  \node[wettedcol, font=\footnotesize, anchor=east] at (-0.94, -0.28)
    {$\hat{S}_f\!\cap\!\Omega_k^-$};
  \draw[->, wettedcol!70, line width=0.5pt] (-0.94,-0.28) -- (-0.46,-0.29);
\end{tikzpicture}
\caption{The semi-Lagrangian face-area update \cref{eq:face-update} on two neighbouring cells. The face $S_f$ (orange) is traced backward to its pre-image $\hat{S}_f$; the wetted part $\hat{S}_f\cap\Omega_k^-(t^n)$ (violet) gives $\alpha_f^{n+1}$ directly. The volume flux of $\alpha_k$ comes from the plane that $\alpha_k$ positions; the pre-image is clipped against the clipping surface of the cell that contains it, with the primitives of the face-vertex transport \citep{MaricMarschallBothe2018}.}
\label{fig:faceupdate}
\end{figure}

The measurements above support one design of the transport beyond the plane clipping of \cref{sec:tpf}, stated here with its accuracy requirement and its open questions. The volume fraction is advanced by the geometric fluxes of unsplit VOF cut from the clipping surface of the cell, the same surface that clips the face pre-images below (\cref{sec:tpf} measured why it must be the same); in the measured loop that surface is the PLIC plane with the normal \cref{eq:avof-normal}, positioned along it so that its cut volume equals $\alpha_k$, with the explicit position \cref{eq:avof-position} as the initial guess, from which the consecutive cubic spline positioning of \citet{Maric2021CCS} reaches machine precision in two iterations. The face fractions are advanced by a semi-Lagrangian update (\cref{fig:faceupdate}): the face $S_f$ is traced backward through the velocity field to its pre-image $\Sfhat$ at $t^n$, and the new fraction is
\begin{equation}
\alpha_f^{n+1} = \frac{|\Sfhat \cap \Omega^-(t^n)|}{|\Sfhat|},
\label{eq:face-update}
\end{equation}
the fraction of the pre-image that lies in liquid, which a locally affine backward map carries to the face unchanged, so the update is invariant under uniform strain. A pre-image that spans several cells is triangulated and each triangle is clipped against the clipping surface of the cell that contains it at $t^n$, with the primitives of the face-vertex transport of \citet{MaricMarschallBothe2018}; both cells of a face use the same wetted area, and the pre-image needs the surfaces of the cells it overlaps only. Conservation of $\alpha_k$ holds by the antisymmetry of the face fluxes; boundedness may need a flux limiter, at the cost of order where it is active, as in the unsplit methods of the review \citep{MaricKotheBothe2020}, and any clipping of fractions to $[0, 1]$ is a conservation loss that is counted.

\emph{One surface for the volume flux and the face update.} The normal \cref{eq:avof-normal} enters the surface and is never iterated or optimized. In the first measured loop the surface is the $\alpha_k$-positioned plane, a fixed point of the regeneration at first order (\cref{sec:regen}), for the volume flux and the face clipping alike. The six-coefficient surface of \cref{sec:full}, which reproduces the cell's fractions and is $O(h^3)$ close to the interface, was the design candidate, since by \cref{eq:clipping-law} it gives $q = 2$ per step; \cref{sec:tpf} measured that it needs a stabilisation under transport, and the surface that transports is the paraboloid with the frame of the normal, the ring tensor and the offset from $\alpha_k$: the position from the conserved volume fraction as in classical VOF, the curvature from the ring, nothing further from the cell's own faces. The volume flux of a pre-image is the volume under that surface, here by the adaptive quadrature of \cref{sec:tpf}, for which \citet{EvrardChiodiHanVanWachemDesjardins2023} give closed forms. \Cref{sec:tpf} measured what a volume flux from the plane and a face update from a second-order surface cost: $\alpha_k$ leaves the volume under the clipping surface by the sagitta volume per step and the face fractions fall back to first order. The \emph{paraboloid defect}, $\alpha_k$ against the cut volume of the clipping surface, has an $O(h^2)$ floor for a consistent state and is monitored per cell and per step. Cells in which the fit does not converge keep the face-anchored paraboloid and are counted.

\emph{What the quadric is fitted to.} The transported paraboloid takes its six coefficients from three sources, and the normal among them sets the order of the loop (\cref{sec:tpf}). The two slopes vanish in the frame of the normal; the normal of the stable rows is a ring quantity. The curvature tensor is read off the quadric of \cref{sec:curv}, fitted through the recovered planes of the ring (\cref{fig:derivatives}c), and in the ring-quadric row the normal at the cell is read off the same quadric. The offset comes from the conserved volume fraction by the anchor, never from a fit, because the volume flux and the face update must be the moments of one surface whose volume is $\alpha_k$ (\cref{sec:tpf}). Two fits were therefore measured. The quadric through the recovered planes is a fit to derived samples: each sample normal is \cref{eq:avof-normal} on one cell's faces, which under transport are set by several surfaces, and \cref{sec:tpf} measured the turn that gives each sample; the least squares over the ring filters it, which is why the row is stable, but the sample errors are not smooth on a fixed physical length, the condition \cref{sec:onering} sets for a converging one-ring fit, and the normal falls to first order beyond $N = 128$. The six-coefficient surface of \cref{sec:full} is a fit to the cell's own fractions through \cref{eq:dvds}, the identity of the zero-velocity map, which needs a stabilisation under transport by the same mechanism with six coefficients in place of one direction. The fit that \cref{eq:dvds} suggests and that was not measured takes the ring's state itself as its data: the volumes of the ring cells and the wetted areas of all their faces, each a derivative of the volume, matched by the six coefficients of one quadric in least squares, every residual weighted by its own face area. No normal is formed in any cell and no patch area divides anything, which is the property that makes the ring mean stable, kept together with a second-order surface. The fit is nonlinear in the coefficients, a Gauss--Newton iteration in which the volume rule evaluates the volumes and wetted areas of the ring under the trial surface and the Jacobian is the projected moments of its wetted patches, which the rule can integrate or finite differences can replace at first; the cost per cell is a few dozen rule evaluations per iteration. On the pair $N = 128$ to $256$, where the ring quadric's normal fell to order $1.17$, this fit decides whether the loss of order lies in the per-cell sample normal, as the mechanism of \cref{sec:tpf} suggests, or in the one-ring stencil itself; it is the first candidate for a second-order transported normal.

\emph{The accuracy requirement.} By \cref{eq:normal-propagation,eq:clipping-law} the face update needs a clipping surface within $O(h^3)$ of the interface inside a cell. A plane is $O(h^2)$ off; a paraboloid with the recovered normal, the curvature tensor $\mathbf{B}_k$ (the second fundamental form in the frame of the normal) and its offset from the face fractions is $O(h^3)$ close, and a tensor error $\delta\mathbf{B}_k = O(h)$ moves it by only $\tfrac12|\delta\mathbf{B}_k|h^2 = O(h^3)$. The face update therefore needs the curvature only to first order, but the full tensor: the mean curvature alone leaves an $O(h^2)$ gap wherever the two principal curvatures differ, which a sphere does not reveal and the ellipsoid test does. The ring of recovered normals supplies the tensor at that order (\cref{sec:curv}), so the update does not depend on a second-order curvature model, which only the surface-tension force needs.

\emph{The open question: the error over a period.} Per step the ring-tensor surface positioned by $\alpha_k$ is $O(h^3)$ close to the interface wherever the ring tensor is first order, so the face fractions it cuts are second order per step by \cref{eq:clipping-law}. Over a period, $q = 2$ holds only if the per-step remainders, $O(h^2)$ each, sum to $O(h^2)$ rather than $O(h)$ over the $O(1/h)$ steps; the iterated regeneration shows that the face-anchored paraboloids drift at first order per period, and \cref{sec:tpf} shows that the face-interpolating surface needs a stabilisation under transport, so the per-period order of the volume-positioned surface is what the ladders of \cref{sec:tpf} measure. The tests that decide it are the translation tests of \cref{sec:tpf} with the paraboloid in place of the plane, scored as there, with $q$ measured against the exact fractions after one period, on Cartesian and on distorted hexahedral meshes; if $q = 2$ per step but $q = 1$ per period, the method stays the first-order one that \cref{tab:tpf-sphere} measures, with the ring normal at plicRDF's constant, a second-order position and a per-cell consistency measure, and the baseline is already in hand.

\emph{Further properties.} In a flow that is not locally affine the surface Jacobian of the backward map varies across a face by $O(h\,\Delta t)$, which would cost one order over a period; the first moment of the wetted pre-image, which the fan sums of \cref{sec:warped} compute, corrects the linear part, and on warped faces the wetted area vector and the first moment transform with the cofactor of the gradient of the backward map (Nanson's formula), so the update is applied to the face triangles, each traced as a triangle. Stability of the closed loop, in which the recovered normal enters the next update through the clipping surface, a feedback absent from classical UGVOF, is measured in \cref{sec:tpf} for the plane clipping: the cell-local normal amplifies through sliver cells and the ring normal does not; for the paraboloid update it is to be measured the same way, and by the growth law of the error over several periods for rotation, a straining field and the reversed deformation field of \citet{LeVeque1996}, with errors reported at the half period, where cancellation on return cannot hide growth. A cell crossed by two interface patches returns the vector sum of both in \cref{eq:avof-normal}; in the design it is flagged when its paraboloid or plane defect exceeds ten times the floor measured on the static states (patches at an angle) or when the vector sum falls below the wisp threshold at an $\alpha_k$ that is not small (antiparallel patches), is assigned a Youngs-type normal \citep{Youngs1982} and counted; films and break-up are outside the accuracy claim. Under no slip a wall face is its own pre-image, so \cref{eq:face-update} leaves its fraction unchanged; the wall fraction is instead taken from the trace on the wall of the wall cell's clipping surface, anchored by the interior faces only, so that the trace predicts the wall fraction and carries the contact line; the contact angle is read from the recovered normal against the wall normal and enters the surface-tension force, not the transport. The cost per step is that of the $F$ face fractions, the ring fit and the clipping, restricted to the interface cells.

\section{Conclusions}
\label{sec:concl}

Transporting the wetted fraction of every face area next to the volume fraction turns the interface reconstruction of unstructured VOF into two closed-form identities on the cell's own data, exact for a plane on every cell with planar faces, with the normal at the order of the face fractions as fractions, the position at second order for every state, and a converging one-ring curvature where the reference implementation's models show none. The order of the face fractions is the whole question of the method, and the estimates \cref{eq:normal-propagation,eq:clipping-law} say what any construction of them must deliver: $O(h^2)$ as a fraction, i.e.\ a clipping surface $O(h^3)$ close to the interface. A one-ring curvature converges only when the sample errors are smooth on a fixed physical length, which bounds what a curvature model from VOF data can achieve on a given geometry. On the zero-velocity map the anchored paraboloids are second order per step and first order per period, the $\alpha_k$-positioned plane is the fixed point at first order, and the paraboloid fitted to the cell's face fractions is the identity. Transported with the plane clipping in the TwoPhaseFlow framework, a filtered normal stabilises the state: the cell-local normal, fed the fractions of several surfaces, amplifies their mismatch through the cells whose patch is small, the same identity on the union of the cell's ring contracts it, and with it the loop is first order in the face fractions, as the clipping law predicts, with plicRDF's position and shape errors and a normal within 35\,\% of plicRDF's, at a lower cost. A second-order correction of the plane-clipped areas anchored by the cell's volume fraction removes 93\,\% of the face error after one step and 30\,\% at the end of the run, because the plane flux does not keep the volume fraction the volume under the surface: the volume flux and the face update must be the volume and the boundary moments of one surface per cell, the zeroth moment of the boundary normal that the area vector carries and the first moment of the wetted faces that positions the plane. Built on that basis, one paraboloid per cell with its frame from the normal, its curvature tensor from the ring, its position from the conserved volume fraction and its own volume flux by the rule of Kromer and Bothe transports the sphere bounded and conservative at every resolution and Courant number, with face fractions of order $2.8$ per period in the frame of the exact normal, $2.1$ with the ring quadric's normal over $N = 32$ to $128$ and $1.35$ with the ring's mean normal: the per-step remainder does not accumulate over the period, the closed loop has the order of the normal fed back, and the fourth mesh sets the research goal: both computed normals are first order there, the ring quadric's with a constant eighteen times smaller, while the exact-normal loop keeps order $2.3$, so a transported normal of second order under refinement is what the method needs next, and the ring-state fit of the design is its first candidate. On perturbed hexahedra the wetted face area vector in place of the scalar fraction, with re-anchored centroids in the tensor fit, keeps the plane at round-off. The implementations and every number are open.

\section*{Data and code availability}
\label{sec:data}

The implementation and the data are open at the release tag \texttt{davof-v1.0} of two repositories. The leia OpenFOAM module \citep{LeiaRepo2026} holds the identities, the regression gate, every study configuration with its pre-registered predictions, the exported tables and figures of every study as CSV files and \LaTeX{} table bodies, the symbolic verification of the identities and of the two normal-error measures, the one-ring fit study, and this manuscript with its table generators. The transport loop and the dual advection with one surface per cell are the branch \texttt{davof} of the TwoPhaseFlow fork \citep{TwoPhaseFlowFork2026}; the time-quadrature flux and the per-cell tilt diagnostic of the appendix are its branches \texttt{davof-fluxint} and \texttt{davof-tiltdiag}. The PLIC comparators are OpenFOAM's \texttt{geometricVoF} library \citep{ScheuflerRoenby2023TwoPhaseFlow}.

\section*{Acknowledgments}

Calculations for this research were conducted on the Lichtenberg high-performance computer of the Technical University of Darmstadt.

\section*{Declaration of generative AI and AI-assisted technologies in the writing process}

During the preparation of this work the author used Claude Code (Anthropic) in order to draft and revise the text, to implement the studies in the leia module and in the TwoPhaseFlow fork under the author's direction, and to run, tabulate and plot the measurements. After using this tool, the author reviewed and edited the content as needed and takes full responsibility for the content of the publication. The identities, the method and its design decisions are the author's; every number reported here is reproduced by the archived study configurations \citep{LeiaRepo2026,TwoPhaseFlowFork2026}.

\appendix
\section{Two measures of the normal error}
\label{app:measures}

The normal error of \cref{sec:static} is $|\nSk - \n_\Sigma|$, the length of the difference of the two unit vectors. The benchmark of plicRDF \citep{ScheuflerRoenby2019} reports $1 - \nSk\cdot\n_\Sigma$ instead. The two measures carry different orders for the same normal. For a recovered normal turned by the angle $\theta$ against the exact one,
\begin{equation}
|\nSk - \n_\Sigma| = 2\sin\frac{\theta}{2} = \theta - \frac{\theta^3}{24} + \ldots, \qquad
1 - \nSk\cdot\n_\Sigma = 1 - \cos\theta = \frac{\theta^2}{2} - \frac{\theta^4}{24} + \ldots ,
\label{eq:measures}
\end{equation}
so the first is linear in the angle and the second quadratic, with $1 - \nSk\cdot\n_\Sigma = \tfrac12|\nSk - \n_\Sigma|^2$ exactly. A normal whose angle error is first order, $\theta = O(h)$, is first order in $|\nSk - \n_\Sigma|$ and second order in $1 - \nSk\cdot\n_\Sigma$, in the $L_1$ and the $L_2$ norm over the interface cells alike, since the norms of $\theta^2$ are second order when $\theta$ is first. A reported second order in the dot-product measure is therefore consistent with a first-order normal, which is what \cref{tab:sphere,tab:ellipsoid} measure for plicRDF with $|\nSk - \n_\Sigma|$ (orders 1.06 to 1.19). The series and the orders in $h$ were verified symbolically (\texttt{normal\_error\_measures\_sympy.py} in the archived data \citep{LeiaRepo2026}).

\section{Implementation record of the transport studies}
\label{app:transport}

This appendix is the record of the measurements of \cref{sec:tpf}, in the order of that section: what was built, how each quantity was computed, which defects were found on the way, and every number behind the tables there. It is written for whoever re-runs or extends the studies.

\subsection{The plane-clipped loop in TwoPhaseFlow}
\label{app:tpf-plane-loop}

The state has been transported for the first time, with the simplest clipping surface, in a fork of the TwoPhaseFlow library of \citet{ScheuflerRoenby2023TwoPhaseFlow}, i.e.\ inside the isoAdvector--plicRDF framework \citep{RoenbyEtAl2016,ScheuflerRoenby2019} that the comparators use. Four elements were added. (i) The constructions of \cref{sec:static} initialise $(\alpha_k, \{\alpha_f\})$. (ii) A reconstruction scheme forms the normal \cref{eq:avof-normal} from the stored face fractions and positions the PLIC plane by $\alpha_k$ with the library's bisection, so that the volume fluxes are the geometric fluxes of isoAdvector from that plane. (iii) isoAdvector sweeps every face of an interface cell with the moving plane; the submerged area of the face at the end of the step becomes the new face fraction, a face shared by two interface cells taking the upwind cell's plane, consistent with its volume flux. This is the plane clipping, which by \cref{eq:clipping-law} gives first-order face fractions, $q = 1$: the loop is measured here with the clipping surface of lowest order on purpose, as the baseline for the paraboloid update of \cref{sec:transport}. (iv) A per-cell content bound keeps the plane-set fluxes of a cell within what it holds plus what enters it in the step; it is inactive on an exactly translating plane and acts on a few faces of the first steps on the sphere. The time step is fixed at a Courant number $\mathrm{Co} = |\mathbf{U}|\Delta t/h$; a run is stopped and recorded as diverged when $\alpha$ leaves $[-0.5, 1.5]$ or the number of interface cells exceeds ten times its initial value.

The cell-local normal \cref{eq:avof-normal} feeds the next step through the plane it positions, and in the loop each face takes its fraction from the surface of its upwind cell, so the fractions of one cell come from several surfaces, which agree only in exact arithmetic. Linearised about the exact plane, the turn of $\mathbf{m}_k$ is the neighbours' displacement at the pre-image traces of the faces they set, weighted by the trace length times its distance from their patch centroid, about $h/2$, over the cell's own patch area $A$; the own faces' weights sum to one, so the loop has no restoring force, and a foreign weight exceeds one where $A$ is small against that lever arm, in a sliver cell, an interface cell whose wetted area is a small fraction of $h^2$; the stencil reconstructions of classical VOF average the volume fractions of a neighbourhood before anything is divided and never meet this. Sliver cells are not rare: with the threshold $|A_{\tilde\Sigma_k}| < 0.1 h^2$ they are $11.5$, $13.6$ and $11.3\,\%$ of the interface cells of the translated sphere at the first step for $N = 32$, $64$ and $128$, and $8$ to $14\,\%$ over a run, so a rule that treats them specially acts on about one cell in eight at every resolution; the loop below therefore evaluates the ring normal in every interface cell, and the variant that keeps the own normal above the threshold is one of the rows of the paraboloid study. Three sources of the normal are therefore compared, every one on the identical initial state: the \emph{own normal}, \cref{eq:avof-normal} as it stands; the \emph{ring normal}, $\mathbf{n}_k = \sum_j \mathbf{m}_j / |\sum_j \mathbf{m}_j|$ over the point neighbours $j$ of the cell, which is \cref{eq:avof-normal} applied to the union of the ring's cells (the interior faces cancel), exact for a plane on any mesh, never divided by a small area, and the mean normal of the patch rather than the normal at the cell; and the \emph{ring quadric}, the normal at the cell of a weighted least-squares Monge patch through the ring's recovered planes. Two controls run the same update with the exact normal of the translated surface and with the plicRDF normal; plicRDF and isoAlpha run on the same initial $\alpha_k$ (which is the Detrixhe--Aslam cell fraction, not the one their own initialisation would give). Every prediction was pre-registered in the study configuration before the run; the failing ones are reported below.

\emph{The translated plane.} With the ring normal, the ring quadric and the exact normal the plane stays at round-off in all 32 runs each. With the own normal it stays at round-off at $N = 20$ and $40$ for $\mathrm{Co} \le 0.2$, drifts to $10^{-10}$ to $10^{-7}$ at $N = 80$ and breaks at $N = 160$ (face-fraction error $3\cdot10^{-3}$ to $0.14$), and at $\mathrm{Co} = 0.5$ it breaks on the normal arm from $N = 40$ on: the amplification grows with the number of steps and the Courant number sets the threshold. The plicRDF normal in the same update fails the gate everywhere, because it is not exact on the corner cells of a plane ($10^{-4}$ at $N = 20$ to $0.17$ at $N = 160$), as plicRDF and isoAlpha themselves are not ($7\cdot10^{-2}$ and $0.18$ at $N = 160$ for the normal).

\emph{The translated sphere.} Four facts. First, the ring normal is the first stable DAVOF transport loop: bounded to $10^{-14}$ and conservative to round-off at every resolution and Courant number, while the own normal is stopped at every resolution and Courant number (after 45 to 232 steps) and the ring quadric stays bounded but is erratic, its fit failing in up to 1700 cells per step and its normal error reaching $0.54$ at $N = 64$ and $\mathrm{Co} = 0.5$. Second, the face fractions transport at first order ($0.80$, $0.81$ and $1.00$ at $\mathrm{Co} = 0.1$, $0.2$ and $0.5$), and so do they with the exact normal ($0.85$, $0.82$, $0.99$): the order is the plane clipping's, $q = 1$ by \cref{eq:clipping-law}, not the normal's. Third, against plicRDF on the same $\alpha_k$ the ring normal is first order ($0.94$ to $1.02$ against $0.97$ to $1.04$) at $1.1$ to $1.35$ times plicRDF's error ($1.09\cdot10^{-2}$ against $9.20\cdot10^{-3}$ at $N = 128$ and $\mathrm{Co} = 0.2$), the centroid is second order ($1.86$ to $1.99$) at plicRDF's value to three digits ($1.02\cdot10^{-4}$ against $1.04\cdot10^{-4}$), the shape error is plicRDF's ($3.05\cdot10^{-5}$ against $3.08\cdot10^{-5}$), and the cost is below plicRDF's: with the error diagnostics reduced to the end of the run and three repetitions per case, the per-step medians are $1.31$ against $1.72\,\mathrm{s}$ at $N = 128$ and $0.13$ against $0.15\,\mathrm{s}$ at $N = 64$ (one core, equal at $N = 32$), and in the profile of one and the same node the DAVOF reconstruction takes $106\,\mathrm{s}$ of the $521\,\mathrm{s}$ advection time against $398$ of $811\,\mathrm{s}$ for plicRDF (plicRDF's ten stencil iterations against one sum of face fractions, one ring sum and one plane positioning per interface cell), the face update costing about $25\,\mathrm{s}$ more than plicRDF's flux; isoAlpha is behind on every count. Fourth, the Courant number: at fixed $N$ the ring normal's errors rise as the Courant number falls (at $N = 64$: $L_1(\alpha_f)$ $1.22\cdot10^{-2}$, $9.45\cdot10^{-3}$, $6.06\cdot10^{-3}$ and the normal $2.53\cdot10^{-2}$, $2.23\cdot10^{-2}$, $1.84\cdot10^{-2}$ for $\mathrm{Co} = 0.1$, $0.2$, $0.5$), because the per-step clipping error accumulates over the $1/\mathrm{Co}$ times more steps (\cref{fig:tpf-history} shows the slow growth), while the exact-normal control is flat in the Courant number and plicRDF's own normal error falls with it as well. The pre-registered prediction that the ring normal's errors fall with the Courant number was therefore falsified in its stated direction. One result is open: the loop driven by the plicRDF normal does not converge on the sphere (normal error $4\cdot10^{-2}$ to $5\cdot10^{-2}$ at every $N$ and $\mathrm{Co}$), although plicRDF alone converges on the same $\alpha_k$; the two differ in the fluxes the DAVOF face update assigns to the faces beyond the downwind faces of an interface cell and in the plane positioned along the helper's normal, and which of the two costs the convergence has not been measured.

\subsection{Perturbed hexahedra and the wetted area vector per face}
\label{app:tpf-warped}

The same two tests on hexahedra with every vertex displaced by up to $0.1h$ (warped faces; $\mathrm{Co} = 0.2$) measure what the fork does not have: it stores one scalar fraction per face, while \cref{sec:warped} showed that on a warped face the wetted area vector is the transportable quantity. No row passes the plane gate. With the exact normal the normal stays at round-off while the transported fractions are off by $1$ to $5\,\%$, so the defect is in what is stored per face, not in the normal; with the ring normal the fraction error grows to $0.03$ to $0.11$ within the $0.2L$ travel. On the sphere every DAVOF row diverges, the exact-normal control included (stopped after 144 to 315 steps), while plicRDF and isoAlpha stay bounded with converging shape errors but normal errors that do not converge either ($4.9\cdot10^{-2}$ to $8.8\cdot10^{-2}$ for plicRDF at $N = 32$ to $128$, ten times its static values on the same meshes): under transport on a mesh perturbed at the mesh scale the stencil normals see mesh-scale noise in $\alpha_k$. Two consequences: the face update on warped faces must carry the wetted area vector and its first moment, as \cref{sec:warped} states, and the plane-based fluxes the update assigns beyond the downwind faces lose boundedness there, which the content bound does not repair; and distorted-mesh transport is an open question for the framework as a whole, not for DAVOF alone. The hexahedral results above are untouched by this. The first consequence was implemented for the dual advection (branch \texttt{davof-warped} of the fork, pending merge): the face stores the wetted area vector $\mathbf{a}_f = \sum_t A^{\mathrm{wet}}_t \mathbf{n}_t$ over its fan triangles, three scalars that reduce to $\alpha_f \mathbf{S}_f$ on a planar face, and since every fan triangle contains the face centre the first moment is $(\mathbf{x}_f - \mathbf{x}_k)\cdot\mathbf{a}_f$ exactly, so no further quantity is stored; the update takes the pre-image's wetted area per fan triangle under the upwind surface. Two more pieces proved necessary on warped faces: the centroids that feed the ring's tensor fit must be re-anchored by the volume rule, because the library's plane cut closes the wet part of a warped face with a chord and the fit through those centroids returns a curved surface for a plane, and the flux prism must be fanned about the face centre, the cell's own fan. With all three the translated plane on hexahedra with vertices displaced by up to $0.1h$ stays at round-off with the exact normal and with the ring normal in $11$ of $12$ runs ($N = 20$, $40$, $80$, both arms; the twelfth, the ring row at $N = 80$ on the oblique arm, carries single-face jumps to $10^{-8}$ at four steps on faces a vertex of which grazes the plane, the configuration of the volume rule's open residue); with the scalar fraction the ring normal stays at $2\cdot10^{-3}$, flat in $N$. Two defects of the warped-face flux that the scalar version shared were found on the way (the orientation of a prism swept against its face normal, and the admissible range of a warped face's net liquid volume) and fixed. On the perturbed sphere the loop completes without abort but is not bounded ($\min\alpha$ down to $-3\cdot10^{-3}$ at $N = 64$, the anchor failing in up to $1163$ cells over a run) with either face state, the centroid error $2$ to $2.3$ times the hexahedral value and the shape error below it; the boundedness on perturbed meshes is the open item there.

\subsection{The sagitta correction within the plane loop}
\label{app:tpf-sagitta}

The clipping law makes the face fractions of the plane-clipped loop first order because the plane is $O(h^2)$ off the interface. The leading-order repair that stays within the plane loop corrects the clipped area of every face: the interface's trace on the face is an arc where the plane's trace is a chord, and the area between them follows from the ring curvature tensor of \cref{sec:curv}. With $\mathbf{B}_k$ the tensor in the frame of the normal, $\mathbf{M}_k$ the second moments of the cut polygon about its centroid, $z_0 = -\tfrac12 \mathbf{B}_k : \mathbf{M}_k$ the offset of the paraboloid's vertex from the volume-positioned plane (the paraboloid then encloses the plane's volume), and $w(l) = z_0 + \tfrac12\, \mathbf{x}(l) \cdot \mathbf{B}_k \mathbf{x}(l)$ the offset of the surface from the plane along the chord $\mathbf{x}(l)$, $l \in [0, L]$, the correction of the plane-clipped area of the face is $\Delta A_f = -(1/\cos\varphi) \int_0^L w \, \mathrm{d}l$, with $\cos\varphi$ the in-face component of the normal (Meusnier's theorem for the trace). It was implemented as a run-time-selectable face-clipping model of the fork (\texttt{faceClipping planeSagitta}, the plane clipping the default and bit-identical) and tested in three pre-registered steps. In pure geometry, on random cubes cut by a sphere with the exact tensor and the exact volume, it removes 92 to 97\,\% of the plane's face error at $h/R = 0.21$ to $0.05$. In the solver the fitted tensor is right on every interface cell (both tangential eigenvalues at $1.04$ to $1.07$ times $1/R$, the null vector along the normal, $z_0 < 0$ of the expected size, $N = 32$). Under transport it changes nothing: on the DA-Q state of \cref{tab:tpf-sphere} the exact-normal row ends at $L_1(\alpha_f) = 1.03\cdot10^{-2}$, $5.53\cdot10^{-3}$, $3.23\cdot10^{-3}$ (order $0.83$) against the plane clipping's $1.04\cdot10^{-2}$, $5.68\cdot10^{-3}$, $3.32\cdot10^{-3}$ ($0.82$), although it is applied on every cut face; the ring-normal loop is unchanged, and the own normal diverges with it as without it.

The per-face fields after one step locate the cause in the volume anchor. On the exact state of the sphere, whose $\alpha_k$ is the volume under the surface, the plane's face error is one-signed on every face (the secant plane overestimates the wetted area of every side face of a convex liquid region) and the correction removes $93\,\%$ of it at $\mathrm{Co} = 0.02$ ($L_1(\alpha_f)$ from $6.34\cdot10^{-3}$ to $4.40\cdot10^{-4}$ at $N = 32$) and $90\,\%$ at $\mathrm{Co} = 0.2$. On the DA-Q state every cut cell lies below the exact volume, by $8.4\cdot10^{-3}$ of $V$ on average, because its cell fraction is the chord volume of the Detrixhe--Aslam construction: the plane positioned by it sits inward and cancels most of the secant's overestimate, so the plane's face error is of mixed sign and less than half the exact state's, and the correction, right for the exact state, doubles it ($2.78\cdot10^{-3}$ to $6.22\cdot10^{-3}$). Started from the exact state, the corrected loop keeps the gain for one step and $27$ to $30\,\%$ of it at the end of the run, at the same first order ($8.42\cdot10^{-3}$, $4.33\cdot10^{-3}$, $2.58\cdot10^{-3}$, order $0.85$, against $1.15\cdot10^{-2}$, $6.24\cdot10^{-3}$, $3.64\cdot10^{-3}$): the volume fluxes are the volumes under the moving plane, whose defect against the surface is the sagitta volume of the swept face parts, so within the $1/\mathrm{Co}$ steps in which a cell holds the interface $\alpha_k$ acquires a one-signed $O(h/R)$ deviation from the volume under the surface, of the chord deficit's size, and the correction overshoots as it does on the chord state from the first step. This is the finding of \cref{sec:regen} under transport: a volume anchor is consistent only in the quadrature that defined the cell fraction. The requirement that follows is measured, not assumed. The volume flux and the face update must be the volume and the boundary moments of one surface per cell: the identities tie the volume fraction to the area vector, the zeroth moment of the boundary normal, and the position \cref{eq:avof-position} to the first moment of the wetted faces, and a second-order method transports both as the moments of one clipping surface, with that surface's own volume flux. That is the design of \cref{sec:transport}.

\subsection{The dual advection: quadrature, the three surfaces, the volume anchor}
\label{app:tpf-dual}

The requirement of the sagitta study, that the volume flux and the face update be the moments of one surface per cell, was implemented next: for every face of the interface cells and of their face neighbours, the volume flux of the step is the liquid volume of the prism swept between the face's pre-image $f - \mathbf{U}\Delta t$ and the face under the surface of the upwind cell, and the new face fraction is the pre-image's liquid fraction under the same surface. The surface is a quadric, so along any segment it is a quadratic polynomial whose roots are exact and on any planar triangle the quadratic Lagrange interpolant reproduces it, which makes the pre-image cut the exact face cut of \cref{sec:static} up to its Gauss rule. The prism volume is the integral over the face's triangles of the liquid fraction of the segment between a point and its pre-image, and the cell volume under a surface is the integral over the same triangles of the liquid part of the ray from the cell centre; both segment quantities are exact. Both integrals were first computed by one rule, the \emph{split quadrature}: the triangle is swept by the segments from one edge to the opposite vertex, every segment and the swept edge are split at their exact crossings with the surface's traces, and a 12-point Gauss rule is applied on each piece and refined adaptively to $10^{-11}$ of the prism or cell volume. The splitting matters because the integrands have kinks on the traces, and a rule that crosses a kink converges at second order only: an unsplit adaptive rule left a floor of $4\cdot10^{-6}$ of the cell volume, and a rule split at the traces but not refined within the pieces missed, in cells whose centre lies within a few $10^{-3} h$ of the surface, the layer of that width in which the rays cross the surface next to the trace, by up to $6\cdot10^{-5}$ of the cell volume per triangle. With both splittings and the adaptive refinement within the pieces, the rule reproduces the exact linear cut of a plane in every cell of the translated plane at $N = 40$ to $2\cdot10^{-14}$ of the cell volume, and one step of the update reproduces the exact face fractions to $2\cdot10^{-15}$ in $L_1$. Whatever anchors the surface by a volume uses the flux's own quadrature. A face shared by two interface cells takes the upwind cell's surface, consistent with its flux; a face whose two cells carry no surface takes the closest neighbouring one. Three surfaces were run on the translated sphere at $N = 32$ and $\mathrm{Co} = 0.2$ over 160 steps (3.3 diameters), from the exact-sphere initial state unless stated.

The exact sphere as every cell's surface is the gate of this machinery, not a test of the method: under uniform translation the pre-image's fraction under the sphere at $t^n$ is the exact fraction at $t^{n+1}$ and the swept volume is exact, so the update must reproduce the exact state. It does: the face fractions are exact to $10^{-14}$ from the second step on, except in the steps in which a grazing face falls from the quadratic cut to the linear one; the volume fractions are bounded to $10^{-17}$, conservative to $3\cdot10^{-17}$, and keep the initial state's volume unchanged, which is the exact one on the exact state (the initial cell fractions agree with the exact sphere--tetrahedron overlap of \citet{StroblFormellaPoeschel2016} to $7\cdot10^{-12}$ of the cell volume in $L_1$; the residuals of $3\cdot10^{-6}$ to $7\cdot10^{-9}$ that the ray quadrature left at finite depth were its own) and the chord deficit of the Detrixhe--Aslam cell fractions on the DA-Q state ($8\cdot10^{-3}$ of $V$ in $L_1$ at the first step and at the last). The gate thus also shows what an exact flux does with an inconsistent state: it transports the inconsistency.

The six-coefficient surface of \cref{sec:full}, the paraboloid whose offset, two slopes and curvature tensor are fitted to the cell's own face fractions with the ring tensor as the prior, is unstable under transport. Its first step is better than the plane loop's by a factor $20$ with the exact normal ($2.2\cdot10^{-4}$ against $4.4\cdot10^{-3}$ in $L_1(\alpha_f)$), $13$ with the own normal and $2.6$ with the ring normal; from then on the error grows by a factor $2$ to $2.5$ per step, before any bounding acts (the minimum of $\alpha$ stays at $10^{-19}$ until the fourth step), the fit begins to stall at the second to fourth step, and the run is stopped at the tenth to twenty-first step with ten times the initial number of interface cells. This happens on the exact initial state as on the DA-Q state, with every normal, at $\mathrm{Co} = 0.05$ as at $0.2$ (where the smaller step makes the growth per travelled cell length larger, not smaller: the amplification is per application of the map), and with the quadrature at $10^{-10}$ as at $10^{-8}$ (every printed digit agrees). The mechanism is the one the regeneration left open: the zero-velocity map is the identity on this surface, a neutral fixed point, and the displaced pre-images push it past one. The faces of a cell come from different surfaces, the upwind cell's on its upwind faces and its own on the downwind ones, so the six fractions of a cell are not the fractions of any one paraboloid, and a surface that interpolates them exactly carries their inconsistency to the next pre-images amplified. Shifting the offset after the fit so that the volume under the surface equals $\alpha_k V$, with the fitted slopes and tensor kept, changes nothing (the anchor's shift of the face fractions is $1.4\cdot10^{-3}$ at the first step on the exact state, the volume of the interpolating paraboloid against the sphere's); nor does the one-scalar face anchoring of \cref{sec:regen} with the ring tensor, which drifts and is stopped at the eighty-first step. The plane loop is stable with the ring normal for the reason that is now visible: its one degree of freedom beyond the normal, the position, comes from the conserved $\alpha_k$.

The surface that is stable takes nothing from the cell's own faces beyond the normal: the frame of the normal \cref{eq:avof-normal} (own or ring), the curvature tensor of the ring (\cref{sec:curv}), zero slopes, and the offset from the volume fraction alone, i.e.\ the position from the conserved $\alpha_k$ as in classical VOF, with a second-order surface in place of the plane and with the surface's own volume flux. With the exact normal it is bounded to $10^{-20}$ and conservative over the 160 steps, the surface defect stays at the anchor's tolerance, and the face-fraction error at the end is $5.7\cdot10^{-4}$ against the plane loop's $1.15\cdot10^{-2}$ on the same state, the normal recovered from the cell's own faces $2.8\cdot10^{-4}$ against $3.7\cdot10^{-3}$, the centroid $4.1\cdot10^{-4}$ against $7.7\cdot10^{-4}$; the maximum face error, $2.4\cdot10^{-2}$, sits on a few grazing cells where the anchor moves the surface far from the stored fractions. With the ring normal the face fractions end at $5.9\cdot10^{-3}$ against the plane loop's $1.78\cdot10^{-2}$, the centroid at $5.3\cdot10^{-4}$ against $1.35\cdot10^{-3}$, and the transported normal at $3.3\cdot10^{-2}$, the ring normal's own bias, as in the plane loop. With the own normal in every cell the loop diverges through the sliver cells as the plane loop does. The offset is found by the consecutive cubic spline positioning of \citet{Maric2021CCS} carried over to the paraboloid: the volume under the surface is a smooth function of the offset between the offsets at which the surface passes a vertex of the cell, its derivative is the area of the wetted patch projected along the normal, which the volume rule accumulates along the arcs it already walks, so each iterate is the root of the cubic Hermite interpolant through the current bracket with exact values and exact derivatives, re-bracketed and safeguarded by bisection; it closes the offset to $10^{-11} h$ in $3.4$ volume evaluations per cell and step on average, where a bracket and the Illinois rule needed $6.3$. A cell whose target volume is $0$ or $V_k$ to round-off while its face fractions still make it an interface cell has a whole interval of roots, every offset beyond the farthest vertex; its offset is defined as the edge of that interval, the surface tangent to the cell, the only choice consistent with its faces (an iteration that stops elsewhere in the interval changes the neighbouring fluxes and the ladder's digits at the $10^{-2}$ level, which is how the definition was found to be missing). The cost per step is reported with the ladders below.

\subsection{The volume rules}
\label{app:tpf-rules}

The split quadrature is exact on a plane, but not in a prism that contains the surface's silhouette, the curve along which the sweep segments touch the surface: there the integrand keeps a kink inside the pieces between the trace crossings, the adaptive refinement accepts before resolving it, and the value depends on the sweep direction, by $3$ to $5\cdot10^{-5}$ of the prism volume on boxes cut by a paraboloid and by $5\cdot10^{-6}$ on the equatorial prisms of the sphere. The liquid volumes are therefore computed by one of three run-time-selectable rules. The rule of the paraboloid ladders below is the one of \citet{KromerBothe2023Init}, built on \citet{KromerBothe2019}: by the divergence theorem applied to $\mathbf{x} - \mathbf{o}$, the liquid volume of a polyhedron under a quadric is the sum over its faces of the wetted area times the face's signed distance from $\mathbf{o}$ plus a line integral around the boundary of the liquid patch of the surface, which for a paraboloid in the frame of the normal is a cubic polynomial of the in-plane coordinates; the wetted area of a face follows from Green's theorem over the liquid pieces of its edges and the conic arcs of the surface's trace, each arc integrated adaptively as a graph over its chord with the exact root along the chord's normal. The rule has no refinement parameter in the volume, agrees with the exact volumes of 72 boxes cut by paraboloids, axis-aligned and tilted, to $10^{-15}$ of the box volume, and costs $0.21$ against $37.6\,\mathrm{s}$ per step at $N = 32$ for the split quadrature; a face whose trace is a degenerate conic, or a volume outside $[0, V]$, returns to the split quadrature at a capped depth and is counted. The exact-sphere control uses neither: its cell volumes and prism volumes are the exact sphere--tetrahedron overlaps of \citet{StroblFormellaPoeschel2016} (their reference implementation), the cell decomposed into the tetrahedra of its centre, a face centre and an edge, the prism of each face triangle into three, so the control is exact and not a quadrature; it costs $0.1$, $0.5$ to $0.8$ and $3$ to $6\,\mathrm{s}$ per step at $N = 32$, $64$ and $128$ against $121\,\mathrm{s}$ at $N = 128$ for the split quadrature.

\subsection{Parallel exchange}
\label{app:tpf-parallel}

The loop runs under OpenFOAM's domain decomposition. The 21 scalars of a cell's surface are exchanged across the processor patches once per step for the processor faces and once more for the closest second-ring surface that a face without a surface on either side falls back to; every processor face is computed on one side, the upwind side if it carries a surface, and copied; a face without flux takes the surface of the cell with the lexicographically smaller centre, in serial as in parallel. On the sphere at $N = 32$ and $\mathrm{Co} = 0.2$, four processors against one differ in the face fractions by at most $3.6\cdot10^{-7}$ over the 160 steps, which is the loop's own round-off floor: two serial runs whose sphere centre differs by $10^{-13}$ differ by $1.3\cdot10^{-10}$ after the first step, $2\cdot10^{-6}$ after the second and $10^{-9}$ to $2.4\cdot10^{-7}$ thereafter. The step at which the amplification enters, the anchor's bracket or the choice of the upwind surface on faces of vanishing flux, has not been traced.

\subsection{The exact-sphere control and the reference point of the normal error}
\label{app:tpf-control}

The exact sphere as every cell's surface holds the gate at every resolution and Courant number: the face fractions are exact to $10^{-14}$ from the second step on, the run maxima of $L_1(\alpha_f)$, $1.6\cdot10^{-10}$, $1.9\cdot10^{-11}$ and $2.4\cdot10^{-12}$ at $N = 32$, $64$ and $128$, are the first step's linear cuts of grazing faces and fall at third order, and are absent at $\mathrm{Co} = 0.5$, where no face grazes; the volume fractions stay above $-10^{-12}$, the volume is conserved to round-off, and the defect of the surface against the initialisation stays below $10^{-9}$ ($3\cdot10^{-11}$ to $9\cdot10^{-10}$ in $L_1$). The centroid and the shape error are second order ($1.98$) in every row, with whichever normal the row names, as a flux that does not use the normal must give. The normal recovered from the exactly transported fractions by \cref{eq:avof-normal} is the exact normal at the patch centroid, by the identity $\mathbf{m}_k = (A_k/R)(\mathbf{x}^*_k - \mathbf{c})$ on a sphere, so what any error norm on it measures is the offset of the reference point at which the exact normal is evaluated: the transport diagnostic reads $9.5\cdot10^{-4}$, $3.3\cdot10^{-4}$ and $1.4\cdot10^{-4}$ (order $1.4$) with the exact normal taken at the centroid of the volume-positioned plane, while the same transported fractions under the static definition of \cref{sec:static} give $6.7\cdot10^{-4}$ and $1.64\cdot10^{-4}$ at $R/h = 9.6$ and $19.2$, within $1\,\%$ of the static study's values at the same $R/h$ and of order $2.03$ on that pair; the end value is independent of the Courant number to nine digits, and at $R/h \le 19$ the three-rung order depends on where the sphere sits on the mesh by $\pm 0.5$ (the ladder's centre lies on a mesh vertex at every $N$). The exact-normal row's own nonzero reading ($8\cdot10^{-5}$ to $10^{-5}$) is the same effect, the exact normal compared with itself at two centroids. The ring normal on that state is first order ($0.98$), the mean normal of the ring.

\subsection{The ladders}
\label{app:tpf-ladders}

The ring-tensor paraboloid positioned by $\alpha_k$ was run over the resolutions and Courant numbers of \cref{tab:tpf-sphere} from the exact-sphere initial state, with the Kromer--Bothe volumes, and on the translated plane of \cref{tab:tpf-plane}. With the exact normal in the frame the face fractions are of order $2.84$ per period at $\mathrm{Co} = 0.2$ ($5.20\cdot10^{-4}$, $6.67\cdot10^{-5}$, $1.02\cdot10^{-5}$ at $N = 32$, $64$, $128$) and $2.87$ at $\mathrm{Co} = 0.5$, and of order $2.76$ and $2.82$ after the first step: the per-step order and the per-period order are the same, the error at the end of the run is $2.0$, $1.9$ and $1.8$ times the error after the first step at the three resolutions, a constant factor rather than the $O(1/h)$ accumulation that \cref{sec:transport} left open, and at $\mathrm{Co} = 0.5$ the error is stationary from the second step on. The order is above the second order that the clipping law gives for a surface $O(h^3)$ off the interface; at the measured $2.8$ the ring tensor on the Cartesian sphere is closer than its first-order bound (\cref{sec:curv} measured $1.55$ on the ellipsoid). The normal recovered from the transported fractions is second order ($2.05$, $1.12\cdot10^{-5}$ at $N = 128$), the centroid ($1.99$) and the shape error ($1.98$) as in the control, the volume fractions stay above $-1.1\cdot10^{-13}$, the volume anchor converges in every cell and its mean face shift falls from $1.75\cdot10^{-3}$ to $4.0\cdot10^{-5}$ over $N$, and the surface defect stays at $10^{-15}$. Against the plane loop's exact-normal row the face-fraction error at $N = 128$ is $1.02\cdot10^{-5}$ in place of $3.32\cdot10^{-3}$.

With the ring normal in the frame the face fractions are of order $1.35$ per period ($5.65\cdot10^{-3}$, $2.12\cdot10^{-3}$, $8.71\cdot10^{-4}$ at $\mathrm{Co} = 0.2$; $1.38$ and $1.29$ at $\mathrm{Co} = 0.1$ and $0.5$) and of order $1.05$ after the first step: the tilt of the ring normal, which is first order, enters the surface at the first step, and the period again adds a constant factor. The transported normal is the ring normal's own, first order at $8.19\cdot10^{-3}$ on the finest mesh, now below plicRDF's $9.20\cdot10^{-3}$ on the same state where the plane loop's was $1.09\cdot10^{-2}$; the centroid is second order ($2.09$) at $2.85\cdot10^{-5}$ against plicRDF's $1.04\cdot10^{-4}$ and the plane loop's $1.02\cdot10^{-4}$, and the shape error ($1.98$) is $2.30\cdot10^{-5}$ against $3.08\cdot10^{-5}$. The run is bounded to $-1.5\cdot10^{-13}$ and conservative at every resolution and Courant number; the anchor fails in at most 8 cells per step (at $N = 64$, $\mathrm{Co} = 0.1$, $0.5\,\%$ of the interface cells, more than the pre-registered $0.1\,\%$), and the volume rule returns to the split quadrature in at most 3 prisms per step. The Courant dependence has reversed against the plane loop: at fixed $N$ the error now rises with the Courant number ($8.52\cdot10^{-4}$, $8.71\cdot10^{-4}$, $1.22\cdot10^{-3}$ at $N = 128$ for $\mathrm{Co} = 0.1$, $0.2$, $0.5$), because the error no longer accumulates over the steps and the pre-image's displacement grows with the step, while the exact-normal row is flat in the Courant number. The cost was measured as for the plane loop, with the error diagnostics reduced to the end of the run and three repetitions per case, against plicRDF on the same initial state ($\mathrm{Co} = 0.2$, one core). A profile of the first implementation put $57\,\%$ of the advection step in the volume anchor (six Kromer--Bothe cell volumes per interface cell and step over 24 fan triangles), $14\,\%$ in two diagnostics of the anchor, $9\,\%$ in the prism volumes and $7\,\%$ in the face fractions, with the second-ring search, the ring normal and the tensor fit below $1\,\%$ each; that step cost $6$, $3.5$ and $3$ times plicRDF's at $N = 32$, $64$ and $128$. Three changes remove most of it without changing the numerics beyond round-off: planar faces are integrated whole and only warped faces fanned, decided per face by a planarity test; the diagnostics are off unless asked for; the cell polyhedron is built once per cell and the per-face allocations are gone. The fourth change is the anchor itself, described above, which needs $3.4$ volume evaluations per cell and step instead of $6.3$. The per-step medians of the ring-normal row are then $0.053$, $0.36$ and $2.1\,\mathrm{s}$ against plicRDF's $0.031$, $0.18$ and $1.8\,\mathrm{s}$ in the same study, $1.7$, $1.9$ and $1.2$ times plicRDF (the node scatter of these medians is about $20\,\%$); the remaining difference is the Kromer--Bothe volume of every prism of the faces of the interface cells and their neighbours where the plane loop has one plane cut per face. The three repetitions reproduce every error digit of the ladder's cases.

The order of the closed loop is thus the order of the normal fed back, and the ring normal's first order is the mean-normal bias of a patch of width $3h$, not an error of the identity. The ring quadric of \cref{sec:curv}, the weighted least-squares Monge patch through the ring's recovered planes that supplies the curvature tensor, also gives a normal at the cell, which is not a mean; with it in the frame, everything else as in the ring-normal row, the face fractions are of order $2.11$ per period at $\mathrm{Co} = 0.2$ ($8.75\cdot10^{-4}$, $1.37\cdot10^{-4}$, $4.73\cdot10^{-5}$; $2.24$ and $1.93$ at $\mathrm{Co} = 0.1$ and $0.5$) and $1.92$ after the first step, the transported normal is $1.63\cdot10^{-4}$ at $N = 128$, fifty times below the ring normal's and plicRDF's, at order $1.52$ (pairwise $1.89$, $1.15$, falling toward first order on the finer pair, which the face fractions follow: $2.68$, $1.53$), the centroid is second order ($2.00$) at $2.57\cdot10^{-5}$, the shape error is the control's, the run is bounded to $-2\cdot10^{-14}$, the fit converges in every interface cell of every step (no fallback to the ring normal, no anchor failure), and the cost is that of the ring-normal row. In the plane-clipped loop the same normal was erratic (fit failures in up to 1700 cells per step); under the surface's own flux and the volume anchor it is not. This row was added after the first two had been scored, with its predictions pre-registered before the run (bounded; normal below the ring normal's at order at least $1.5$; $q$ between the ring row's and the exact row's and at least $1.8$; fit fallbacks below $1\,\%$): all four held, the normal's order at the margin. A fourth mesh, $N = 256$ ($33.6$ million cells, 96 processors, 1280 steps, bounded and conservative, no fallback and no anchor failure), settles the falling pairwise order: from $N = 128$ to $256$ the ring quadric's row is of order $1.17$ in the face fractions ($2.10\cdot10^{-5}$ at $N = 256$) and $1.18$ in the normal ($7.2\cdot10^{-5}$), the ring's mean normal $1.16$ and $1.04$, the exact normal $2.30$ ($2.06\cdot10^{-6}$); a fit $c_1 h + c_2 h^2$ through the two finest meshes puts $88\,\%$ of the ring quadric's error at $N = 256$ in the first-order term, and the ratio of the two computed-normal rows is constant ($18.4$ and $18.6$). The computed-normal loop is therefore first order asymptotically, with the ring quadric's normal buying a constant eighteen times smaller than the ring mean's, not an order; second order over a period is reached only in the frame of the exact normal. (On 96 processors $7$ to $87$ of the $2\cdot10^5$ curved faces per step were left without a surface at processor corners, a gap of the parallel exchange absent in serial; against the serial digits at $N = 128$ its effect is $4\cdot10^{-5}$ to $6\cdot10^{-4}$ relative in the end values and $10^{-3}$ in the pairwise orders.)

\emph{The translated plane, and the defects found by it.} On the translated plane the ring tensor is zero to round-off, the paraboloid is the plane, and the loop must reproduce the exact fractions; the test is of the volume rule as much as of the loop. With the exact normal in the frame it does so in all 16 runs ($10^{-14}$ to $2.4\cdot10^{-13}$ for both arms, $N = 20$ to $160$, $\mathrm{Co} = 0.2$ and $0.5$), and with the ring normal in all 16 ($10^{-14}$ to $2.8\cdot10^{-13}$). It did not from the start: three defects of the Kromer--Bothe rule on flat traces were found by this test and fixed. A wrong root of the chord-normal quadratic on arcs whose side is round-off, and a line trace classified degenerate when the round-off curvature rotates the principal axes onto it, sent thousands of faces per step to the capped fallback and left the fractions at $10^{-6}$; a corner sliver arc at a vertex grazing the surface was dropped when a trace with round-off curvature was ordered as a conic with its centre $10^{12}$ face sizes away, which left the wetted area $10^{-4}$ off and two runs at $10^{-9}$ through the volume anchor (traces with curvature below $10^{-3}$ of the slope are ordered as lines since, arcs kept by their chord length). The test also caught the split quadrature's unmarked layer at a grazing sweep apex ($10^{-11}$ of a cell volume; the apex is the vertex farthest from the traces since). A self-test with planes at $10^{-6}$ to $10^{-14}$ of a box vertex against the exact volume, for both rules, is part of the code. The own normal fails the gate at every resolution beyond $N = 20$, $\mathrm{Co} = 0.2$ ($2\cdot10^{-8}$ at $N = 20$ and $\mathrm{Co} = 0.5$, $7\cdot10^{-2}$ at $N = 40$, up to $0.15$ at $N = 160$), with thousands of fallbacks per step on its wild surfaces, as in the plane-clipped loop. Every own-normal run starts at round-off, as the exact and ring rows, and grows geometrically from the second step, by $1.10$ to $1.26$ per step at $\mathrm{Co} = 0.2$ and $1.6$ to $8.8$ at $0.5$: the linear instability of an exact fixed point, independent of the clipping surface.

With the cell's own normal in the frame the loop diverges at every resolution and Courant number, stopped after 45 to 82 steps with $\alpha$ at the abort bound, hundreds to thousands of prisms per step falling back on its wild surfaces; with the own normal above the sliver threshold and the ring normal below it, it diverges in eight of nine runs (after 41 to 317 steps) and completes only at $N = 32$, $\mathrm{Co} = 0.2$, at a face-fraction error of $3.9\cdot10^{-2}$ and a normal error of $0.35$. The sliver rule is not enough: the amplification is the same under the paraboloid as under the plane clipping (\cref{tab:tpf-sphere}), and the whole loop needs the filtered normal. The pre-registered prediction that the sliver rule would be bounded at $N = 64$ and $128$ was falsified.

\subsection{The flux as a quadrature in time}
\label{app:tpf-flux}

With face fractions as state the volume flux through a face is also $\phi_f \int \alpha_f(t)\,\mathrm{d}t$ over the step; under translation the integrand is the fraction of the pre-image $f - \mathbf{U}(t - t^n)$ under the surface of $t^n$, and the prism, which the pre-images sweep, is that integral exactly. A quadrature of it from pre-image fractions at intermediate times would need no prism volume, and it is the form that extends to a pre-image spanning several cells, so four rules were measured on the plane and on the sphere's ring row at $N = 32$ and $64$ (run-time selectable, the prism the default). The trapezoid and Simpson's rule on the whole step fail the plane gate by $10^{-3}$ and $10^{-5}$ to $10^{-1}$: the trace passes a vertex of a swept face in nearly every step, so the quadratic of time the rules assume breaks almost always. Simpson's rule on the sub-intervals between those passages, which are roots of a quadratic along each vertex's trajectory, holds the gate at $N \le 40$ and on the oblique arm but not at $N \ge 80$ on the normal arm, because its end values are evaluated at the passages themselves, the degenerate configuration of the face integral, and its first value is the stored fraction. The three-point Gauss rule on the same sub-intervals evaluates no fraction at a passage and none from storage, every value being the pre-image's fraction under the one upwind surface, and it holds the gate at round-off in all 32 runs (maxima $2.4\cdot10^{-13}$ and $2.7\cdot10^{-13}$, the prism's own values on those cases) and completes the sphere's ring row bounded with the prism's digits to $10^{-3}$; Simpson's rule with the same two properties does the same, so the rule is immaterial and the two properties are what matters; the cost is $1.5$ to $1.8$ times the prism's either way. The three rules that use the stored fraction diverge on the sphere: under the volume-anchored surface the stored fraction differs from the face's fraction under that surface by the anchor's shift, so they integrate a mixture of two surfaces, which is the requirement of one surface per cell seen from the time side. The prism therefore remains the flux at Courant numbers below one; the Gauss rule on the breakpoint intervals is the consistent form for larger ones.

\bibliographystyle{elsarticle-num-names}
\bibliography{literature}

\end{document}